\documentclass[twocolumn]{article}

\usepackage[a4paper, total={7in, 9in}]{geometry}

\usepackage{graphicx} 
\usepackage[hidelinks,colorlinks=true,allcolors=blue]{hyperref}
\usepackage{url}
\usepackage[format=plain,
            labelfont={bf,it},
            textfont=it]{caption}
\usepackage{subcaption} 
\usepackage{amsmath}
\usepackage{amssymb}
\usepackage{amsthm}
\usepackage[dvipsnames]{xcolor}
\usepackage{dsfont}
\usepackage{algpseudocode}
\usepackage{algorithm}
\usepackage{mathtools}
\usepackage{multirow}
\usepackage{makecell}
\usepackage{adjustbox}
\usepackage{abstract}

\makeatletter
\newcommand*{\addFileDependency}[1]{
  \typeout{(#1)}
  \@addtofilelist{#1}
  \IfFileExists{#1}{}{\typeout{No file #1.}}
}
\makeatother

\theoremstyle{plain}
\newtheorem{theorem}{Theorem}[section]
\newtheorem{proposition}[theorem]{Proposition}

\theoremstyle{definition}
\newtheorem{definition}[theorem]{Definition}
\newtheorem{assumption}[theorem]{Assumption}

\newtheorem{remark}[theorem]{Remark}

\usepackage{bbold}

\newcommand\independent{\protect\mathpalette{\protect\independenT}{\perp}}
\def\independenT#1#2{\mathrel{\rlap{$#1#2$}\mkern2mu{#1#2}}}

\DeclareMathOperator*{\sign}{sign}

\DeclareMathOperator*{\E}{\mathbb{E}}

\usepackage[
    backend=biber,
    style=numeric-comp, 
    uniquename=mininit,
    sorting=none
  ]{biblatex}

\title{\textbf{Calibrating confounding strength in sensitivity models for weighting estimators: a comparative review and a new method}}

\author{
  Jean-Baptiste Baitairian\textsuperscript{1,2,3,*}
  \and
  Bernard Sebastien\textsuperscript{1}
  \and
  Rana Jreich\textsuperscript{1}
  \and
  Sandrine Katsahian\textsuperscript{2,3,4}
  \and
  Agathe Guilloux\textsuperscript{2}
}

\date{%
    \normalsize \textsuperscript{1}Sanofi, Gentilly, France\\%
    \textsuperscript{2}Inria, Université Paris Cité, Inserm, HeKA (UMR 1346), F-75015 Paris, France\\%
    \textsuperscript{3}Université Paris Cité, Paris, France\\%
    \textsuperscript{4}CIC-EC 1418 - Paris HEGP, Paris, France\\[1ex]%
    \textsuperscript{*}Corresponding author: jean-baptiste.baitairian@inria.fr\\[2ex]%
}

\newcommand{\abstractText}{\noindent
Causal inference is only valid when its underlying assumptions are satisfied, one of the most central being the ignorability or unconfoundedness assumption. However, this hypothesis is often unrealistic in observational studies, as some confounding variables may remain unobserved. To address this limitation, sensitivity models for Inverse Probability Weighting (IPW) estimators, known as Marginal Sensitivity Models, have been introduced, allowing for a controlled relaxation of ignorability. A substantial body of literature has emerged around these models, aiming to derive sharp and robust bounds for both binary and continuous treatment effects. A key element of these approaches is the specification of a sensitivity parameter, referred to as the ``confounding strength", which quantifies the extent of deviation from ignorability. Yet, determining an appropriate value for this parameter is challenging, and the final interpretation of sensitivity analyses can be unclear. We believe these difficulties represent major obstacles to the adoption of such methods in practice. Therefore, after introducing sensitivity analyses for IPW estimators, we review different strategies to estimate or lower bound the confounding strength, introduce a new method leveraging negative controls, provide a decision tree with guidelines to choose a suitable approach, and compare the methodologies in an in-depth simulation study.
}

\begin{document}


\twocolumn[
  \begin{@twocolumnfalse}
    \maketitle
    \begin{abstract}
      \abstractText
      \newline
      \newline
    \end{abstract}
  \end{@twocolumnfalse}
]



\section{Introduction}

Estimating the effect of a treatment or an exposure on a particular outcome is an important task in many domains, such as medicine, economics or finance. Examples of interest include the study of the effect of tranexamic acid on mortality in patients with traumatic bleeding~\parencite{colnet2024causal}, the influence of education on earnings~\parencite{card1999causal}, or the effect of the concentration of a pollutant in the air on cardiovascular mortality rate~\parencite{bahadori2022end}. In medicine, to evaluate the effect of a new drug, \textit{Randomized Controlled Trials} (RCTs) are usually considered the \textit{gold standard}~\parencite{hariton2018randomised}. As the treatment allocation is random, the distributions of patient covariates between treated and control arms are balanced. Therefore, estimated treatment effects on these data are robust to potential unobserved confounders, i.e.\ variables that have an effect both on the treatment and the outcome, and that can lead to a spurious causal effect between the two. However, due to their cost, duration, and low external validity~\parencite{benson2017comparison}, researchers are now turning their attention towards \textit{observational} or, more generally, \textit{real-world} data. Two important uses of these data include mimicking clinical trials (target trial emulation) or designing synthetic control arms \parencite{thorlund2020synthetic}. Nonetheless, observational studies cannot substitute RCTs in the evaluation of new drugs, as these treatments are not yet available on the market and therefore cannot be assessed through observational data. Real-world data typically offer bigger sample sizes, higher external validity and lower costs, as they are usually routinely collected, for instance, as Electronic Health Records (EHRs) in hospitals~\parencite{sauer2022leveraging}. Since the 21\textsuperscript{st} Century Cures Act, the United States Food and Drug Administration (FDA) has even been requested to develop a program to evaluate how real-world evidence could support medical product approvals~\parencite{fda2018framework}. Successively, the European Medicines Agency (EMA) proposed a guide on real-world evidence for regulatory decision-making~\parencite{ema2024guide}.

Causal inference methods are a powerful tool when one wants to estimate treatment effects with observational data. Their validity, however, relies on a set of assumptions. One of them is \textit{ignorability}, also called \textit{unconfoundedness}, or \textit{exogeneity}~\parencite{rubin1974estimating}. In the Neyman-Rubin potential outcome framework~\parencite{neyman1990application, rubin1974estimating}, this assumption posits that, conditionally on observed baseline covariates, treatment assignment is independent from the potential outcomes of a unit (see, later, Assumption~\ref{ass:X-ignorability}). Said differently, ignorability states that the treatment is assigned randomly among units that are similar in terms of observed confounders. In the binary treatment case, a non-exhaustive list of estimands that are computed under this assumption includes the \textit{Average Treatment Effect} (ATE), the \textit{Conditional Average Treatment Effect} (CATE), the \textit{expected potential outcome} (also called mean response) for the treated or for the control, and the \textit{Individual Treatment Effect} (ITE). For example, \textit{Inverse Probability Weighting} (IPW) or \textit{Augmented IPW} (AIPW) estimation approaches can be used to compute the ATE \parencite{hirano2003efficient, imbens2003sensitivity, glynn2010introduction}. In the continuous treatment case, one is usually interested in expected potential outcomes for all treatment values, namely the \textit{Average Potential Outcome} (APO), sometimes called \textit{dose-response function}, or the \textit{Conditional Average Potential Outcome} (CAPO). These quantities can be estimated thanks to IPW-like estimators~\parencite{imai2004causal, hirano2004propensity, kennedy2017non, kallus2018policy, colangelo2020double}, \textit{Bayesian Additive Regression Trees} (BART)~\parencite{hill2011bayesian}, or \textit{Adversarial CounterFactual Regression} (ACFR)~\parencite{kazemi2024adversarially}.

Departures from the ignorability assumption were already described in the seminal work of \textcite{cornfield1959smoking} on smoking and lung cancer, which demonstrated how unobserved confounding can distort causal conclusions. Indeed, ignorability is often unrealistic in observational studies, where unmeasured confounders and other types of biases may substantially affect estimated treatment effects. To address this challenge, researchers have proposed using \textit{sensitivity models} that relax ignorability by quantifying the possible impact of unobserved confounding with a parameter denoted $\Gamma$ that measures its strength on the treatment.
Rather than providing a single point estimate—--such as an IPW estimate of the ATE--—these approaches belong to the broader literature on partially identified sets, i.e., sets of treatment effect values that remain consistent with the chosen sensitivity model. In this review, we focus on sensitivity models that bound odds ratios of propensity scores in the binary treatment case--—namely the \textit{Marginal Sensitivity Model} (MSM)~\parencite{tan2006distributional, zhao2019sensitivity, dorn2022sharp, oprescu2023b, dorn2024doubly, tan2024model} and \textit{Rosenbaum’s Sensitivity Model} (RSM)~\parencite{rosenbaum2002observational, yadlowsky2022bounds}—--and bounded ratios of generalized propensity scores in the continuous treatment case, namely the \textit{Continuous Marginal Sensitivity Model} (CMSM)~\parencite{jesson2022scalable, baitairian2025sharp}. While we do not cover them here, attempts to unify these frameworks have recently been made through \textit{Generalized Sensitivity Models}~\parencite{bonvini2022sensitivity, frauen2024sharp}.

Unfortunately, these models are still underused, even if sensitivity analyses are often encouraged when using observational data, which makes them essential, even compulsory, to ensure the reliability of the conclusions. Still, the reader can refer to \textcite{beenstock2021plea}, \textcite{beenstock2021selection}, \textcite{jethani2022new}, \textcite{lopez2023trends}, and \textcite{ben2023varying} who performed sensitivity analyses with the MSM in fields such as criminology or health science. One reason for the limited use of these models that we identified, and that was also identified before us~\parencite{cinelli2020making}, is the difficulty to estimate with reliability the sensitivity parameter $\Gamma$ that bounds the ratio in the MSM or the CMSM. In particular, in the binary treatment case, this parameter represents the strength a confounder should have on the treatment to change the odds of being treated by at most a factor of $\Gamma$ among individuals that share similar observed characteristics. Regrettably, the majority of the aforementioned papers usually present only parts of the methods to estimate $\Gamma$, do not apply them, or use the most simple but flawed ones.

\paragraph{Contributions} This review helps practitioners in estimating or lower bounding the sensitivity parameter $\Gamma$, to make the MSM and CMSM more accessible and practical. Our key contributions include:
\begin{enumerate}
    \item A decision tree (Figure~\ref{fig:decision_tree}) and clear guidelines (Table~\ref{tab:methods_summary}) that explain when to use each estimation method, along with pros and cons;
    \item To our knowledge, the first working implementation of $\Gamma$ estimation using negative controls~\parencite{lipsitch2010negative}, turning a previously theoretical idea into a usable tool;
    \item A suggestion to extend the robustness value~\parencite{cinelli2020making} to simultaneous sensitivity analyses~\parencite{hsu2013calibrating}, inspired by the E-value used with risk ratios~\parencite{vanderweele2017sensitivity}.
\end{enumerate}

\paragraph{Outline} Section~\ref{sec:problem_setting} introduces notations and key concepts. Then, Section~\ref{sec:basic_approaches} presents basic strategies for choosing and interpreting the sensitivity parameter, together with important caveats. In Section~\ref{sec:advanced_approaches}, more advanced methods that aim to estimate, or at least lower bound, the sensitivity parameter using observed confounders, randomized controlled trials, or negative control outcomes are exposed. These advanced approaches are numerically compared in Section~\ref{sec:experiments_non-param} through an extensive simulation study. Finally, Section~\ref{sec:conclusion} concludes with a comparative table of the reviewed methods and a decision tree, intended as a practical guide to selecting the most suitable approach for a given application.

\section{Problem setting} \label{sec:problem_setting}

\subsection{General notations and assumptions} \label{sec:notations_and_assumptions}

In the following, we adopt the \textit{Neyman-Rubin potential outcome framework}~\parencite{neyman1990application, rubin1974estimating} and report to the books of~\textcite{neal2020introduction} and \textcite{wager2024causal} for background on the causal inference concepts used here.

We denote by $\mathbf{X} \in \mathcal{X} \subseteq \mathbb{R}^{p_\mathbf{X}}$, where $p_\mathbf{X} \geq 1$, the vector of observed confounders, $T \in \mathcal{T}$ the treatment or exposition, and $Y(t) \in \mathcal{Y} \subseteq \mathbb{R}$ the potential outcome for a treatment value $t \in \mathcal{T}$. If the treatment is binary, $\mathcal{T} = \{0,1\}$ and the two potential outcomes are $Y(0)$ and $Y(1)$, with the convention that $t=0$ represents control and $t=1$ represents treated. If the treatment is continuous, $\mathcal{T} \subseteq \mathbb{R}$ and, for a particular value of treatment $t \in \mathbb{R}$, the potential outcome is denoted $Y(t)$. We make the usual \textit{Stable Unit Treatment Value Assumption} (SUTVA), i.e.\ consistency of the treatment/no multiple versions ($Y = Y(T)$) and non-interference between units (the outcome of one unit is unaffected by the treatment assigned to other units). In the binary treatment case, it is usually written $Y = T Y(1) + (1-T) Y(0)$. Unless stated otherwise, estimated quantities are denoted with a hat $\hat \cdot$.

For a fixed (binary or continuous) treatment or exposition value $t \in \mathcal{T}$ and vector of confounders $\mathbf{x} \in \mathcal{X}$, we are interested in the (conditional) expected potential outcomes
\begin{align*}
    \theta(t) \coloneq \E[Y(t)] \text{ and } \theta(t, \mathbf{x}) \coloneq \E[Y(t) | \mathbf{X}=\mathbf{x}].
\end{align*}
In the binary treatment case, $\theta(t=0) = \E[Y(0)]$ and $\theta(t=1) = \E[Y(1)]$ are, respectively, the expected potential outcomes for the control and for the treated units, and $\theta(t=0, \mathbf{x}) = \E[Y(0) | \mathbf{X}=\mathbf{x}]$ and $\theta(t=1, \mathbf{x}) = \E[Y(1) | \mathbf{X}=\mathbf{x}]$ are their equivalent conditionally on $\mathbf{X}=\mathbf{x}$.
The Average Treatment Effect (ATE) and the Conditional ATE (CATE) are then defined as
\begin{align*}
    & \mathrm{ATE} \coloneq \theta(1) - \theta(0) = \E[Y(1) - Y(0)] \text{ and } \\
    & \mathrm{CATE}(\mathbf{x}) \coloneq \theta(1, \mathbf{x}) - \theta(0, \mathbf{x}) = \E[Y(1) - Y(0) | \mathbf{X}=\mathbf{x}].
\end{align*}
In the continuous treatment case, we are usually interested in the Average Potential Outcome (APO), or dose-response function, and Conditional APO (CAPO) defined as $\mathrm{APO}(t) \coloneq \theta(t)$ and $\mathrm{CAPO}(t, \mathbf{x}) \coloneq \theta(t, \mathbf{x})$.

The expected potential outcome $\theta(t)$ and conditional expected potential outcome $\theta(t, \mathbf{x})$ (and all derived quantities) can only be estimated under the following ignorability assumption for binary~\parencite{rubin1974estimating} or continuous treatments~\parencite{hirano2004propensity}.
\begin{assumption}[$\mathbf{X}$-ignorability] \label{ass:X-ignorability}
    For binary and continuous treatments, $\forall t \in \mathcal{T}, \, Y(t) \independent T \mid \mathbf{X}$.
\end{assumption}
Assumption~\ref{ass:X-ignorability} posits that, conditionally on observed confounders $\mathbf{X}$, treatment assignment can be \textit{ignored}. Therefore, enough covariates must be measured to capture the dependency between $T$ and the potential outcomes. Under this assumption, with binary treatments, the expected potential outcome for the treated $\theta(1)$ can be rewritten
\begin{equation}  \label{eqn:conf_theta_1}
    \theta(1) = \E \biggl[ \frac{TY}{e(\mathbf{X})} \biggr],
\end{equation}
where $e(\cdot) \coloneq \mathbb{P}(T=1|\mathbf{X}=\cdot)$ is called the (nominal) \textit{propensity score}~\parencite{rosenbaum1983central}.
Denote $\mathcal{D} = \{ (\mathbf{X}_i, T_i, Y_i) \}_{i=1}^{n}$ an i.i.d.\ sample of size $n$ drawn from a common distribution $\mathbb{P}^\mathrm{obs}$. With binary treatments ($T_i \in \{0,1\}$), an IPW estimator $\hat \theta(1)$ of the true $\theta(1)$ can be defined,
\begin{align} \label{eqn:conf_estim_theta_1}
    \hat \theta(1) = \frac{1}{n}\sum_{i=1}^n \frac{T_i Y_i}{\hat e(\mathbf{X}_i)},
\end{align}
where $\hat e(\cdot)$ is an estimate of $e(\cdot)$ obtained, for example, via logistic regression. With continuous treatments ($T_i \in \mathbb{R}$ and $t \in \mathbb{R}$), an IPW-like estimator $\hat \theta_h(t)$ of $\theta(t)$ has been proposed by~\textcite{kallus2018policy}:
\begin{align} \label{eqn:conf_estim_theta_h_t}
    \hat \theta_h(t) = \frac{1}{n} \sum_{i=1}^n \frac{K_h(T_i-t) Y_i}{\hat f(T=T_i | \mathbf{X}=\mathbf{X}_i)},
\end{align}
where $\hat{f}$ is an estimator of the density of $T$ conditionally on $\mathbf{X}=\mathbf{x}$, $t \mapsto f(T=t | \mathbf{X}=\mathbf{x})$, known as \textit{Generalized Propensity Score} (GPS)~\parencite{hirano2004propensity}, and $K_h$ is defined as $K_h(s) = K(s/h) / h$, where $h>0$ is a \textit{bandwidth} and $K$ is a \textit{kernel}. The purpose of $K_h$ is to localize the estimation around the treatment of interest $t$. See Appendix~\ref{app:assumptions} for details on the kernel.

As in \textcite{rosenbaum1983central}, we assume the classical positivity of the propensity scores, also called overlap, or common support, to ensure consistency of the IPW estimators from Equations~\eqref{eqn:conf_theta_1}, \eqref{eqn:conf_estim_theta_1} and \eqref{eqn:conf_estim_theta_h_t}. Assumption~\ref{ass:positivity} states that, conditionally on $\mathbf{X}=\mathbf{x}$, it is always possible to find units that received the binary or continuous treatment $t$ in the data.
\begin{assumption}[Positivity] \label{ass:positivity}
    For binary treatments, $\forall \mathbf{x} \in \mathcal{X}, \, 0 < e(\mathbf{x}) < 1$.

    \noindent For continuous treatments, $\forall (\mathbf{x}, t) \in \mathcal{X} \times \mathcal{T}, \, 0 < f(T=t | \mathbf{X}=\mathbf{x})$.
\end{assumption}
This can be considered a rather strong assumption, thus we refer the reader to \textcite{ju2019adaptive, leger2022causal} for examples of articles that tackle the issue of violations or near violations of positivity.

As previously noted, ignorability (Assumption~\ref{ass:X-ignorability}) is rarely satisfied in practice, since treatment assignment may depend on unobserved confounders $\mathbf{U} \in \mathcal{U} \subseteq \mathbb{R}^{p_\mathbf{U}}$, with $p_\mathbf{U} \geq 1$, which can bias the estimation of treatment effects.
An example of this situation can be found in \textcite{messerli2012chocolate}, where the author studied the link between chocolate consumption (the exposition) on cognitive function, and especially on the number of Nobel laureates per country (the outcome), and found a strong positive correlation. However, as correlation does not imply causation, he suspected unobserved confounders, such as differences in socioeconomic status between countries, to have an effect both on the exposition and the outcome. Therefore, it is more reasonable to consider that, had we observed confounders $\mathbf{U}$, we would have captured all common causes of $T$ and $Y(t)$ not included in $\mathbf{X}$~\parencite{imbens2003sensitivity, kallus2021causal}. This is formalized in the following assumption.
\begin{assumption}[($\mathbf{X}, \mathbf{U}$)-ignorability] \label{ass:XU-ignorability}
    For binary and continuous treatments, $\forall t \in \mathcal{T}, \, Y(t) \independent T \mid \mathbf{X}, \mathbf{U}$.
\end{assumption}
Under this new ($\mathbf{X}, \mathbf{U}$)-ignorability assumption, estimators~\eqref{eqn:conf_estim_theta_1} and \eqref{eqn:conf_estim_theta_h_t} become biased. Indeed, in the binary treatment case, had we observed $\mathbf{U}$, the \textit{true} propensity score would be $e(\mathbf{X}=\cdot, \mathbf{U}=\cdot) \coloneq \mathbb{P}(T=1|\mathbf{X}=\cdot, \mathbf{U}=\cdot)$, and we would have the following proposition (see Appendix~\ref{sec:proof_true_IPW_theta1} for a proof):
\begin{proposition} \label{prop:true_IPW_theta1}
    Under Assumptions~\ref{ass:positivity}, \ref{ass:XU-ignorability}, and SUTVA,
    \begin{equation} \label{eqn:true_ipw_theta_1}
        \theta(1) = \E \biggl[ \frac{TY}{e(\mathbf{X}, \mathbf{U})} \biggr],
    \end{equation}
    where $e(\mathbf{X}, \mathbf{U}) = \mathbb{P}(T=1|\mathbf{X}, \mathbf{U})$.
\end{proposition}
Note that we use the same notation $e$ here as for the nominal propensity score, which is a slight abuse of notations because it is a function of $\mathbf{X}$ only, whereas the true (unknown) propensity score depends on both $\mathbf{X}$ and $\mathbf{U}$. In the following, we will assume that all propensity scores (nominal, true, and their estimates) follow positivity (Assumption~\ref{ass:positivity}).

Nevertheless, as $\mathbf{U}$ is by definition never observed, we will not try to get \textit{a single} point estimation of our treatment effects, as in Equation~\eqref{eqn:true_ipw_theta_1}, but rather to get \textit{a set} of point estimates when the true propensity score, i.e.\ $e(\mathbf{X}, \mathbf{U})$ in the binary treatment case, is constrained to be ``close to" the nominal propensity score, i.e.\ $e(\mathbf{X})$ in the binary treatment case. This closeness is measured via sensitivity models that we introduce in Section~\ref{sec:sensitivity_models}. Then, in Section~\ref{sec:bounds}, we explain how they can be employed to get the lower and upper bounds of the aforementioned set of point estimates.

\subsection{Sensitivity models and sensitivity parameter \texorpdfstring{$\Gamma$}{}} \label{sec:sensitivity_models}

A sensitivity model can be used to evaluate the sensitivity of a result to the violation of an assumption. In our case, we would like to assess the sensitivity of the treatment effect estimates to a violation of ignorability (Assumption~\ref{ass:X-ignorability}). Many sensitivity models have been proposed in the literature for treatment effect estimation. \textcite{zhao2019sensitivity} give a non-exhaustive list of such models. In this review, we are interested in two sensitivity models that were originally developed for binary treatments, namely the \textit{Marginal Sensitivity Model} (MSM)~\parencite{tan2006distributional} and \textit{Rosenbaum's Sensitivity Model} (RSM)~\parencite{rosenbaum2002observational}, and their continuous counterpart, namely the \textit{Continuous Marginal Sensitivity Model} (CMSM)~\parencite{jesson2022scalable}.

\subsubsection{Sensitivity models for binary treatments} \label{sec:sensitivity_models_binary}

The Marginal Sensitivity Model~\parencite{tan2006distributional} assumes that units who appear similar in terms of observed confounders $\mathbf{X}$ may differ in their odds of being treated by at most a factor of $\Gamma \geq 1$, the confounding strength, if an unobserved confounder $\mathbf{U}$ was measured. To make reading easier in the following, we introduce the odds ratio function: $\mathrm{OR}(a,b) \coloneq (a / (1-a)) \, / \, (b / (1-b))$.
\begin{definition}[MSM, formulation with $\mathbf{U}$] \label{def:MSM_U}
    Let $\Gamma \geq 1$ be a fixed sensitivity parameter. Under positivity (Assumption~\ref{ass:positivity}), the MSM is defined as the set $\mathrm{MSM}_\mathbf{U}(\Gamma)$ of propensity scores such that
    \begin{align*}
        \mathrm{MSM}_\mathbf{U}(\Gamma) \coloneq \{& e(\cdot, \cdot); \; \forall (\mathbf{x}, \mathbf{u}) \in \mathcal{X} \times \mathcal{U}, \\
        & \Gamma^{-1} \leq \mathrm{OR}(e(\mathbf{x}, \mathbf{u}), e(\mathbf{x})) \leq \Gamma \}.
    \end{align*}
\end{definition}
In the special case $\Gamma = 1$, the MSM coincides with the ignorability assumption (Assumption~\ref{ass:X-ignorability}). A detailed discussion of the interpretation of $\Gamma$ in the general setting is provided in Section~\ref{sec:interpretation_gamma}.

In Appendix~\ref{app:other_sensitivity_models}, another formulation of the MSM is given. Indeed, as underlined by \textcite{zhao2019sensitivity}, \textcite{robins2002covariance} points out (Section 3) that it suffices to consider the case where the unobserved confounders $\mathbf{U}$ are replaced with one of the two potential outcomes $Y(0)$ or $Y(1)$.

On the other hand, Rosenbaum's Sensitivity Model~\parencite{rosenbaum2002observational} is more general than the MSM. Instead of only involving a true propensity score $e(\mathbf{x}, \mathbf{u})$, and a nominal propensity score $e(\mathbf{x})$, it involves two different propensity scores $e(\mathbf{x}, \mathbf{u}_1)$ and $e(\mathbf{x}, \mathbf{u}_2)$. The RSM is more adapted when the data are paired or grouped with matching, whereas the MSM is more appropriate when dealing with IPW estimators~\parencite{zhao2019sensitivity}. We refer the reader to Appendix~\ref{app:other_sensitivity_models} for formal definitions of the RSM and theoretical links between the RSM and the MSM, as we will mostly focus on the MSM in the rest of the paper.

\subsubsection{Sensitivity models for continuous treatments}

Under absolute continuity assumptions, \textcite{jesson2022scalable} extended the MSM to the continuous case using generalized propensity scores instead of binary propensity scores. A formulation in terms of unobserved confounders $\mathbf{U}$ is given in this section, while the original definition in terms of potential outcomes $Y(t)$ is recalled in Appendix~\ref{app:other_sensitivity_models}. In the following, we assume that, for all $(\mathbf{x}, \mathbf{u}) \in \mathcal{X} \times \mathcal{U}$, the conditional density $t \mapsto f(T=t | \mathbf{X}=\mathbf{x},\mathbf{U}=\mathbf{u})$ is absolutely continuous with respect to $t \mapsto f(T=t | \mathbf{X}=\mathbf{x})$.
\begin{definition}[CMSM, formulation with $\mathbf{U}$] \label{def:CMSM_U}
    Let $\Gamma \geq 1$ be a fixed sensitivity parameter. The CMSM is defined as the set $\mathrm{CMSM}_{\mathbf{U}}(\Gamma)$ of generalized propensity scores such that
    \begin{align*}
        & \mathrm{CMSM}_\mathbf{U}(\Gamma) \coloneq \Biggl\{ f(T=t | \mathbf{X}=\cdot, \mathbf{U}=\cdot); \\
        & \forall (\mathbf{x}, \mathbf{u}) \in \mathcal{X} \times \mathcal{U}, \, \Gamma^{-1} \leq \frac{f(T=t | \mathbf{X}=\mathbf{x}, \mathbf{U}=\mathbf{u})}{f(T=t | \mathbf{X}=\mathbf{x})} \leq \Gamma \Biggr\}.
    \end{align*}
\end{definition}
Definitions~\ref{def:CMSM_U} and \ref{def:CMSM_Y} can be viewed as the natural extensions of Definitions~\ref{def:MSM_U} and \ref{def:MSM_Y} to the continuous treatment setting. The key distinction is that the MSM is formulated in terms of odds ratios of propensity scores, whereas the CMSM relies on likelihood ratios of generalized propensity scores. A major advantage of sensitivity analyses based on these models is that they require only minimal assumptions. In particular, as underlined by \textcite{vanderweele2017sensitivity}, they do not impose the restrictive assumption of a binary unmeasured confounder~\parencite{cornfield1959smoking, bross1966spurious, rosenbaum1983assessing} or a unique unmeasured confounder~\parencite{bross1966spurious, schlesselman1978assessing, rosenbaum1983assessing, lin1998assessing, imbens2003sensitivity}, which makes them more broadly applicable.

\begin{figure*}[h]
    \centering
    \includegraphics[width=0.8\linewidth]{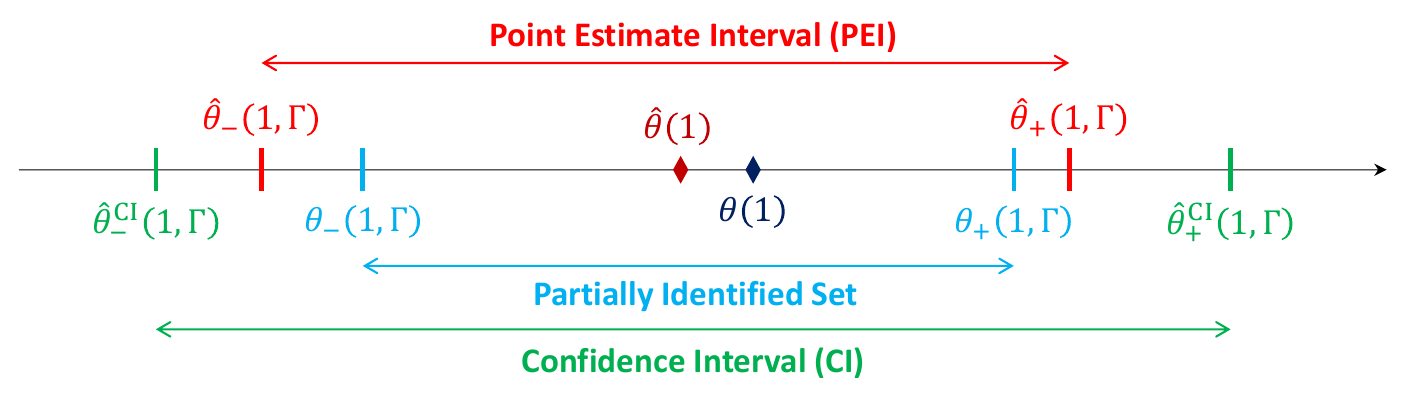}
    \caption{Illustration of the quantities involved in the sensitivity analysis for $\theta(1)$ (in dark blue), as defined by Equation~\eqref{eqn:true_ipw_theta_1}. $\hat \theta(1)$ (in dark red), as defined by Equation~\eqref{eqn:conf_estim_theta_1}, is the estimation of $\theta(1)$ under $\mathbf{X}$-ignorability. $\theta(1)$ is bounded by $\theta_-(1, \Gamma)$ and $\theta_+(1, \Gamma)$, the Partially Identified set (PI set) (in blue), as defined by Equation~\eqref{eqn:theta_1_PIS}. Estimating the PI set gives the Point Estimate Interval (PEI) $[\hat{\theta}(1, \Gamma)_-, \hat{\theta}(1, \Gamma)_+]$ (in red), as defined by Equation~\eqref{eqn:theta_1_PEI}. Finally, the PEI can be used to get a Confidence Interval (CI) (in green). When $\Gamma = 1$, the PI set reduces to $\theta(1)$ and the PEI reduces to $\hat{\theta}(1)$.}
    \label{fig:sa_bounds_illustration}
\end{figure*}

\subsubsection{Interpretation of the sensitivity parameter \texorpdfstring{$\Gamma$}{}} \label{sec:interpretation_gamma}

In the previous sensitivity models, it is legitimate to search for a plausible interpretation of the sensitivity parameter $\Gamma$, in order to be able to communicate on the results of a sensitivity analysis, or to choose a correct estimate $\hat \Gamma$ for the study. Whether the treatment is binary or continuous, $\Gamma$ measures the degree of violation of $\mathbf{X}$-ignorability, or equivalently, the strength of confounding:
\begin{itemize}
    \item If $\Gamma = 1$, the sensitivity model is equivalent to assuming $\mathbf{X}$-ignorability (Assumption~\ref{ass:X-ignorability}). In other words, in the binary treatment case, $e(\mathbf{x}, \mathbf{u}) \equiv e(\mathbf{x})$ and, in the continuous treatment case, $f(T=t | \mathbf{X}=\mathbf{x}, \mathbf{U}=\mathbf{u}) \equiv f(T=t | \mathbf{X}=\mathbf{x})$, as if there were no unobserved confounders.
    \item As $\Gamma$ increases, it quantifies how much $e(\mathbf{x}, \mathbf{u})$ differs from $e(\mathbf{x})$ in $\mathrm{MSM}_\mathbf{U}(\Gamma)$, or how much $f(T=t | \mathbf{X}=\mathbf{x}, \mathbf{U}=\mathbf{u})$ differs from $f(T=t | \mathbf{X}=\mathbf{x})$ in $\mathrm{CMSM}_\mathbf{U}(\Gamma)$. In other words, the assumed influence of unobserved confounders on treatment grows, leading to a departure from the standard ignorability assumption.
\end{itemize}

In the binary treatment case, $\Gamma$ has a natural interpretation. For example, in $\mathrm{MSM}_\mathbf{U}(\Gamma)$, it represents the maximum factor by which units who appear similar in terms of observed confounders $\mathbf{X}$ would differ in their odds of being treated if unobserved confounders $\mathbf{U}$ were omitted. However, even if such statements are important, the choice of an appropriate sensitivity parameter $\Gamma$ can remain obscure.

The interpretation in the continuous case is less straightforward, but seeing densities as infinitesimal probabilities leads to similar reasoning as for the binary treatment setting. \textcite{jesson2022scalable} sought to give two interpretations. The first one is based on the Kullback-Leibler (KL) divergence and treats the absolute value of the logarithm of $\Gamma$ as an upper bound on the KL divergence between $f(T=t | \mathbf{X}=\mathbf{x}, \mathbf{U}=\mathbf{u})$ and $f(T=t | \mathbf{X}=\mathbf{x})$ (see also \textcite{zhao2019sensitivity} for a similar link with the KL divergence in the binary setting). The second one is based on the ``proportion of unexplained range in $Y$ assumed to come from unobserved confounders after observing $\mathbf{X}$ and $T$" (Section~3.1 from~\cite{jesson2022scalable}). See Section~\ref{sec:critical_value_domain_knowl} for a mathematical formulation of this second point.

\subsection{Bounds on treatment effect estimates with unobserved confounders} \label{sec:bounds}

In the potential presence of unmeasured confounders, and to assess the robustness of their causal estimates, researchers have proposed bounding the classical point estimates of treatment effects (ATE, CATE, APO, CAPO, etc.)---which are only valid under $\mathbf{X}$-ignorability---to obtain partially identified sets that remain valid under the sensitivity models introduced in Section~\ref{sec:sensitivity_models}. We now present this approach.

As an example, recall that, according to Proposition~\ref{prop:true_IPW_theta1}, under Assumption~\ref{ass:XU-ignorability}, the expected potential outcome for the treated $\theta(1)$ can be identified by
\begin{equation*}
    \theta(1) = \E \biggl[ \frac{TY}{e(\mathbf{X}, \mathbf{U})} \biggr],
\end{equation*}
but, by definition, $\mathbf{U}$ is unobserved. \textcite{zhao2019sensitivity} therefore bounded the unknown true propensity score $e(\mathbf{X}, \mathbf{U})$ using the sensitivity model from Definition~\ref{def:MSM_U}. Later, \textcite{dorn2022sharp} refined these bounds with an additional constraint on the propensity scores, called Quantile Balancing (QB) constraint. We recall here this second formulation.\footnote{The QB constraint involves conditional quantiles, hence its name. We recall here another formulation given in \textcite{dorn2022sharp} that does not directly involve these quantiles.}

First, let us rewrite the MSM in terms of random variables (r.v.):
\begin{align*}
    \mathrm{MSM}(\Gamma) \coloneq \{ & E \text{ r.v. with values in } (0,1); \\
    & \Gamma^{-1} \leq \mathrm{OR}(E, e(\mathbf{X})) \leq \Gamma \text{ with proba. 1} \}.
\end{align*}
$E$ must also satisfy the QB constraint to ``look like" a propensity score. Thus, the lower and upper bounds for $\theta(1)$ are then respectively defined as
\begin{align} \label{eqn:theta_1_PIS}
    & \theta(1, \Gamma)_- = \underset{E \in \mathrm{MSM}(\Gamma)}{\min} \E[TY/E] \text{ s.t. $\E[T/E|\mathbf{X}] = 1$} \quad \text{and} \nonumber \\
    & \theta(1, \Gamma)_+ = \underset{E \in \mathrm{MSM}(\Gamma)}{\max} \E[TY/E] \text{ s.t. $\E[T/E|\mathbf{X}] = 1$},
\end{align}
for a particular value of the sensitivity parameter $\Gamma$.

The optimization problems from Equation~\eqref{eqn:theta_1_PIS} have analytical solutions given, respectively, by
\begin{align*}
    E_- & = \biggl( 1 + \frac{1-e(\mathbf{X})}{e(\mathbf{X})} \Gamma^{- \sign(Y - Q_{1-\gamma}(\mathbf{X}, 1))} \biggr)^{-1} \quad \text{and} \\
    E_+ & = \biggl( 1 + \frac{1-e(\mathbf{X})}{e(\mathbf{X})} \Gamma^{\sign(Y - Q_{\gamma}(\mathbf{X}, 1))} \biggr)^{-1},
\end{align*}
with $\gamma = \Gamma / (1 + \Gamma)$ and $Q_{\nu}(\mathbf{X}, T)$, the $\nu$-quantile of the distribution of $Y$ conditionally on $\mathbf{X}$ and $T$. Notice that, while $E_\pm(\mathbf{X}, T, Y, \Gamma)$, we dropped this dependency in the notations.

\textcite{oprescu2023b, dorn2024doubly, tan2024model} consecutively improved the previous results, in particular, by providing bounds that are more robust to model misspecifications. Under the RSM (Definitions~\ref{def:RSM_U} and \ref{def:RSM_Y}), \textcite{rosenbaum2002observational, yadlowsky2022bounds} proposed similar bounding approaches. In the continuous treatment case, \textcite{jesson2022scalable, baitairian2025sharp} suggested bounds for the APO and CAPO under the CMSM (Definitions~\ref{def:CMSM_U} and \ref{def:CMSM_Y}).

Using an i.i.d. sample $\mathcal{D}$, as introduced in Section~\ref{sec:notations_and_assumptions}, we can estimate $\theta(1, \Gamma)_-$ and $\theta(1, \Gamma)_+$ to get a \textit{Point Estimate Interval} (PEI) $[\hat{\theta}(1, \Gamma)_-, \, \hat{\theta}(1, \Gamma)_+]$, such that
\begin{align} \label{eqn:theta_1_PEI}
    & \hat{\theta}(1, \Gamma)_- = \frac{1}{n} \sum_{i=1}^n T_i Y_i / \hat{E}_{-, i} \quad \text{and} \nonumber \\
    & \hat{\theta}(1, \Gamma)_+ = \frac{1}{n} \sum_{i=1}^n T_i Y_i / \hat{E}_{+, i},
\end{align}
where $\hat{E}_{-, i}$ and $\hat{E}_{+, i}$ are estimators of $E_-$ and $E_+$ for individual $i$, with $e$ and $Q_\nu$ replaced with $\hat{e}$ and $\hat{Q}_\nu$, respectively.

In practice, Equation~\eqref{eqn:theta_1_PIS} is solved for a normalized or stabilized IPW (SIPW) estimator, i.e.\ the expectation from Equation~\eqref{eqn:theta_1_PIS} is divided by $\E[T/E]$ or $\E[T/e(\mathbf{X})]$~\parencite{dorn2022sharp}. \textcite{zhao2019sensitivity} recall the advantage of using the SIPW estimator compared to the IPW estimator. In particular, the SIPW estimator is sample bounded and so are the estimated bounds. Therefore, for all $\Gamma \geq 1$, including $\Gamma = +\infty$,
\begin{align*}
    \bigl[ \hat{\theta}(1, \Gamma)_-, \, \hat{\theta}(1, \Gamma)_+ \bigr] \subseteq \Bigl[ \underset{i; T_i = 1}{\min} Y_i, \, \underset{i; T_i = 1}{\max} Y_i \Bigr].
\end{align*}

Finally, confidence intervals (CIs) are obtained via the \textit{percentile bootstrap} method (see~\cite{zhao2019sensitivity, dorn2022sharp, jesson2022scalable}), or using Wald-type confidence intervals (see, for instance,~\cite{dorn2024doubly}). We denote confidence intervals with a "CI" in exponent.
For instance, $\hat{\theta}(1, \Gamma)_-^\mathrm{CI}$ and $\hat{\theta}(1, \Gamma)_+^\mathrm{CI}$ are, respectively, the lower and upper bounds of the confidence interval obtained on the point estimate interval $[\hat{\theta}(1, \Gamma)_-, \, \hat{\theta}(1, \Gamma)_+]$.
Figure~\ref{fig:sa_bounds_illustration} gives a visual representation of the different quantities that are involved in the sensitivity analysis.

In the following, we will refer indistinctly to the CI or the PEI as \textit{sensitivity bounds}. It is straightforward to notice that, the more $\Gamma$ increases, the more the set of possible propensity scores that are compatible with the sensitivity model grows, and the minimum and maximum values in this set get, respectively, lower and larger. Therefore, as $\Gamma$ grows, the PEI increases and, for $1 \leq \Gamma_1 \leq \Gamma_2$, we have the following inclusion
\begin{equation*}
    [ \theta(1, \Gamma_1)_-, \, \theta(1, \Gamma_1)_+ ] \subseteq [ \theta(1, \Gamma_2)_-, \, \theta(1, \Gamma_2)_+ ].
\end{equation*}
For this reason, we are interested in the lowest possible $\Gamma$ associated with $\mathbf{U}$ because, if the true $\Gamma$ was equal to $\Gamma_1$, then $\Gamma_2$ would also satisfy the sensitivity model but would lead to more conservative bounds.

As it is the case with any parameter, the question of the choice of the right sensitivity parameter $\Gamma$ arises when one wants to use the previous methods to get bounds on the treatment effect. We noticed that the difficulty to estimate $\Gamma$ was probably the main reason that explained why the aforementioned sensitivity models were underused. In Sections~\ref{sec:basic_approaches} and \ref{sec:advanced_approaches}, we therefore discuss different tools to \textit{calibrate} the sensitivity parameter from these models.

\section{Caveats and basic approaches} \label{sec:basic_approaches}

\subsection{Why trying to estimate \texorpdfstring{$\Gamma$}{} could not be reasonable} \label{sec:not_reasonable}

\begin{figure*}[ht]
    \centering
    \includegraphics[width=\linewidth]{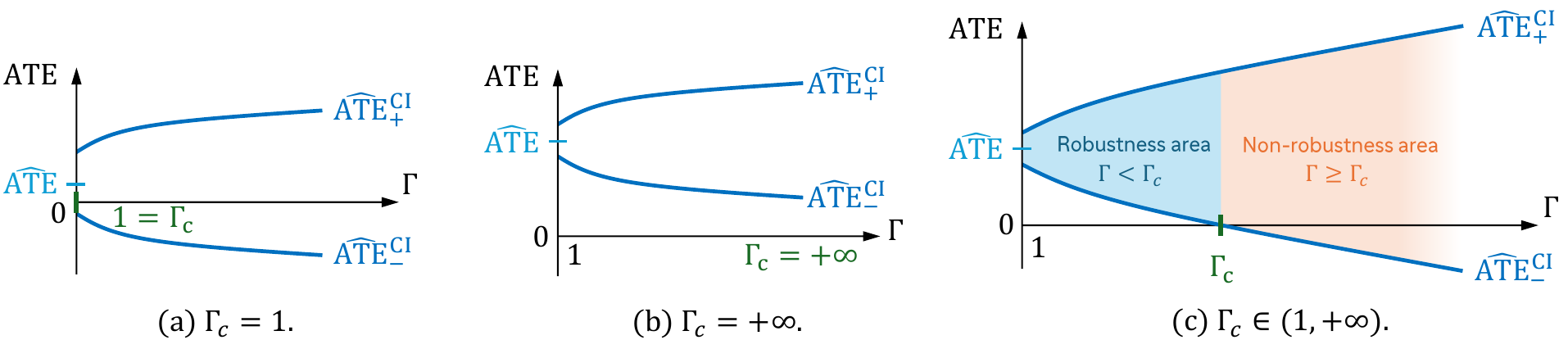}
    \caption{Different configurations of sensitivity analysis for the ATE. Figure~\ref{fig:sa_critical_value_ATE}a depicts the case where the critical value $\Gamma_c = 1$ and the confidence intervals (in blue) contain the null effect for all values of $\Gamma$ (no robustness to unobserved confounders). Figure~\ref{fig:sa_critical_value_ATE}b shows the opposite case where $\Gamma_c = +\infty$ and no confidence interval contains the null effect (robustness to unobserved confounders of any strength). Figure~\ref{fig:sa_critical_value_ATE}c corresponds to the intermediate case, where $\Gamma_c \in (1, +\infty)$. If the estimated $\hat{\Gamma}$ is lower than $\Gamma_c$, the conclusions are robust to unobserved confounders (blue area), but if it is larger, the conclusions may not be robust to possible unobserved confounders (orange area).}
    \label{fig:sa_critical_value_ATE}
\end{figure*}

The first question to consider when we aim at estimating or choosing a sensitivity parameter $\Gamma$ is whether or not this could be a reasonable or feasible task. By trying to estimate $\Gamma$, we are making the untestable assumption that $\Gamma < + \infty$~\parencite{kallus2018confounding} (see, for example, Figure~\ref{fig:sa_critical_value_ATE}c). However, if absolutely nothing was known about the unobserved confounders, then the only right option would be to choose $\Gamma = + \infty$, which would lead to extremely conservative sensitivity bounds. Nonetheless, we could still conclude about the robustness of the conclusions to unobserved confounders in two particular cases: if the so-called \textit{critical value}, as presented in the next subsection, is 1 or $+\infty$.

If one still decides to estimate $\Gamma$, \textcite{scharfstein1999adjusting, robins2002covariance} underlined the fact that the sensitivity parameter that appears in sensitivity models from Section~\ref{sec:sensitivity_models_binary} depends on the observed confounders $\mathbf{X}$, and carefulness is needed if one tries to use the results from another study to infer the value of $\Gamma$, i.e.\ the same observed covariates should be considered.

Moreover, \textcite{robins2002covariance} explained that $\Gamma$ is a \textit{paradoxical measure}, i.e.\ the magnitude of $\Gamma$ can increase as we decrease the amount of hidden bias by measuring some of the unmeasured confounders. Said differently, imagine that we had observed $\mathbf{X} = (X^{(1)}, \dots, X^{(10)})$, a vector of observed confounders of size $p_\mathbf{X} = 10$, and that we would not have observed $\mathbf{U} = (U^{(1)}, U^{(2)})$, a vector of unobserved confounders of size $p_\mathbf{U} = 2$, and consider the sensitivity model $\mathrm{MSM}_\mathbf{U}(\Gamma)$ from Definition~\ref{def:MSM_U}. The odds ratio from this sensitivity model involves a true propensity score that depends on both the observed and unobserved confounders $\mathbf{X}$ and $\mathbf{U}$, namely $e(\mathbf{X}, \mathbf{U})$, and a nominal propensity score that only depends on the observed confounders, namely $e(\mathbf{X})$. Imagine now that new data are available and that we are able to observe $U_2$. Our new vector of observed confounders becomes $\mathbf{X}^\prime = (X^{(1)}, \dots, X^{(10)}, U^{(2)})$, while the vector of unobserved confounders reduces to $\mathbf{U}^\prime = U_1$. The true propensity score $e(\mathbf{X}^\prime, \mathbf{U}^\prime)$ is still involving both observed and unobserved confounders, and it stays equal to $e(\mathbf{X}, \mathbf{U})$, but the nominal propensity score $e(\mathbf{X}^\prime)$ now depends on all previous observed confounders from the vector $\mathbf{X}$ \textit{and} on $U_2$, and it is not equal to $e(\mathbf{X})$. Therefore, between the first and the second case, the sensitivity parameter $\Gamma$ has changed, and nothing tells us that it has decreased by observing one additional confounder. In conclusion, a common expectation that additional information on unobserved confounders should lower the sensitivity parameter may be misleading and lead to misinterpretation.

\begin{figure*}[ht]
    \centering
    \includegraphics[width=\linewidth]{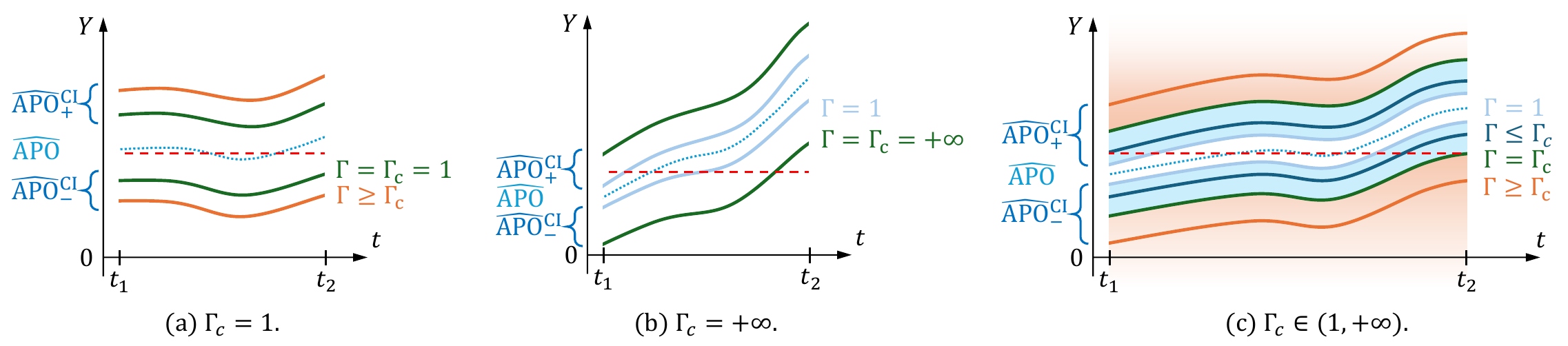}
    \caption{Different configurations of sensitivity analysis for the APO. The dotted line in light blue is the estimated APO curve. The red dotted line corresponds to the null effect. The critical value $\Gamma_c$ is defined as the lowest value of $\Gamma$ for which all intervals between expositions $t_1$ and $t_2$ contain the null effect (dark green curve). Figure~\ref{fig:sa_critical_value_APO}a depicts the case where the critical value $\Gamma_c = 1$ and the confidence intervals between $t_1$ and $t_2$ contain the null effect for all values of $\Gamma$ (no robustness to unobserved confounders). Figure~\ref{fig:sa_critical_value_APO}b shows the opposite case where $\Gamma_c = +\infty$ and the null effect is not in all confidence intervals between $t_1$ and $t_2$ (robustness to unobserved confounders of any strength). Figure~\ref{fig:sa_critical_value_APO}c corresponds to the intermediate case, where $\Gamma_c \in (1, +\infty)$. If the estimated $\hat{\Gamma}$ is lower than $\Gamma_c$, the conclusions are robust to unobserved confounders (blue curves), but if it is larger, the conclusions may not be robust to possible unobserved confounders (orange curves).}
    \label{fig:sa_critical_value_APO}
\end{figure*}

\subsection{Critical value \texorpdfstring{$\Gamma_c$}{}, range of \texorpdfstring{$\Gamma$}{}, and use of domain knowledge} \label{sec:critical_value_domain_knowl}

Having thought about the different questions that arise from Section~\ref{sec:not_reasonable} and deciding to estimate $\Gamma$, a user of the different methods described in Section~\ref{sec:bounds} could start by computing the sensitivity bounds corresponding to the treatment effect of interest (ATE, CATE, APO, CAPO...) for different values of the sensitivity parameter $\Gamma$~\parencite{tan2006distributional, kallus2018confounding, zhao2019sensitivity, jesson2021quantifying, dorn2022sharp, yadlowsky2022bounds, jesson2022scalable, dorn2024doubly}, and use domain knowledge~\parencite{robins2002covariance, oprescu2023b, frauen2024sharp} to choose a plausible value of $\Gamma$ that aligns with expert a priori.

To know if a value of $\Gamma$ is plausible, a practitioner using model from Definition~\ref{def:MSM_U} should ask the question:
\begin{quote}
    \textit{``Could the odds of being treated change by as much as this particular value of $\Gamma$ if similar individuals differed by an unobserved confounder?"}
\end{quote}
This is analogous to the seminal discussion of \textcite{cornfield1959smoking} about the existence of a potential unobserved confounder of strength 9 on the risk ratio scale when studying the effect of smoking on lung cancer.

This method, although simple, may be inefficient when no expert a priori is available, if the expert a priori is biased, or in the continuous treatment setting, because giving an intuition about the magnitude of change in a density ratio can be less intuitive than the same task for an odds ratio. However, in this last case, \textcite{jesson2022scalable} suggest to report values of $\Gamma$ for which a certain proportion $\rho$ of unexplained range in $Y$ is assumed to come from $\mathbf{U}$ after observing $\mathbf{x}$ and $t$. For the CAPO, this proportion is defined as
\begin{align*}
    \rho(t, \mathbf{x}, \Gamma) \coloneq \frac{\hat{\theta}(t, \mathbf{x}, \Gamma)_+ - \hat{\theta}(t, \mathbf{x}, \Gamma)_-}{\hat{\theta}(t, \mathbf{x}, \infty)_+ - \hat{\theta}(t, \mathbf{x}, \infty)_-},
\end{align*}
where $\hat{\theta}(t, \mathbf{x}, \infty)_\pm$ is the limit of the upper or lower bound for $\widehat{\mathrm{CAPO}}(t, \mathbf{x})$ when $\Gamma \to \infty$. Typical values of $\rho$ for which $\Gamma$ can be reported include $0.25$, $0.50$, $0.75$ and $0.90$. This approach has some similarities with the partial $R^2$ method of \textcite{imbens2003sensitivity, cinelli2020making} presented in Section~\ref{sec:param_approaches}.

Among all the values that $\Gamma$ can take, one is of particular interest: the \textit{critical}, or \textit{cutoff}, \textit{value} $\Gamma_c$~\parencite{tan2006distributional, vanderweele2017sensitivity, jesson2021quantifying, yadlowsky2022bounds, dorn2022sharp, dorn2024doubly, frauen2024sharp, de2024hidden, baitairian2025sharp}. This value is defined as the minimum confounding strength for which the null effect, or any value of interest (such as in non-inferiority trials), is in the bounds of the sensitivity analysis (PEI or CI, depending on the choice of the user).

In Figures~\ref{fig:sa_critical_value_ATE} and \ref{fig:sa_critical_value_APO}, we present different configurations of sensitivity analyses when $\Gamma_c$ evolves between 1 and $+ \infty$. In the binary treatment case, the critical value:
\begin{itemize}
    \item is 1 (Figure~\ref{fig:sa_critical_value_ATE}a) if the interval already contains the null effect for $\Gamma = 1$, in which case, the sensitivity analysis shows that the estimation of treatment effect is \textit{not robust} to unobserved confounders and the conclusions of a study could be altered by unobserved confounders of any strength;
    \item is $+\infty$ by convention (Figure~\ref{fig:sa_critical_value_ATE}b) if the bounds never contain the null effect, in which case the estimation is \textit{strongly robust} to unobserved confounders because the conclusions of a study will never be altered by unobserved confounders, no matter what value they take;
    \item can lie strictly between 1 and $+\infty$ (Figure~\ref{fig:sa_critical_value_ATE}c). In this case, either we are not able to estimate $\Gamma$ because of a lack of data, so we must discuss if $\Gamma_c$ is a plausible value for the unknown confounding strength, or we are able to estimate $\Gamma$ (using methods presented further), so we can compare our estimate $\hat \Gamma$ to $\Gamma_c$:
    \begin{itemize}
        \item if $\hat \Gamma < \Gamma_c$, then the conclusions of a study are reasonably \textit{robust} to unobserved confounders, because the sensitivity interval does not contain the null effect;
        \item if $\hat \Gamma \geq \Gamma_c$, then the conclusions of a study are \textit{not robust} to unobserved confounders, because the sensitivity interval contains the null effect, so it is impossible to say if an effect of the treatment/exposition on the outcome exists or not.
    \end{itemize}
\end{itemize}

In the continuous treatment case, a subtlety is to define properly the null effect. One suggestion is given in \textcite{baitairian2025sharp}, where the null effect is defined as the case where the relationship between the outcome and the treatment is a constant, i.e., the outcome is not influenced by the treatment. In particular, the authors choose the special case where this constant is equal to the mean of the observed outcomes. Then, the critical value $\Gamma_c$ is defined as the minimum confounding strength for which all sensitivity intervals between two predefined values of treatment/exposition $t$, say $t_1$ and $t_2$, contain the null effect (Figure~\ref{fig:sa_critical_value_APO}c). In that case, $\Gamma_c$ becomes a function of $t_1$ and $t_2$. Of course, $\Gamma_c = 1$ corresponds to the case where the null effect is in all sensitivity intervals between $t_1$ and $t_2$, no matter the confounding strength $\Gamma$, and the conclusion is the same as in the binary case (Figure~\ref{fig:sa_critical_value_APO}a). Similarly, $\Gamma_c = +\infty$ corresponds to the case where the null effect is never inside all the sensitivity intervals between $t_1$ and $t_2$, no matter the confounding strength $\Gamma$ (Figure~\ref{fig:sa_critical_value_APO}b).

\section{Advanced approaches} \label{sec:advanced_approaches}

Beyond the approaches described in Section~\ref{sec:basic_approaches}, more advanced methods can be employed to estimate the sensitivity parameter $\Gamma$. These methods offer the advantage of being non-parametric, i.e.\ they require no assumptions about the distributions of treatment, outcome, or observed/unobserved confounders, nor the specification of a parametric model linking them. However, a brief discussion of parametric methods to estimate confounding strength is deferred to Section~\ref{sec:param_approaches}.

\subsection{Informal benchmarking}

\begin{algorithm*}[h]
    \caption{Informal benchmarking (leave-one-out procedure)} \label{alg:informal_benchmarking}
    \begin{algorithmic}
        \Require Observed confounders $\{ \mathbf{X}_i \}_{i=1}^n = \bigl\{ \bigl(X_i^{(1)}, \ldots, X_i^{(p_\mathbf{X})} \bigr) \bigr\}_{i=1}^n$. For binary treatments, an estimate $\hat e$ of the propensity score. For continuous treatments, an estimate $\hat f(T|\mathbf{X})$ of the generalized propensity score, and the treatment of interest $t$.
        \For{$i \in \{1, \ldots, p_\mathbf{X}\}$}
            \For{$j \in \{1, \ldots, n\}$}
                \If{treatment is binary}
                    \State Compute $\hat r_{i,j} \coloneq \mathrm{OR}(\hat e(\mathbf{X}_j), \hat e(\mathbf{X}_j^{(-i)}))$.
                    \ElsIf{treatment is continuous}
                    \State Compute $\hat r_{i,j} \coloneq \hat f(T=t|\mathbf{X}=\mathbf{X}_j) / \hat f(T=t|\mathbf{X}=\mathbf{X}_j^{(-i)})$.
                \EndIf
            \EndFor
            \State Compute $\hat \Gamma_i^+ \coloneq \underset{j}{\max}(\hat r_{i,j})$ and $\hat \Gamma_i^- \coloneq 1 / \underset{j}{\min}(\hat r_{i,j})$.
            \State Compute $\hat \Gamma_i \coloneq \max(\hat \Gamma_i^+, \hat \Gamma_i^-)$.
        \EndFor
        \State Compute $\hat \Gamma_{\mathrm{low}} \coloneq \underset{i}{\min}(\hat \Gamma_i)$ and $\hat \Gamma_{\mathrm{high}} \coloneq \underset{i}{\max}(\hat \Gamma_i)$.
    \end{algorithmic}
\end{algorithm*}

To estimate the confounding strength $\Gamma$ of unobserved confounders $\mathbf{U}$, one could use already observed confounders $\mathbf{X} = (X^{(1)}, \ldots, X^{(p_\mathbf{X})})$, assuming $\Gamma$ is of the same magnitude as the $p_\mathbf{X}$ confounding strengths associated to each observed confounder. This method, known as \textit{informal benchmarking}, exists in two versions: a \textit{leave-one-out} procedure, where only one observed confounder at a time is considered unobserved, and a \textit{leave-multiple-out} procedure, where groups of observed confounders are considered unobserved. We refer the reader to Table~\ref{tab:methods_references} in appendix for an extensive list of references about these two approaches.

\begin{algorithm*}[h]
    \caption{Lower bound $\hat \Gamma_\mathrm{LB}^\mathrm{RCT}$ on $\Gamma$ via RCT~\parencite{de2024hidden}} \label{alg:via_RCT}
    \begin{algorithmic}
        \Require Target population $\diamond \in \{ \mathrm{rct}, \mathrm{obs}^\prime \}$, RCT data $\mathcal{D}_\mathrm{rct}$, observational data $\mathcal{D}_\mathrm{obs}$, significance level $\alpha$, list of confounding strengths $\{ \Gamma_1, \dots, \Gamma_m \}$.
        \State With $\mathcal{D}_\mathrm{rct}$ and according to the target population $\diamond$, estimate $\widehat{\mathrm{ATE}}$ and its empirical variance $\hat \sigma^2$ using Equation~\eqref{eqn:ATE_target_pop_obs}.
        \For{$\Gamma \in \{ \Gamma_1, \dots, \Gamma_m \}$}
            \State With $\mathcal{D}_\mathrm{obs}$, estimate the sensitivity analysis bounds $\widehat{\mathrm{CATE}}_-(\mathbf{X}, \Gamma)$ and $\widehat{\mathrm{CATE}}_+(\mathbf{X}, \Gamma)$ for confounding strength $\Gamma$, and average over the target population $\diamond$ to get the bounds for the ATE, denoted $\widehat{\mathrm{ATE}}_-(\Gamma)$ and $\widehat{\mathrm{ATE}}_+(\Gamma)$. Estimate the empirical variances of the CATE bounds with respect to the target population $\diamond$, denoted $\hat \sigma_-^2$ and $\hat \sigma_+^2$.
            \State Compute the test statistics:
            \begin{align*}
                \hat T_\Gamma^+ & = \frac{\widehat{\mathrm{ATE}}_+(\Gamma) - \widehat{\mathrm{ATE}}}{\hat \sigma_\diamond^+}, \quad \text{where} \quad \hat \sigma_\diamond^+ = \sqrt{\hat \sigma^2 + \hat \sigma_+^2 + 2 \hat \sigma \hat \sigma_+ \mathds{1}\{\diamond = \mathrm{rct}\}} \\
                \hat T_\Gamma^- & = \frac{\widehat{\mathrm{ATE}} - \widehat{\mathrm{ATE}}_-(\Gamma)}{\hat \sigma_\diamond^-}, \quad \text{where} \quad \hat \sigma_\diamond^- = \sqrt{\hat \sigma^2 + \hat \sigma_-^2 + 2 \hat \sigma \hat \sigma_- \mathds{1}\{\diamond = \mathrm{rct}\}}
            \end{align*}
            \State Compute $\hat \phi_\diamond(\Gamma, \alpha) = \mathds{1}\{\min(\hat T_\Gamma^+, \hat T_\Gamma^-) < z_{\alpha/2} \}$, where $z_{\alpha/2}$ is the $\alpha/2$-quantile of the standard normal distribution.
        \EndFor
        \State Compute $\hat \Gamma_\mathrm{LB}^\mathrm{RCT} = \underset{\Gamma \in \{ \Gamma_1, \dots, \Gamma_m \}}{\min}\{ \Gamma:\: \hat \phi_\diamond(\Gamma, \alpha) = 0 \}$.
    \end{algorithmic}
\end{algorithm*}

More formally, denote $\mathbf{X}^{(-i)}$ the vector of observed confounders $\mathbf{X}$ where $X^{(i)}$ was removed. The leave-one-out procedure consists in estimating the odds ratio from Definition~\ref{def:MSM_U}, or the density ratio from Definition~\ref{def:CMSM_U}, by replacing $\mathbf{U}$ with $X^{(i)}$ and $\mathbf{X}$ with $\mathbf{X}^{(-i)}$, for each $i$ in $\{ 1, \ldots, p_\mathbf{X} \}$. For instance, with the MSM, we would be interested in estimating the minimum and maximum values taken by $\mathrm{OR}(e(\mathbf{X}^{(-i)}, X^{(i)}), e(\mathbf{X}^{(-i)})) = \mathrm{OR}(e(\mathbf{X}), e(\mathbf{X}^{(-i)}))$. The procedure is formalized in Algorithm~\ref{alg:informal_benchmarking} for the MSM and the CMSM. Notice that, to avoid overfitting, $\hat e$ and $\hat f(T|\mathbf{X})$ can be obtained by cross-fitting.

On the other hand, the idea behind the leave-multiple-out procedure is to account for the multidimensionality of $\mathbf{U}$. For this reason, we recommend performing both leave-one-out \textit{and} leave-multiple-out informal benchmarking (with all possible combinations of observed confounders), to get a complete overview of the plausible values for the unknown confounding strength $\Gamma$.

Once we have obtained estimates of the confounding strengths associated to each observed confounder or group of observed confounders, we recommend reporting all of them. Among all of these estimations, the lowest one, $\hat \Gamma_{\mathrm{low}}$, and the highest one, $\hat \Gamma_{\mathrm{high}}$ are of particular interest. Indeed, assuming the true unknown $\Gamma$ is included in $[\hat \Gamma_{\mathrm{low}}, \hat \Gamma_{\mathrm{high}}]$, we could then consider that, ``at best", the true confounding strength associated to $\mathbf{U}$ is equal to $\hat \Gamma_{\mathrm{low}}$, and could not be lower, and that, ``at worst", it is equal to $\hat \Gamma_{\mathrm{high}}$, and it could not be greater. In practice, users of the informal benchmarking method are more prone to report only $\hat \Gamma_{\mathrm{high}}$, because, as the true $\Gamma$ is unknown, they prefer overestimating it. For this reason, we define the estimate $\hat \Gamma_{\mathrm{IB}}$ of $\Gamma$ via informal benchmarking as $\hat \Gamma_{\mathrm{IB}} \coloneq \hat \Gamma_{\mathrm{high}}$.

In some cases, it is possible to improve informal benchmarking with domain knowledge. If one knows that a particular observed covariate $X^{(i)}$ is a strong cause of both treatment assignment and outcome, one can expect the confounding strength associated to $\mathbf{U}$ to be weaker than the confounding strength associated to $X^{(i)}$. For example, \textcite{lee2024sensitivity} analyzed the German Breast Cancer Study Group dataset~\parencite{schumacher1994randomized} that evaluates the long-term effectiveness of the tamoxifen therapy in patients with node-positive breast cancer. They indicate that, according to an experienced radiologist, ``the degree of possible unmeasured confounding is expected to be weaker than tumor size", and that, according to~\textcite{early1998tamoxifen}, ``tumor size is a significant prognostic factor for long-term survival". After adjusting on other observed prognostic factors, the confounding strength associated to tumor size was approximately $1.2$. Therefore, they concluded that the true confounding strength of $\mathbf{U}$ was necessarily lower than $1.2$.

However, although very used, informal benchmarking suffers from some drawbacks. First, it relies on the untestable positivity assumption of the propensity scores. If this assumption was violated, the estimate $\hat \Gamma_{\mathrm{IB}}$ could be arbitrarily big or small. Moreover, the results are sensitive to the estimation method of the propensity score, e.g.\ logistic regression or random forests. Also, as this method is based on observed confounders, enough must be available to lead to an accurate estimate. Finally, as mentioned in Section~\ref{sec:not_reasonable}, it is important to keep in mind that the confounding strength $\Gamma$ associated to $\mathbf{U}$ depends on the observed confounders $\mathbf{X}$ used in the analysis. Therefore, trying to estimate $\Gamma$ by removing observed confounders could possibly give misleading results. The analysis conducted by~\textcite{lee2024sensitivity} should therefore be subject to caution because, if they had estimated the confounding strength of tumor size after adjusting on other observed prognostic factors \textit{and} on $\mathbf{U}$ (which is, in practice, impossible), they would certainly have obtained a value different from $1.2$ (higher or lower). See also the discussion in Section~6.2 from~\textcite{cinelli2020making} on how informal benchmarking with an unobserved confounder $\mathbf{U}$ ``exactly like" $\mathbf{X}$ (same distribution) would lead to incorrect conclusions.

\subsection{Leveraging Randomized Controlled Trials (RCTs)} \label{sec:via_rct}

Another approach to estimate $\Gamma$ is to leverage external data. In cases where observational data stem from post-marketing drug surveillance—--namely, the monitoring of a drug’s safety and efficacy following market authorization—--\textcite{yadlowsky2022bounds} suggested leveraging the Randomized Controlled Trial (RCT) that supported the drug’s approval to obtain an unbiased estimate of the binary treatment effect on the same primary outcome. This approach enables the assessment of the confounding strength $\Gamma$ attributable to unobserved confounders within the observational dataset. Based on this idea, \textcite{de2024hidden} proposed a testing procedure using the sensitivity bounds introduced in \textcite{dorn2022sharp}, \textcite{dorn2024doubly}, or \textcite{oprescu2023b} for the ATE and CATE to get an asymptotically valid lower bound on the sensitivity parameter $\Gamma$ from the MSM (Definition~\ref{def:MSM_U}).

Under the potential outcome framework presented in Section~\ref{sec:notations_and_assumptions}, where they assume that the potential outcomes are bounded, the authors suppose they have two datasets available: one RCT, $\mathcal{D}_\mathrm{rct}$, of size $n_\mathrm{rct}$, and one observational dataset, $\mathcal{D}_\mathrm{obs}$, of size $n_\mathrm{obs}$, coming, respectively, from the unknown distributions $\mathbb{P}_\mathrm{full}^\mathrm{rct}$ and $\mathbb{P}_\mathrm{full}^\mathrm{obs}$ over $(\mathbf{X}, \mathbf{U}, Y(0), Y(1), Y, T)$. However, recall that only $(\mathbf{X}, Y, T)$ is observed. We will use the notations ``$\mathrm{rct}$" for quantities related to RCT data, and ``$\mathrm{obs}$" for quantities related to observational data. For example, we denote $\mathbb{P}_\mathbf{X}^\mathrm{rct}$ and $\mathbb{P}_\mathbf{X}^\mathrm{obs}$, the distributions of the observed confounders $\mathbf{X}$, respectively, in the RCT and in the observational data.

The authors want to test if the full unobserved distribution $\mathbb{P}_\mathrm{full}^\mathrm{obs}$ has confounding strength at most $\Gamma$, which is their null hypothesis $H_0(\Gamma)$. They design their testing procedure using a simple fact: under $H_0(\Gamma)$, the bounds of the PEI computed with confounding strength $\Gamma$ by the method from, for instance, \textcite{dorn2022sharp}, should contain the unbiased estimate of the treatment effect computed on the RCT. Therefore, for multiple values of $\Gamma$, the testing procedure should check if the unbiased estimate obtained on the RCT---here, an estimate of the ATE, denoted $\widehat{\mathrm{ATE}}$---is well included in the PEI $[\widehat{\mathrm{ATE}}_-(\Gamma), \widehat{\mathrm{ATE}}_+(\Gamma)]$, if it is close to the bounds and was only outside the interval by chance, or if it is outside and far from the interval.

\begin{figure*}[h]
    \centering
    \includegraphics[width=\linewidth]{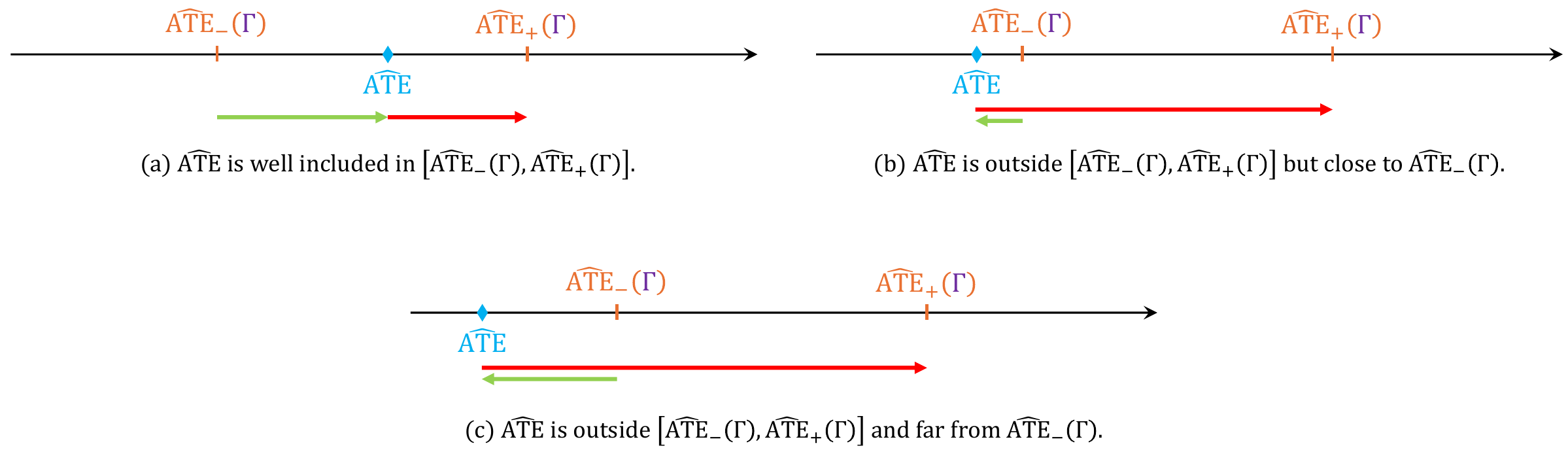}
    \caption{Three main cases when computing $\hat \phi_\diamond(\Gamma, \alpha)$. The green arrow represents the numerator of the test statistics $\hat T_\Gamma^-$ and the red arrow, the numerator of the test statistics $\hat T_\Gamma^+$. In Figures~\ref{fig:via_rct}a and \ref{fig:via_rct}b, $H_0(\Gamma)$ is not rejected, whereas in Figure~\ref{fig:via_rct}c, $H_0(\Gamma)$ is rejected. The significance level $\alpha$, and therefore, the quantile $z_{\alpha/2}$ determine what ``close to" means in \ref{fig:via_rct}b.}
    \label{fig:via_rct}
\end{figure*}

Three assumptions that are common in the literature for combining RCTs with observational data~\parencite{colnet2024causal} are needed to allow the testing procedure to be asymptotically valid. The first one, \textit{internal validity} (along with \textit{positivity}), is usually satisfied with RCTs.
\begin{assumption}[Internal validity and positivity] \label{ass:internal_validity}
    In the RCT, the treatment $T$ is assigned randomly, independently from the observed covariates $\mathbf{X}$ and potential outcomes $Y(0)$ and $Y(1)$. Moreover, $\pi \coloneq \mathbb{P}_T^\mathrm{rct}(T=1) \in (0,1)$.
\end{assumption}
The two other assumptions, \textit{support inclusion} and \textit{transportability}, implicate both the observational data and the RCT. Support inclusion ensures that we can always find individuals in the observational data that have the same observed characteristics as the ones from the RCT data, but some characteristics from the observational data may not be found in the RCT. For example, this is the case when the age in the population from the RCT is between 60 and 75, but between 50 and 80 in the observational data.
\begin{assumption}[Support inclusion] \label{ass:support_inclusion}
    The support of the distribution of the observed confounders in the RCT is included in the support of the distribution of the observed confounders in the observational data, i.e.\ $\mathrm{supp}(\mathbb{P}_\mathbf{X}^\mathrm{rct}) \subseteq \mathrm{supp}(\mathbb{P}_\mathbf{X}^\mathrm{obs})$.
\end{assumption}
On the other hand, transportability ensures that the results from the RCT could be transposed to the observational population, if we had access to the full distributions on the RCT and on the observational data. In slightly different contexts, this assumption can sometimes be also referred to as \textit{generalizability} \parencite{colnet2024causal}.
\begin{assumption}[Transportability] \label{ass:transportability}
    The CATE is invariant across studies, i.e.\ for all $\mathbf{x} \in \mathrm{supp}(\mathbb{P}_\mathbf{X}^\mathrm{rct})$, $\mathrm{CATE}(\mathbf{x})$ computed with respect to $\mathbb{P}_\mathrm{full}^\mathrm{rct}$ and $\mathbb{P}_\mathrm{full}^\mathrm{obs}$ are equal.
\end{assumption}
Because of support inclusion (Assumption~\ref{ass:support_inclusion}), two versions of the same testing procedure are proposed by the authors, depending on the target population of interest. We will first focus on the case where the target population is the observed population from the RCT, $\mathbb{P}_\mathbf{X}^\mathrm{rct}$. In this situation, the ATE can simply be estimated on $\mathcal{D}_\mathrm{rct}$ via the following formula:
\begin{equation} \label{eqn:ATE_target_pop_RCT}
    \widehat{\mathrm{ATE}} = \frac{1}{n_\mathrm{rct}} \sum_{(\mathbf{X}_i, T_i, Y_i) \in \mathcal{D}_\mathrm{rct}} \biggl( \frac{T_i}{\hat \pi} - \frac{1-T_i}{1-\hat \pi} \biggr) Y_i.
\end{equation}

\begin{algorithm*}[h]
    \caption{Lower bound $\hat \Gamma_\mathrm{LB}^\mathrm{NC}$ on $\Gamma$ via negative control outcomes} \label{alg:via_NCO}
    \begin{algorithmic}
        \Require Observational data $\mathcal{D}_\mathrm{obs} = \{ (\mathbf{X}_i, T_i, W_i^{(1)}, \dots, W_i^{(q)}) \}_{i=1}^n$, list of confounding strengths $\{ \Gamma_1, \dots, \Gamma_m \}$.
        \For{$i \in \{ 1, \dots, q \}$}
            \For{$\Gamma^{(i)} \in \{ \Gamma_1, \dots, \Gamma_m \}$}
                \State Compute the sensitivity analysis bounds for the treatment effect of $T$ on $W^{(i)}$ for confounding strength $\Gamma^{(i)}$. For example, for the $\mathrm{ATE} = \E[W(1)-W(0)]$, compute $\widehat{\mathrm{ATE}}_-(\Gamma^{(i)})$ and $\widehat{\mathrm{ATE}}_+(\Gamma^{(i)})$.
            \EndFor
            \State Compute the critical value $\hat \Gamma_c^{(i)}$ corresponding to negative control $i$, i.e.\ the minimum confounding strength for which the null effect is included in the sensitivity bounds. For example, for the $\mathrm{ATE}$, compute
            \begin{equation} \label{eqn:lb_negative_control_i}
                \hat \Gamma_c^{(i)} = \underset{\Gamma \in \{ \Gamma_1, \dots, \Gamma_m \}}{\min} \Bigl\{ \Gamma : \: 0 \in \bigl[ \widehat{\mathrm{ATE}}_-(\Gamma), \widehat{\mathrm{ATE}}_+(\Gamma) \bigr] \Bigr\}.
            \end{equation}
        \EndFor
        \State Compute $\hat \Gamma_\mathrm{LB}^\mathrm{NC} = \underset{i \in \{ 1, \dots, q \}}{\max}\{ \hat \Gamma_c^{(i)} \}$.
    \end{algorithmic}
\end{algorithm*}

Then, we can use Algorithm~\ref{alg:via_RCT} with the target population $\diamond = \mathrm{rct}$ to perform the statistical test for several values of $\Gamma$ and get a lower bound on the true confounding strength. For these different values of $\Gamma$, sensitivity bounds for the CATE are computed on the observational data using, for example, the method from \textcite{oprescu2023b}, and are averaged over the population of the RCT to get bounds for the ATE. Depending on the method, bounds for the ATE can also be obtained directly, without computing bounds for the CATE first.

Afterwards, two test statistics are computed: one for the lower bound, $\hat T_\Gamma^-$, and one for the upper bound, $\hat T_\Gamma^+$. These test statistics evaluate the position of the bounds with respect to the unbiased estimate $\widehat{\mathrm{ATE}}$. Finally, the testing procedure outputs a single number (0 or 1), defined as
\begin{equation*}
    \hat \phi_\mathrm{rct}(\Gamma, \alpha) = \mathds{1}\{\min(\hat T_\Gamma^+, \hat T_\Gamma^-) < z_{\alpha/2} \},
\end{equation*}
where $z_{\alpha/2}$ is the $\alpha/2$-quantile of the standard normal distribution. For example, if $\alpha = 0.05$, $z_{\alpha/2}$ is approximately equal to $-1.96$.

If $\hat \phi_\mathrm{rct}(\Gamma, \alpha) = 1$, the null hypothesis $H_0(\Gamma)$ is \textit{rejected}. We can identify three main situations, summarized in Figure~\ref{fig:via_rct}:
\begin{itemize}
    \item if $\widehat{\mathrm{ATE}}$ is \textit{well included} in the sensitivity bounds, the two test statistics are large enough (positive) compared with $z_{\alpha/2}$ (negative), and $\min(\hat T_\Gamma^+, \hat T_\Gamma^-) \geq z_{\alpha/2}$, so $\hat \phi_\mathrm{rct}(\Gamma, \alpha) = 0$. Thus, $H_0(\Gamma)$ is not rejected (Figure~\ref{fig:via_rct}a);
    \item if $\widehat{\mathrm{ATE}}$ is outside the sensitivity bounds but \textit{close to} one of them, one of the two test statistics becomes negative but is still larger than $z_{\alpha/2}$, so $\min(\hat T_\Gamma^+, \hat T_\Gamma^-) \geq z_{\alpha/2}$, and $\hat \phi_\mathrm{rct}(\Gamma, \alpha) = 0$. Thus, $H_0(\Gamma)$ is not rejected (Figure~\ref{fig:via_rct}b). The ``close to" condition is typically defined through the significance level $\alpha$ of the test;
    \item if $\widehat{\mathrm{ATE}}$ is not included and \textit{far from} the sensitivity interval, one of the two test statistics becomes negative and lower than the threshold $z_{\alpha/2}$, so $\min(\hat T_\Gamma^+, \hat T_\Gamma^-) < z_{\alpha/2}$, and then $\hat \phi_\mathrm{rct}(\Gamma, \alpha) = 1$. Thus, $H_0(\Gamma)$ is rejected (Figure~\ref{fig:via_rct}c).
\end{itemize}

Finally, the asymptotically valid lower bound $\hat \Gamma_\mathrm{LB}^\mathrm{RCT}$ is defined as the lowest $\Gamma$ such that the test does not reject the null hypothesis. In practice, the testing procedure is performed with a list of increasing confounding strengths $\{ \Gamma_1, \dots, \Gamma_m \}$, starting from $\Gamma_1 = 1$.

A second version of the testing procedure can be obtained by changing the target population. Indeed, by support inclusion, another choice could be to take the population from the observational data that lies in the support of the population from the RCT, i.e.\ $\mathbb{P}_\mathbf{X}^\mathrm{obs^\prime} = \mathbb{P}_\mathbf{X}^\mathrm{obs} \, | \, \mathbf{X} \in \mathrm{supp}(\mathbb{P}_\mathbf{X}^\mathrm{rct})$. In this case, the unbiased estimate of the ATE becomes
\begin{align} \label{eqn:ATE_target_pop_obs}
    \widehat{\mathrm{ATE}} = \frac{1}{n_\mathrm{rct}} \sum_{(\mathbf{X}_i, T_i, Y_i) \in \mathcal{D}_\mathrm{rct}} \biggl( \frac{T_i}{\hat \pi} - \frac{1-T_i}{1-\hat \pi} \biggr) Y_i \hat w(\mathbf{X}_i),
\end{align}
where $\hat w$ corrects the distribution shift between $\mathbb{P}_\mathbf{X}^\mathrm{rct}$ and $\mathbb{P}_\mathbf{X}^\mathrm{obs^\prime}$. Of course, if the target population is $\mathbb{P}_\mathbf{X}^\mathrm{rct}$, $\hat w(\mathbf{X}_i)$ is equal to 1 and we return to Equation~\eqref{eqn:ATE_target_pop_RCT}. See Appendix~A.2 from \textcite{de2024hidden} to estimate $\hat w$ in the particular case of \textit{nested trial designs}, i.e.\ designs where a two-stage nested sampling procedure is used. See also \textcite{colnet2024causal} for more information on nested and non-nested trial designs. When the target population is $\mathbb{P}_\mathbf{X}^\mathrm{obs^\prime}$, \textcite{de2024hidden} compute the sensitivity bounds using either the method from \textcite{dorn2022sharp}, or the method from \textcite{dorn2024doubly}.

Even if it is appealing, one major difficulty with the approach presented by \textcite{de2024hidden} is to find an RCT that corresponds to the observational dataset, as discussed by \textcite{yadlowsky2022bounds}. Nevertheless, we can still cite three relevant examples of such cases, respectively, two with nested designs and one with non-nested design: the \href{https://www.whi.org/}{\textit{Women's Health Initiative}} study, a study on \textit{Medicaid} by \textcite{degtiar2023conditional}, and a study on the \href{https://www.traumabase.eu/en_US}{\textit{Traumabase}} observational dataset and \textit{CRASH-3} RCT~\parencite{crash2019effects} on the effect of tranexamic acid on mortality in patients with traumatic bleeding by \textcite{colnet2024causal}

To get a tight lower bound on $\Gamma$, \textcite{de2024hidden} discovered through their synthetic experiments that the potential outcomes $Y(0)$ and $Y(1)$ must be highly correlated with the unobserved confounder $\mathbf{U}$. However, this is not something we can control or check, so the lower bound could be loose if $\mathbf{U}$ is not informative on the potential outcomes. On that ground, \textcite{de2024hidden} suggest adding information on this correlation in the sensitivity model to improve the estimation of the lower bound.

The tightness of the lower bound can also be improved when the sample size of the observational data is large. In particular, when the target population is $\mathbb{P}_\mathbf{X}^\mathrm{obs^\prime}$, the power of the statistical test increases when $n_\mathrm{obs}$ is large compared to $n_\mathrm{rct}$.

Additionally, \textcite{de2024hidden} point out that the proposed procedure cannot detect confounding bias that cancels out on average, which is the case with average treatment effects.

Finally, as stated above, when the target population is $\mathbb{P}_\mathbf{X}^\mathrm{rct}$, the ratio $\hat w$ equals 1 in the estimator of the ATE~\eqref{eqn:ATE_target_pop_obs}, whereas it requires some additional assumptions, such as nested trial design, to be estimated when the target population is $\mathbb{P}_\mathbf{X}^\mathrm{obs^\prime}$, but these assumptions are not always fulfilled in practice. Another set of assumptions is internal validity (Assumption~\ref{ass:internal_validity}), which can be considered quite realistic, but also transportability and support inclusion (Assumptions~\ref{ass:support_inclusion} and \ref{ass:transportability}), which could be violated by the data, leading to a biased estimate of $\Gamma_\mathrm{LB}^\mathrm{RCT}$.

\subsection{Leveraging negative controls} \label{sec:via_negative_controls}

\begin{figure*}[h]
    \centering
    \includegraphics[width=\linewidth]{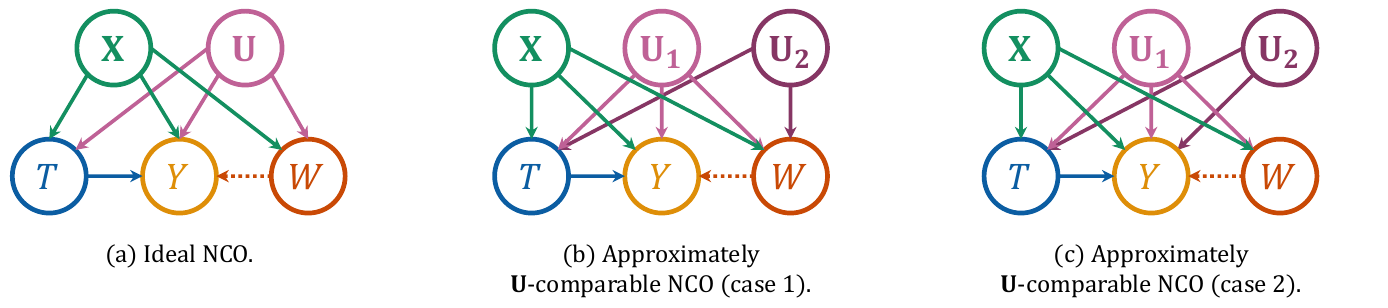}
    \caption{Examples of causal diagrams of a negative control outcome (NCO) $W$. An arrow directed from $A$ to $B$ indicates that $A$ causes $B$. Dashed edges may be absent. An ideal NCO $W$ (a) is such that the set of common causes $\mathbf{U}$ between $T$ and $Y$ and between $T$ and $W$ is the same. An approximately $\mathbf{U}$-comparable NCO $W$ (b, c) is such that an additional unobserved confounder $\mathbf{U}_2$ biases the relation between $T$ and $W$ (case 1) or between $T$ and $Y$ (case 2). In all cases, $T$ does not cause $W$.}
    \label{fig:negative_control_outcome}
\end{figure*}

Negative controls~\parencite{lipsitch2010negative} are a tool used in epidemiology to detect confounding bias. Two types of negative controls can be defined: negative control exposures and negative control outcomes. In this review, we only consider negative control outcomes to estimate a lower bound $\hat \Gamma_\mathrm{LB}^\mathrm{NC}$ on the true confounding strength $\Gamma$.

To quote \textcite{lipsitch2010negative}, the ``purpose of a negative control is to reproduce a condition that cannot involve the hypothesized causal mechanism, but is very likely to involve \textit{the same sources of bias} that may have been present in the original association." For example, if we study the effect of influenza vaccination ($T$) on influenza hospitalization ($Y$), then hospitalization for injury could play the role of a negative control outcome $W$, as it has, in all likelihood, no link with influenza vaccination.

An ideal negative control outcome is an outcome $W$ such that the set of common causes of treatment $T$ and outcome $Y$ is identical to the set of common causes of $T$ and $W$, and such that \textit{$T$ does not cause $W$}, so $W$ is independent from $T$. As a reminder, $Y$ cannot be a negative control outcome, in that it is caused by $T$. See Figure~\ref{fig:negative_control_outcome}a for an illustration. In practice, $(T,Y)$ and $(T,W)$ only partially share the same causes and, therefore, only partially share the same unobserved confounders. \textcite{lipsitch2010negative} called such negative control \textit{approximately $\mathbf{U}$-comparable}. For example, the relation between $T$ and $W$ may be confounded by two unobserved confounders, $\mathbf{U}_1$ and $\mathbf{U}_2$, whereas the association of interest between $T$ and $Y$ may only be confounded by $\mathbf{U}_1$ (Figure~\ref{fig:negative_control_outcome}b).

Negative controls can be used directly to detect or correct for unobserved confounders. See, for example, \textcite{lipsitch2010negative, tchetgen2014control, sofer2016negative, miao2018confounding, shi2020multiply, shi2020selective}.

\textcite{kallus2018confounding} suggest using negative controls to get a lower bound on $\Gamma$, exploiting the fact that $T$ is not supposed to have an influence on the negative control outcome $W$. If an effect between $T$ and $W$ is observed, then the confounding strength must be sufficiently large to cancel that effect out. In practice, if the sensitivity bounds for a measure of treatment effect on outcome $W$ (and not $Y$), as defined in Section~\ref{sec:bounds}, do not contain the null effect under $\mathbf{X}$-ignorability, then, a lower bound $\hat \Gamma_\mathrm{LB}^\mathrm{NC}$ can be obtained by increasing the confounding strength, starting from 1, until the sensitivity bounds contain this null effect. In other words, $\hat \Gamma_\mathrm{LB}^\mathrm{NC}$ is the critical value of these sensitivity bounds. Therefore, if the sensitivity bounds already contain the null effect under $\mathbf{X}$-ignorability, then the estimated lower bound on $\Gamma$ is 1.

We propose Algorithm~\ref{alg:via_NCO} to summarize the procedure when a collection of $q$ negative control outcomes $\{ (W_i^{(1)}, \dots, W_i^{(q)}) \}_{i=1}^n$ is available in the same dataset as $\mathbf{X}$ and $T$. To our knowledge, it is the first time an algorithm to get a lower bound on the sensitivity parameter from the MSM or CMSM with negative control outcomes is described and implemented (see Section~\ref{sec:experiments_non-param} for experimental results).

To get a valid lower bound on $\Gamma$, the negative control outcome $W$ should meet all required assumptions. However, as said before, in practice, we can only find approximately $\mathbf{U}$-comparable negative controls. If the relations between $T$ and $Y$ and between $T$ and $W$ are confounded as in Figure~\ref{fig:negative_control_outcome}b ($\mathbf{U}_1$ and $\mathbf{U}_2$ are confounders of the ($T$, $W$) relation, but only $\mathbf{U}_1$ is a confounder of the ($T$, $Y$) relation), then our procedure could detect confounders, namely $\mathbf{U}_2$, even if the relation of interest between $T$ and $Y$ is not confounded by $\mathbf{U}_2$, thus potentially \textit{overestimate} the lower bound $\hat \Gamma_\mathrm{LB}^\mathrm{NC}$. If the relations between $T$ and $Y$ and between $T$ and $W$ are confounded as in Figure~\ref{fig:negative_control_outcome}c ($\mathbf{U}_1$ and $\mathbf{U}_2$ are confounders of the ($T$, $Y$) relation, but only $\mathbf{U}_1$ is a confounder of the ($T$, $W$) relation), then our procedure could overlook some confounders, namely $\mathbf{U}_2$, and \textit{underestimate} the lower bound $\hat \Gamma_\mathrm{LB}^\mathrm{NC}$.

To illustrate approximately $\mathbf{U}$-comparable negative control outcomes, consider the example given by \textcite{lipsitch2010negative}. If $T$ represents vaccination for influenza, and $Y$ represents influenza hospitalization, hospitalization from injury can be used as a negative control outcome $W$. Indeed, $T$ is not likely to cause $W$ but the set of common causes of $T$ and $Y$ and of $T$ and $W$ could be similar. A possible unobserved confounder of the $(T,Y)$ relationship but not of the $(T,W)$ relationship (Figure~\ref{fig:negative_control_outcome}c) could be an aversion to vaccination that makes a patient less likely to get the influenza vaccine. Therefore, this particular unobserved confounder could not be detected using this negative control.

Moreover, to get an accurate lower bound on $\Gamma$, as much negative control outcomes as possible are needed, each negative control $i$ providing one estimate $\hat \Gamma_c^{(i)}$ (Equation~\eqref{eqn:lb_negative_control_i}). However, finding many good negative control outcomes is not always possible, so it can limit the quality of the estimation. Nevertheless, contrary to the method presented in Section~\ref{sec:via_rct}, where RCTs are only defined for binary treatments, the method presented in this section can be used for both binary and continuous treatment effects. Another advantage is that at least one negative control outcome satisfying few assumptions is required, whereas in Section~\ref{sec:via_rct}, a complete RCT with same treatment and outcome meeting stronger assumptions (Assumptions~\ref{ass:internal_validity} to \ref{ass:transportability}) must be found. Finally, we show in Section~\ref{sec:experiments_non-param} that $\hat \Gamma_\mathrm{LB}^\mathrm{NC}$ can be a tight lower bound when the sample size $n$ is large, and when the correlation between $\mathbf{U}$ and $W$ is high.

\begin{figure}[t]
    \centering
    \includegraphics[width=\linewidth]{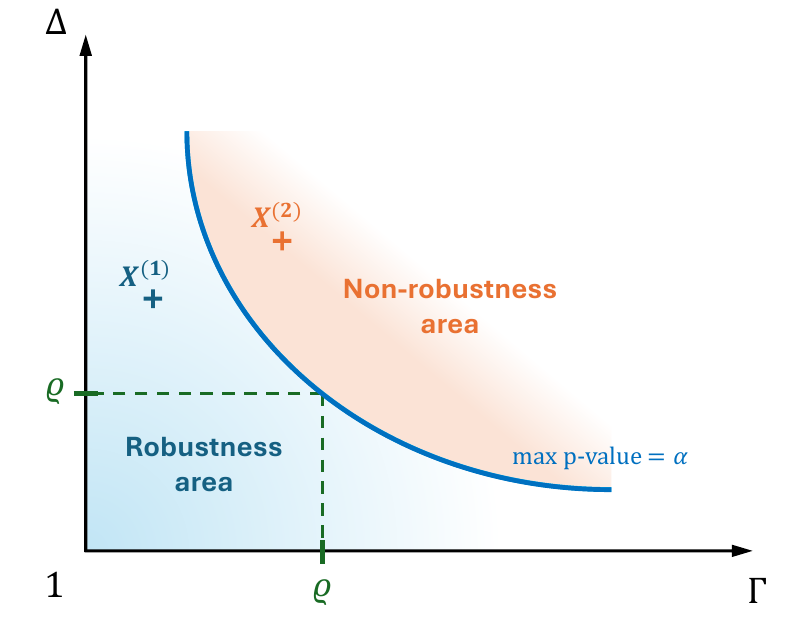}
    \caption{Simultaneous sensitivity analysis. The contour plot (blue curve) corresponds to the case where the maximum p-value over $\mathcal{U}$ is equal to the significance level $\alpha$. The robustness area (in blue) is located below the contour plot. Above the contour plot (in orange), the results are not robust to unobserved confounders. The point of the blue curve where $\Gamma = \Delta$ is used to define the robustness value $\varrho$. After adjustment, if the unobserved confounder $\mathbf{U}$ was judged similar to some observed confounder $X^{(1)}$ falling into the robustness area, the conclusions of the analysis could be considered robust to unobserved confounders. If it was judged similar to some observed confounder $X^{(2)}$ falling into the non-robustness area, the conclusions of the analysis could be considered non robust to unobserved confounders.}
    \label{fig:double_sensitivity_analysis}
\end{figure}

\subsection{A note on parametric approaches and sensitivity contour plots} \label{sec:param_approaches}

In the previous subsections, only non-parametric/model-free methods have been presented, which, we believe, is an advantage when no a priori about the functional form of $\mathbf{X}$, $T$ or $Y$ exists. Nevertheless, we give an overview of other methods that assign models, for example, to the treatment $T$ or to the outcome $Y$, but we do not study them in detail in the next sections.

\textcite{hsu2013calibrating, zhang2020calibrated} use double/simultaneous sensitivity analyses: one sensitivity model relates $\mathbf{U}$ to the treatment $T$ (as in Definition~\ref{def:MSM_U}) and the other one relates $\mathbf{U}$ to the outcome $Y$. The sensitivity parameter of the first one is denoted $\Gamma$ and the sensitivity parameter of the second one is denoted $\Delta$. \textcite{hsu2013calibrating} draw contour plots of $\Delta$ versus $\Gamma$ for a cutoff value defined by some maximum p-value over $\mathcal{U}$ for testing the null hypothesis of no treatment effect with significance level $\alpha$, and evaluate if \textit{unobserved} confounders of certain strengths could exist based on the relative position of \textit{observed} confounders with respect to the cutoff value (Figure~\ref{fig:double_sensitivity_analysis}). The contour plot defines two areas: a robustness area and a non-robustness area, where results are, respectively, robust and non-robust to unobserved confounders. If a practitioner judges that a potential unobserved confounder $\mathbf{U}$ is similar to an observed confounder $X^{(1)}$ located in the robustness area, then the conclusions are considered robust to unobserved confounders. On the other side, if the practitioner thinks that the potential unobserved confounder $\mathbf{U}$ is similar to an observed confounder $X^{(2)}$ located in the non-robustness area, then the conclusions are considered non-robust to unobserved confounders. \textcite{zhang2020calibrated} later addressed some limitations of the approach proposed by \textcite{hsu2013calibrating}. These methods are inspired by \textcite{imbens2003sensitivity} and \textcite{cinelli2020making} who use two partial $R^2$ (or coefficients of partial determination), one for the treatment, $R^2_T$, and one for the outcome, $R^2_Y$, instead of sensitivity models. Please also refer to \textcite{rosenbaum2009amplification} and \textcite{soriano2023interpretable} for a related approach called \textit{amplification}.

One interesting point about these methods is that it allows to define a \textit{robustness value} (RV)~\parencite{cinelli2020making}. For a fixed cutoff value, such as the null effect for the ATE, the RV was first defined as the value $R^2_\mathrm{RV}$ for which $R^2_T$ is equal to $R^2_Y$, and, by extension, could be defined as the value $\varrho$ for which $\Delta$ is equal to $\Gamma$ on the contour plot of this cutoff value (Figure~\ref{fig:double_sensitivity_analysis}). The RV presents the advantage of being ``independent" from the observed confounders, in the sense that we do not need to report the pairs $(\Gamma, \Delta)$ adjusted on each observed confounder to conclude if the analysis is robust or not.

The RV can be compared with the E-value~\parencite{vanderweele2017sensitivity}, which is an equivalent measure on the risk ratio scale that is also covariate-independent. If a study reports a high RV, then the conclusions drawn from this study (e.g., the ATE is positive) are likely to be robust to unobserved confounders. If the RV is low, then the conclusions are not robust and unobserved confounders of low strength could change the direction of the treatment effect. The evaluation of what a high or low RV is can only be made when many studies have reported one. This is why \textcite{vanderweele2017sensitivity} and \textcite{cinelli2020making} suggest routinely reporting the E-value or the RV.

\section{Numerical comparison} \label{sec:experiments_non-param}

We compared the three advanced methods presented in Section~\ref{sec:advanced_approaches}, namely informal benchmarking, lower bound estimation via RCTs and lower bound estimation via negative controls, on synthetic data. See \textcite{de2024hidden} for a real data application of the estimation of a lower bound on $\Gamma$ using RCTs. To be able to compare the three methods, we only tackled the case of binary treatments, as RCTs are not defined for continuous treatments, and, for simplicity, we focused on the MSM from Definition~\ref{def:MSM_U}. In the following, we are interested in estimating $\Gamma$ or a lower bound on $\Gamma$ when the treatment effect is the ATE. We did not compare the methods on real data as it is complex to find a dataset with enough covariates to perform informal benchmarking, an associated RCT with same treatment, outcome and covariates, and negative control outcomes at the same time.

\subsection{Simulation setup}

To compare the three methods, we extended the simulation setup from \textcite{de2024hidden} by increasing the dimensionality of $\mathbf{X}$ and $\mathbf{U}$. In our experiments, unless stated otherwise, we considered $p_\mathbf{X} = 5$ observed confounders and $p_\mathbf{U} = 2$ unobserved confounders. We used multidimensional observed confounders to be able to apply informal benchmarking and we chose
\begin{equation*}
    \mathbf{X}_\mathrm{rct} \sim \mathcal{U}(-0.9, 0.9)^{p_\mathbf{X}} \quad \text{and} \quad \mathbf{X}_\mathrm{obs} \sim \mathcal{U}(-1, 1)^{p_\mathbf{X}},
\end{equation*}
where $\mathcal{U}(a, b)^p$ is the uniform distribution on $(a, b)$ of dimension $p$, to satisfy support inclusion (Assumption~\ref{ass:support_inclusion}). For $j$ in $[\![1, p_\mathbf{U}]\!]$, conditionally on $\mathbf{X}=\mathbf{x}$, we designed each unobserved confounder $U_j$ as
\begin{equation*}
    U_j|\mathbf{X}=\mathbf{x} \sim \lambda G + (1 - \lambda) \beta_j^T \mathbf{x} = \mathcal{N} \bigl( (1-\lambda) \beta_j^T \mathbf{x}, \, \lambda^2 \bigr),
\end{equation*}
with $G \sim \mathcal{N}(0, 1)$, $0 < \lambda < 1$, and $\beta_j = (\beta_{j,1}, \dots, \beta_{j,p_\mathbf{X}}) \in \mathbb{R}^{p_\mathbf{X}}_+$. For the observational data, we chose the distribution of $T$ conditionally on $\mathbf{X}$ and $\mathbf{U}$ to be a Bernoulli satisfying the MSM with a true sensitivity parameter $\Gamma^\star$. In the observational data, we designed the true propensity score $e^\mathrm{obs}(\mathbf{X}, \mathbf{U}) = \mathbb{P}^\mathrm{obs}(T=1|\mathbf{X}, \mathbf{U})$ such that it marginalizes to the nominal propensity score $e^\mathrm{obs}(\mathbf{X}) = \mathbb{P}^\mathrm{obs}(T=1|\mathbf{X}) = \mathrm{logistic}(\delta^T \mathbf{X} + 0.5)$, with $\delta \in \mathbb{R}^{p_\mathbf{X}}$. We explain how this is done in Appendix~\ref{app:complete_simulation_setup}. In the RCT data, we simply defined the true propensity score as $e^\mathrm{rct}(\mathbf{X}, \mathbf{U}) = \mathbb{P}^\mathrm{rct}(T=1|\mathbf{X}, \mathbf{U}) = 1/2$ to satisfy internal validity and positivity (Assumption~\ref{ass:internal_validity}).

Denoting $p_W$ the number of negative control outcomes, the outcome $Y$ and negative control outcome $W_k$, for $k \in [\![1, p_W]\!]$, were defined via the following linear models
\begin{align*}
    Y(T) & = (2T-1) \theta_1^T \mathbf{X} + \theta_2^T \mathbf{U} + (2T-1) \times \mathrm{ATE}^\star/2 + \varepsilon_Y, \\
    W_k & = \theta_{3, k}^T \mathbf{X} + \theta_{4, k}^T \mathbf{U} + \varepsilon_W,
\end{align*}
where $\mathrm{ATE}^\star$ is the true ATE, $(\theta_1, \theta_2)$ and $(\theta_{3,k}, \theta_{4,k})$ are in $\mathbb{R}^{p_\mathbf{X}} \times \mathbb{R}^{p_\mathbf{U}}$, $\varepsilon_Y \sim \mathcal{N}(0, \sigma_Y^2)$ and $\varepsilon_W \sim \mathcal{N}(0, \sigma_W^2)$. In our experiments, $p_W = 2$ and $\mathrm{ATE}^\star = 0.25$.

The complete simulation setup is given in Appendix~\ref{app:complete_simulation_setup}. In the following, correlations were measured empirically with the Pearson correlation coefficient and were considered moderate when their absolute value is between 0.3 and 0.7, low below 0.3, and high over 0.7.

\subsection{Implementation details}

Our code was developed in \texttt{R} version 4.3.2~\parencite{rstatisticalsoftware}. For licensing reasons, only a modified implementation is publicly available at \href{https://github.com/Sanofi-Public/confounding-strength-estimation}{https://github.com/Sanofi-Public/confounding-strength-estimation} under a non-commercial license. In the next sections, we present our implementation choices regarding each method.

\subsubsection{Sensitivity analysis via Quantile Balancing (QB)} \label{sec:impl_details_qb}

To perform a sensitivity analysis for the ATE and estimate a lower bound on $\Gamma$ via RCTs and negative control outcomes, we re-implemented the Quantile Balancing (QB) method from \textcite{dorn2022sharp}.\footnote{See \href{https://www.tandfonline.com/doi/abs/10.1080/01621459.2022.2069572}{Supplemental Material} from \textcite{dorn2022sharp} for the original implementation.} More precisely, we used the stabilized AIPW estimator, as presented in their paper, where the outcome regression was estimated with a linear model (\texttt{lm} function), the propensity scores were estimated with a logistic regression (\texttt{glm} function), and the conditional quantiles were estimated with the \texttt{rq} function from the \texttt{quantreg} package~\parencite{quantreg}. These three nuisance parameters were computed using 5-fold cross-fitting to avoid overfitting.

To estimate the sensitivity bounds, we used the weighted quantile regression approach presented in Appendix~A.1 from \textcite{dorn2022sharp}. Confidence intervals were obtained using the percentile bootstrap method (see~\cite{zhao2019sensitivity}) with a number of bootstrap samples $B = 500$ and a significance level $\alpha = 0.05$. Finally, the sensitivity analyses were performed for 20 equally spaced values of $\Gamma$ between 1 and 20.

\begin{table*}[h!]
    \centering
    \caption{Different experimental settings. $\rho_{A,B}$ is the correlation between $A$ and $B$ (measured by Pearson correlation). The types of NCO correspond to the ones presented in Figure~\ref{fig:negative_control_outcome}. The ``Interval" column corresponds to the type of sensitivity interval used by Algorithms~\ref{alg:via_RCT} and \ref{alg:via_NCO}. NCO: Negative Control Outcome.}
    \label{tab:simul_cases}
    \begin{tabular}{|c|c|c|c|c|c|c|c|c|c|}
        \hline
        \textbf{Exp.\ \#} & \textbf{Sec.} & \textbf{Type of NCO} & $\rho_{\mathbf{X}, \mathbf{U}}$ & $\rho_{\mathbf{U}, Y(0)}$ and $\rho_{\mathbf{U}, W}$ & $n_\mathrm{rct}$ & $n_\mathrm{obs}$ & $p_\mathbf{X}$ & $\Gamma^\star$ & \textbf{Interval} \\
        \hline
        \hline
        \hyperref[fig:sa_1]{1} & & Ideal & Uncorrelated & Moderate & 2000 & 2000 & 5 & 8 & PEI \\
        \hyperref[fig:sa_2]{2} & \ref{sec:effect_rho_UY_rho_UW} & Ideal & Uncorrelated & Low & 2000 & 2000 & 5 & 8 & PEI \\
        \hyperref[fig:sa_3]{3} & & Ideal & Uncorrelated & High & 2000 & 2000 & 5 & 8 & PEI \\
        \hline
        \hyperref[fig:sa_4]{4} & & Ideal & Low & Moderate & 2000 & 2000 & 5 & 8 & PEI \\
        \hyperref[fig:sa_5]{5} & \ref{sec:effect_rho_XU} & Ideal & Low & High & 2000 & 2000 & 5 & 8 & PEI \\
        \hyperref[fig:sa_6]{6} & & Ideal & High & High & 2000 & 2000 & 5 & 8 & PEI \\
        \hline
        \hyperref[fig:sa_7]{7} & \multirow{4}{*}{\ref{sec:effect_gamma_n_px}} & Ideal & Uncorrelated & Moderate & 2000 & 2000 & 5 & 5 & PEI \\
        \hyperref[fig:sa_8]{8} & & Ideal & Uncorrelated & Moderate & 2000 & 2000 & 5 & 20 & PEI \\
        \hyperref[fig:sa_9]{9} & & Ideal & Uncorrelated & Moderate & 500 & 500 & 5 & 8 & PEI \\
        \hyperref[fig:sa_10]{10} & & Ideal & Uncorrelated & Moderate & 2000 & 2000 & 10 & 8 & PEI \\
        \hline
        \hyperref[fig:sa_11]{11} & & Case 1 & Uncorrelated & Moderate & 2000 & 2000 & 5 & 8 & PEI \\
        \hyperref[fig:sa_12]{12} & \ref{sec:effect_nco} & Case 2 & Uncorrelated & Moderate & 2000 & 2000 & 5 & 8 & PEI \\
        \hyperref[fig:sa_13]{13} & & Case 1 & Low & Moderate & 200 & 2000 & 5 & 8 & PEI \\
        \hline
        \hyperref[fig:sa_14]{14} & \ref{sec:effect_type_interval} & Ideal & Uncorrelated & Moderate & 2000 & 2000 & 5 & 8 & CI \\
        \hline
    \end{tabular}
\end{table*}

\subsubsection{Informal benchmarking}

We implemented the leave-one-out and leave-multiple-out procedures of informal benchmarking, as described in Algorithm~\ref{alg:informal_benchmarking}, and estimated the propensity scores either via logistic regression (\texttt{glm} function) or via random forests (\texttt{probability\_forest} function from the \texttt{grf} package~\parencite{grf}) using 5-fold cross-fitting.

When using random forests, we tuned the hyperparameters via 5-fold cross-validation using a log loss. As we set $p_\mathbf{X} = 5$ in our experiments, we report the results of the leave-multiple-out procedure when all possible combinations of 1 to $p_\mathbf{X} - 1 = 4$ observed confounders were omitted from the data. See Appendix~\ref{app:add_impl_details} for more implementation details about informal benchmarking.

\subsubsection{Lower bound via RCTs} \label{sec:impl_details_rct}

We implemented the method from \textcite{de2024hidden} as described in Algorithm~\ref{alg:via_RCT}. For the original implementation, see \textcite{de2024confounder}.

We restricted the study to the case where the target population is $\mathbb{P}^{\mathrm{obs}^\prime}_\mathbf{X}$ as the statistical power of the test increases when the sample size of the observational data is large compared to the sample size of the RCT, according to \textcite{de2024hidden}, which is not the case when the target population is $\mathbb{P}^{\mathrm{rct}}_\mathbf{X}$. This choice also allowed the method to be comparable to the others.

The tests were performed with a significance level of 0.05 and 25 equally spaced values of $\Gamma$ between 1 and 13 (except for $\Gamma^\star = 20$, where we chose 30 equally spaced values between 1 and 20). We estimated the ATE on the RCT data using Equation~\eqref{eqn:ATE_target_pop_obs} and computed the distribution shift correction $\hat{w}$ as described in Appendix~A.2 from \textcite{de2024hidden} using logistic regression to fit probability models. The sensitivity bounds were then estimated on the observational data using the QB method as explained in Section~\ref{sec:impl_details_qb}. We used the PEI as our sensitivity bounds but the CI could have also been chosen, although confidence intervals lead to underestimated lower bounds because they are larger than the PEIs (see Section~\ref{sec:effect_type_interval}). Finally, the empirical standard deviations $\hat{\sigma}$, $\hat{\sigma}_-$ and $\hat{\sigma}_+$ were estimated on 100 bootstrap samples.

\subsubsection{Lower bound via negative controls}

We implemented the method as described in Algorithm~\ref{alg:via_NCO}. The sensitivity bounds were estimated on the observational data using the QB method as explained in Section~\ref{sec:impl_details_qb}. As stated earlier, we used the PEI as our sensitivity bounds but the CI could have also been chosen. Finally, the grid search was performed over 25 equally spaced values of $\Gamma$ between 1 and 13 (except for $\Gamma^\star = 20$, where we chose 30 equally spaced values between 1 and 20).

To generate data that correspond to approximately $\mathbf{U}$-comparable negative control outcomes, as described in Figure~\ref{fig:negative_control_outcome}, we simulated $p_\mathbf{U} = 3$ unobserved confounders. We removed one unobserved confounder---without loss of generality, the third one---from the generation process of $Y$ in case 1 (Figure~\ref{fig:negative_control_outcome}b), and one from the generation process of $W$ in case 2 (Figure~\ref{fig:negative_control_outcome}c).

\subsection{Simulation experiments}

We evaluated the three methods across various scenarios, as outlined in Table~\ref{tab:simul_cases}, to determine conditions under which one method outperformed the others. We varied the type of negative control outcome (NCO), the correlation between $\mathbf{X}$ and $\mathbf{U}$, $\rho_{\mathbf{X}, \mathbf{U}}$, between $\mathbf{U}$ and $Y(0)$, $\rho_{\mathbf{U}, Y(0)}$, and between $\mathbf{U}$ and $W$, $\rho_{\mathbf{U}, W}$, the sample size of the observational data and the RCT, the dimension $p_\mathbf{X}$ of $\mathbf{X}$, the true confounding strength $\Gamma^\star$, and the type of sensitivity interval (PEI or CI). 

In each experiment, we began with the same random seed of 1 to avoid variability due to randomness between setups, we generated 20 Monte-Carlo samples and estimated $\Gamma$ or a lower bound on $\Gamma$ with each method. For each procedure, the minimum, median, and maximum estimations are recorded in Table~\ref{tab:estimated_gamma}.
For each experiment, the values of the parameters are supplied in Appendix~\ref{app:details_sim_data_and_results}, along with visual representations of the results.

\begin{table*}[ht]
    \centering
    \caption{Estimated confounding strength and lower bound on confounding strength (minimum, median and maximum) on 20 Monte-Carlo samples for each experiment in Table~\ref{tab:simul_cases}. $\Gamma^\star$ is the true unknown confounding strength to estimate. We highlight the results in bold when they are larger than $\Gamma^\star$. IB: Informal Benchmarking. LB: lower bound. RCT: Randomized Controlled Trials. NCO: Negative Control Outcomes.}
    \label{tab:estimated_gamma}
    \begin{tabular}{|c|ccc|ccc|ccc|ccc|c|}
      \hline
      \textbf{Exp.} & \multicolumn{3}{|c|}{\textbf{IB: logistic}} & \multicolumn{3}{|c|}{\textbf{IB: random forests}} & \multicolumn{3}{|c|}{\textbf{LB via RCT}} & \multicolumn{3}{|c|}{\textbf{LB via NCO}} & \\
      \textbf{\#} & Min & Median & Max & Min & Median & Max & Min & Median & Max & Min & Median & Max & $\Gamma^\star$ \\
      \hline
      \hline
      1 & 2.49 & 3.78 & 6.33 & 2.18 & 2.49 & 3.02 & 2.50 & 4.00 & 5.50 & 3.50 & 4.00 & 5.00 & 8 \\
      2 & 2.49 & 3.78 & 6.33 & 2.18 & 2.49 & 3.02 & 1.00 & 1.00 & 1.50 & 1.50 & 1.50 & 2.00 & 8 \\
      3 & 2.49 & 3.78 & 6.33 & 2.18 & 2.49 & 3.02 & 5.50 & 7.00 & \textbf{8.50} & 7.00 & \textbf{8.50} & \textbf{11.00} & 8 \\
      \hline
      4 & 2.49 & 3.78 & 6.33 & 2.18 & 2.49 & 3.02 & 1.00 & 1.00 & 1.50 & 1.50 & 1.50 & 2.00 & 8 \\
      5 & 2.49 & 3.78 & 6.33 & 2.18 & 2.49 & 3.02 & 5.50 & 7.00 & \textbf{8.50} & 7.00 & \textbf{8.50} & \textbf{11.00} & 8 \\
      6 & 2.49 & 3.78 & 6.33 & 2.18 & 2.49 & 3.02 & 1.00 & 1.75 & \textbf{12.00} & 6.50 & 7.00 & \textbf{8.50} & 8 \\
      \hline
      7 & 2.40 & 3.82 & \textbf{6.52} & 2.16 & 2.44 & 2.83 & 2.00 & 3.00 & 4.50 & 3.00 & 3.50 & 4.00 & 5 \\
      8 & 2.32 & 3.90 & 5.47 & 1.98 & 2.47 & 2.93 & 4.28 & 7.55 & 11.48 & 4.28 & 6.24 & 10.17 & 20 \\
      9 & 1.91 & 3.51 & 6.64 & 1.21 & 1.30 & 1.45 & 1.50 & 3.00 & 6.00 & 3.50 & 4.00 & 6.50 & 8 \\
      10 & 4.46 & \textbf{8.52} & \textbf{13.14} & 2.16 & 2.69 & 3.33 & 2.50 & 3.25 & 5.00 & 3.00 & 4.00 & 5.50 & 8 \\
      \hline
      11 & 2.53 & 3.51 & 4.99 & 1.99 & 2.35 & 2.78 & 2.50 & 3.00 & 3.50 & 4.50 & 5.00 & 6.50 & 8 \\
      12 & 2.53 & 3.51 & 4.99 & 1.99 & 2.35 & 2.78 & 4.00 & 4.50 & 6.00 & 3.00 & 3.00 & 4.00 & 8 \\
      13 & 2.53 & 3.51 & 4.99 & 1.99 & 2.35 & 2.78 & 2.00 & 2.50 & 3.00 & 4.00 & 4.50 & 5.50 & 8 \\
      \hline
      14 & 2.27 & 3.56 & 6.04 & 2.10 & 2.42 & 3.02 & 1.50 & 2.50 & 4.00 & 3.00 & 3.75 & 4.50 & 8 \\
      \hline
    \end{tabular}
\end{table*}

\begin{figure}
    \centering
    \includegraphics[width=\linewidth]{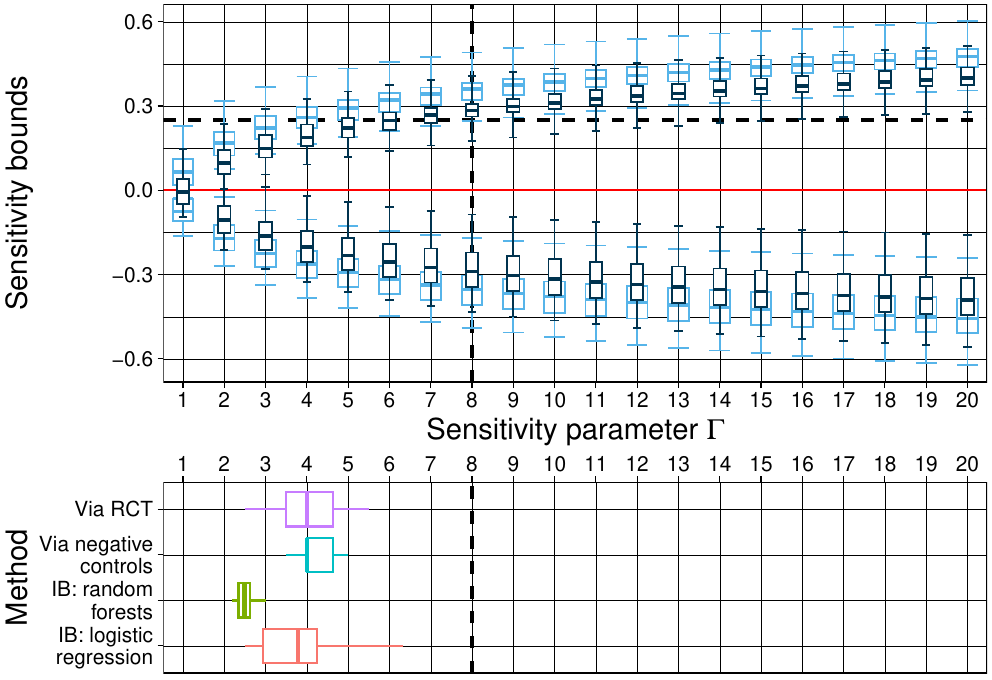}
    \caption{Sensitivity analysis n°1. The sensitivity bounds for the ATE as a function of the sensitivity parameter $\Gamma$ are represented above. 20 Monte-Carlo samples were used. Boxplots in light blue correspond to the CIs, whereas boxplots in dark blue correspond to the PEIs. The red horizontal line indicates the null effect. The dotted horizontal line is the true ATE, $\mathrm{ATE}^\star = 0.25$, and the dotted vertical line is the true sensitivity parameter $\Gamma^\star = 8$ to estimate. Below, for each method, boxplots show the estimated sensitivity parameters on the same 20 Monte-Carlo samples. In this representation, the boxplots include the outliers, i.e.\ they go from the minimum to the maximum estimated value. IB: Informal Benchmarking.}
    \label{fig:sa_1}
\end{figure}

\begin{table*}[ht]
    \centering
    \caption{Percentage of inclusion of the null effect and of the true ATE, $\mathrm{ATE}^\star$, in the CIs after performing sensitivity analyses via QB when $\Gamma = 1$, $\hat{\Gamma}$ (median), and $\Gamma^\star$ on 20 Monte-Carlo samples for the first three experiments and for each method in Table~\ref{tab:simul_cases}.}
    \label{tab:robustness}
    \begin{tabular}{|c|cccccc|cccccc|}
        \hline
        \textbf{Exp.} & \multicolumn{6}{|c|}{\textbf{Null effect}} & \multicolumn{6}{|c|}{$\mathrm{\mathbf{ATE}}^\star$} \\
        \textbf{\#} & $\Gamma = 1$ & $\hat{\Gamma}_\mathrm{IB,\,log}$ & $\hat{\Gamma}_\mathrm{IB,\,rf}$ & $\hat{\Gamma}_\mathrm{LB}^\mathrm{RCT}$ & $\hat{\Gamma}_\mathrm{LB}^\mathrm{NC}$ & $\Gamma = \Gamma^\star$ & $\Gamma = 1$ & $\hat{\Gamma}_\mathrm{IB,\,log}$ & $\hat{\Gamma}_\mathrm{IB,\,rf}$ & $\hat{\Gamma}_\mathrm{LB}^\mathrm{RCT}$ & $\hat{\Gamma}_\mathrm{LB}^\mathrm{NC}$ & $\Gamma = \Gamma^\star$ \\ 
        \hline
        \hline
        1 & 85 & 100 & 100 & 100 & 100 & 100 & 0 & 55 & 15 & 55 & 55 & 95 \\
        2 & 0 & 10 & 5 & 0 & 0 & 40 & 90 & 100 & 100 & 90 & 95 & 100 \\
        3 & 0 & 10 & 0 & 100 & 100 & 100 & 0 & 0 & 0 & 60 & 95 & 90 \\
        \hline
    \end{tabular}
\end{table*}

\subsubsection{Effect of the correlation between \texorpdfstring{$\mathbf{U}$}{} and \texorpdfstring{$Y(0)$}{}, and between \texorpdfstring{$\mathbf{U}$}{} and \texorpdfstring{$W$}{}} \label{sec:effect_rho_UY_rho_UW}

In Experiments~1, 2 and 3 from Table~\ref{tab:simul_cases}, we varied the degree of correlation between $\mathbf{U}$ and $Y(0)$, and between $\mathbf{U}$ and $W$ while keeping other parameters constant. Note that we chose to study the correlation between $\mathbf{U}$ and $Y(0)$, as in \textcite{de2024hidden}, but we could also have chosen $Y(1)$ instead. We took ideal NCOs and uncorrelated $\mathbf{X}$ and $\mathbf{U}$ to compare the results more easily, and we varied the correlations $\rho_{\mathbf{U}, Y(0)}$ and $\rho_{\mathbf{U}, W}$ between low, moderate and high levels. 

Results in Table~\ref{tab:estimated_gamma} indicate that informal benchmarking (IB) was unaffected by variations of correlation between $\mathbf{U}$ and $Y(0)$. This is expected as IB only relies on the observed confounders $\mathbf{X}$ and treatment $T$. Moreover, IB with logistic regression exhibited marginally better performance than IB with random forests, probably because the nominal propensity score $e(\mathbf{X})$ has a logistic form.

On the other side, the lower bound via RCT improved and approached $\Gamma^\star$ when $\rho_{\mathbf{U}, Y(0)}$ increased. This is consistent with findings from \textcite{de2024hidden} and can be explained by the fact that we better capture $\mathbf{U}$ through $Y$ when the correlation is high.

Finally, the lower bound via NCO improved when $\rho_{\mathbf{U}, W}$ increased but, as opposed to the lower bound via RCT, this method can overestimate $\Gamma$ more significantly when the correlation is too large (median of 8.50 when $\Gamma^\star = 8$ versus median of 7.00 with the RCT). This improvement can be explained in the same way as before: we better capture $\mathbf{U}$ through $W$ when the correlation between the two is high. Note that the correlation between $\mathbf{U}$ and $Y(0)$ had no impact on the method using NCO because it does not rely on the values of $Y$.

Table~\ref{tab:robustness} and Figures~\ref{fig:sa_1}, \ref{fig:sa_2} and \ref{fig:sa_3} show that the true ATE is included in 90\% of the CIs under ignorability ($\Gamma = 1$) when the correlation $\rho_{\mathbf{U}, Y(0)}$ is low (Experiment~2), but in 0\% of the CIs when the correlation increases (Experiments~1 and 3). This bias can be explained by the fact that $\mathbf{X}$ and $\mathbf{U}$ are uncorrelated. Even if the correlation between $\mathbf{U}$ and $Y(0)$ increases, we only observe $\mathbf{X}$. As the role of $\mathbf{U}$ in $Y(0)$ increases but note the role of $\mathbf{X}$, the bounds from the sensitivity analysis become biased. Table~\ref{tab:robustness} also indicates that, if $\Gamma$ was correctly estimated ($\Gamma = \Gamma^\star$), then 100\% of the CIs would contain the true ATE in Experiment~2. However, in 40\% of the cases, we would not be able to conclude about the sign of the ATE because 40\% of the CIs would contain the null effect. Under ignorability, 90\% of the CIs contain the true ATE and we would always conclude---with a bias due to unobserved confounders---about the sign of the ATE because no interval contains the null effect. See Table~\ref{tab:robustness_complete} in appendix for the complete results on the 14 experiments.

\begin{table*}[h]
    \centering
    \caption{Overview of methods for estimating the confounding strength $\Gamma$, including their respective advantages and limitations. PS: propensity score.}
    \label{tab:methods_summary}
    \begin{adjustbox}{max width=\textwidth}
    \begin{tabular}{|c|c|c|c|}
        \hline
        \textbf{Method} & \textbf{When?} & \textbf{Pros} & \textbf{Cons} \\
        \hline
        \hline
        \textbf{\makecell{No\\ estimation}} & \makecell{Few knowledge about potential\\ unobserved confounders,\\ few observed confounders,\\ no external data,\\ small sample size.} & Safest solution. & \makecell{No conclusion reached\\ about the treatment effect\\ unless the critical value\\ is 1 or $+\infty$.} \\
        \hline
        \textbf{\makecell{Domain\\ knowledge}} & \makecell{Good a priori about $\Gamma$.\\ Answering the question\\ from Section~\ref{sec:critical_value_domain_knowl} is easy.} & No computations needed. & Prone to bias. \\
        \hline
        \textbf{\makecell{Informal\\ benchmarking}} & \makecell{High number of\\ observed confounders $\mathbf{X}$\\ (more than 10\\ in our experiments).\\ Well-specified\\ propensity score.} & \makecell{Easy to understand\\ and to implement.\\ Few assumptions.\\ Compatible with continuous\\ treatments.} & \makecell{Only based on\\ observed confounders $\mathbf{X}$\\ and treatment $T$.\\ Possible violation\\ of positivity.\\ Sensitive to the\\ estimation of the PS.} \\
        \hline
        \textbf{\makecell{Lower bound\\ via RCT}} & \makecell{High sample size\\ (RCT and observational data),\\ high correlation between\\ $\mathbf{U}$ and $Y(0)$ or $Y(1)$.} & \makecell{Leverages the observed confounders $\mathbf{X}$,\\ the treatment $T$ and the outcome $Y$.\\ Based on external data instead of\\ potentially biased a priori.} & \makecell{Need of an RCT with\\ same treatment and outcome.\\ The RCT must share\\ some observed covariates with\\ the observational data.\\ Does not detect\\ confounders that cancel\\ out on average. \\ Multiple assumptions.} \\
        \hline
        \textbf{\makecell{Lower bound\\ via NCO}} & \makecell{High sample size,\\ high correlation\\ between $\mathbf{U}$ and $W$,\\ good NCO (close to ideal).} & \makecell{Easier to find a NCO\\ than an RCT. Based on\\ additional data instead of\\ potentially biased a priori.\\ Compatible with continuous\\ treatments.} & \makecell{NCOs are usually\\ approximately $\mathbf{U}$-comparable,\\ which can bias the estimation.} \\
        \hline
        \textbf{\makecell{Sensitivity \\ contour plot}} & \makecell{Enough observed confounders $\mathbf{X}$,\\ model assigned to the\\ treatment and to the outcome,\\ or even to the confounders.} & \makecell{Visually intuitive. Defines a\\ covariate-independent\\ robustness value.} & \makecell{Model-based.} \\
        \hline
    \end{tabular}
    \end{adjustbox}
\end{table*}

\subsubsection{Effect of the correlation between \texorpdfstring{$\mathbf{X}$}{} and \texorpdfstring{$\mathbf{U}$}{}} \label{sec:effect_rho_XU}

In Experiments~4 and 5, we added a low correlation between $\mathbf{X}$ and $\mathbf{U}$, and moved $\rho_{\mathbf{U}, Y(0)}$ and $\rho_{\mathbf{U}, W}$ from moderate to high levels. Note that the results from Experiment~4 are the same as the ones from Experiment~2, where $\mathbf{X}$ and $\mathbf{U}$ were uncorrelated and $\rho_{\mathbf{U}, Y(0)}$ and $\rho_{\mathbf{U}, W}$ were low, and results from Experiment~5 are identical to the ones from Experiment~3.

In Experiment~6, the correlation between $\mathbf{X}$ and $\mathbf{U}$ increased and this caused the median estimate via RCT to drop as compared with Experiment~5, even if $\rho_{\mathbf{U}, Y(0)}$ and $\rho_{\mathbf{U}, W}$ were high. Indeed, as the correlation between $\mathbf{X}$ and $\mathbf{U}$ increases, observing $\mathbf{X}$ becomes similar to observing $\mathbf{U}$, which can be interpreted as a return to the classical ignorability assumption (Assumption~\ref{ass:X-ignorability}), i.e.\ when the true $\Gamma = \Gamma^\star$ should be equal to 1. On the other hand, the estimation via NCO seems more robust to a change in the correlation between $\mathbf{X}$ and $\mathbf{U}$. This is supported by Experiments~5 and 6, in which the estimation declined but remained near the true value $\Gamma^\star$.

Between Experiments~3, 5 and 6, $\rho_{\mathbf{X}, \mathbf{U}}$ increased but the estimation via IB did not change. However, a variation in the estimation could have been expected because $T$ depends on $\mathbf{X}$ and $\mathbf{U}$. This was not observed because of our simulation setting (see Remark~\ref{rem:corr_XU_and_T} in appendix).

\subsubsection{Effects of the true confounding strength \texorpdfstring{$\Gamma^\star$}{}, of the sample size and of the dimension \texorpdfstring{$p_\mathbf{X}$}{}} \label{sec:effect_gamma_n_px}

In Experiments~1, 7 and 8, we changed the true confounding strength $\Gamma^\star$ while keeping other parameters constant. In Experiment~1, $\Gamma^\star = 8$, in Experiment~7, $\Gamma^\star = 5$ and in Experiment~8, $\Gamma^\star = 20$. First, IB stayed almost unaffected by a change in the value of $\Gamma^\star$. This can be explained by the fact that $\Gamma^\star$ has no impact on $\mathbf{X}$ and only little impact on $T$ in our simulation setting. On the contrary, lower bounds on $\Gamma$ via RCT and NCO decreased when the true confounding strength was reduced, and increased when the true confounding strength was augmented, which shows that these methods can adapt to variations of the true sensitivity parameter.

In Experiments~1 and 9, we altered the sample sizes of the RCT and of the observational data from 2000 (Experiment~1) to 500 (Experiment~9). IB with logistic regression estimator showed robustness to a decrease in the sample size, probably because the nominal propensity score $e(\mathbf{X})$ has a logistic form, but the range of the estimated values became slightly larger. This phenomenon is explained by a tendency to higher uncertainty when the sample size is reduced. However, estimations via IB with random forests dropped, most likely because the model was not well specified in comparison with logistic regression. The estimation via RCT also gave a lower median (4.00 when the sample size was 2000 versus 3.00 when the sample size was 500), a behavior that was described by \textcite{de2024hidden}. In addition, \textcite{de2024hidden} showed that, when the target population is $\mathbb{P}^{\mathrm{obs}^\prime}_\mathbf{X}$ as in our experiments, the lower bound on $\Gamma^\star$ is looser when $n_\mathrm{obs} / n_\mathrm{rct}$ gets smaller. This experimental result is not shown again in the present paper. Finally, the estimation via NCO does not substantially suffer from a reduction in the sample sizes but, as expected, the dispersion in the estimates increases slightly.

In Experiments~1 and 10, we increased the dimension of $\mathbf{X}$ from $p_\mathbf{X} = 5$ to 10. As observed, the estimations of a lower bound via RCT and NCO, and the estimation with IB using random forests do not notably change. However, IB using logistic regression performs better and the median estimate gets close to the true confounding strength when $p_\mathbf{X}$ is larger. Nevertheless, note that the range of the estimates is quite large (from 4.46 to 13.14). This result suggests that, when the propensity score estimator is correctly specified and many observed confounders are available in the data, IB can perform well.

\subsubsection{Effect of the type of negative control outcome} \label{sec:effect_nco}

In Experiments~1, 11 and 12, we compared different types of NCOs: ideal and approximately $\mathbf{U}$-comparable, as presented in Figure~\ref{fig:negative_control_outcome}. We recall that, as explained in Section~\ref{sec:via_negative_controls}, we expect the lower bound via NCO to be overestimated in case 1 (where an additional unobserved confounder biases the ($T$, $W$) causal relation but not the ($T$, $Y$) causal relation) and to be underestimated in case 2 (where an additional unobserved confounder biases the ($T$, $Y$) causal relation but not the ($T$, $W$) causal relation). This phenomenon was indeed observed in the experiments: in comparison with the ideal case (Experiment~1), the estimation via NCO in case 1 (Experiment~11) slightly increased, and the estimation in case 2 (Experiment~12) dropped. Observe that, for Experiment~11, the slight variations in the other methods were only due to randomness because a third unobserved confounder was included.

In Experiment~13, we aimed to design a realistic setup where the NCOs were approximately $\mathbf{U}$-comparable (case 1), $\mathbf{X}$ and $\mathbf{U}$ were weakly correlated, $\rho_{\mathbf{U}, Y(0)}$ and $\rho_{\mathbf{U}, W}$ were moderate, the RCT had a small sample size ($n_\mathrm{rct} = 200$) and the observational data had a bigger sample size ($n_\mathrm{rct} = 2000$), with potentially unobserved confounders of strength $\Gamma^\star = 8$. Results with IB are quite similar to the other experiments and estimations via NCOs are also close to the results from Experiment~1. However, the lower bound on $\Gamma^\star$ via RCT is quite loose even if the correlation between $\mathbf{U}$ and $Y(0)$ was moderate. As we saw before, a low sample size, here the sample size of the RCT, and a non-null correlation between $\mathbf{X}$ and $\mathbf{U}$ lead to underestimated lower bounds.

In practice, if good negative control outcomes can be found, the lower bound via NCO could be better than the lower bound with RCT, as RCTs often have a low sample size. Of course, if the additional unobserved confounder ($U_2$ in Figure~\ref{fig:negative_control_outcome}) had a strong effect on $W$ or on $Y$, the estimation via NCO could also be strongly affected.

\subsubsection{Effect of the type of interval (PEI or CI)} \label{sec:effect_type_interval}

In Experiments~1 and 14, we compared the impact of using the CI as the sensitivity interval instead of the PEI with the methods leveraging RCTs and NCOs. Confidence intervals are, in practice, larger than PEIs, so they can lead to underestimated lower bounds on $\Gamma^\star$. In our experiments, we indeed observed that the estimations decreased when the CIs were used instead of the PEIs. Moreover, validity of the test from \textcite{de2024hidden} was obtained with the PEI and not the CI. This motivated us to use the PEI instead of the CI in the previous experiments.

\begin{figure}[h]
    \centering
    \includegraphics[width=\linewidth]{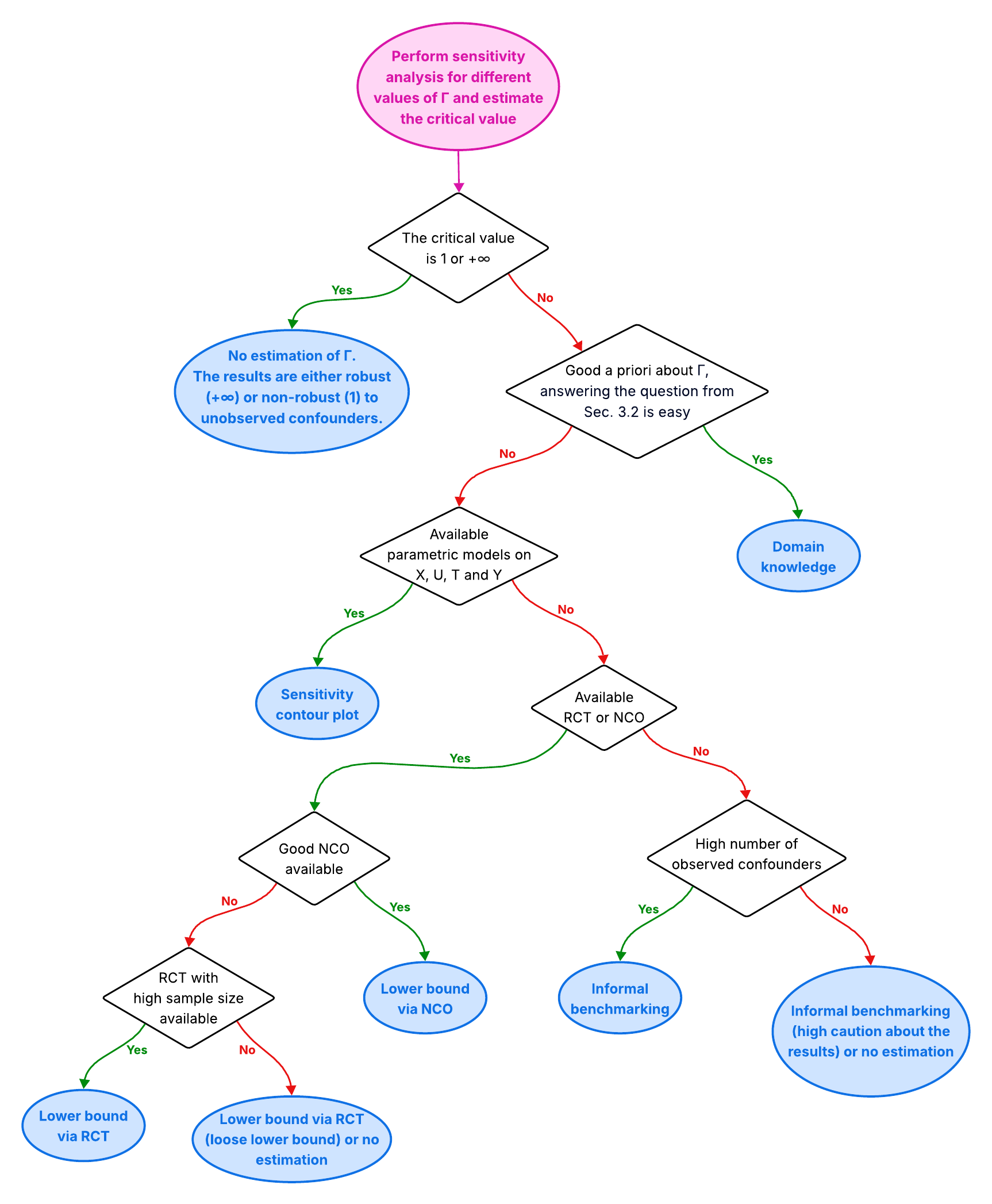}
    \caption{Decision tree to choose the method for estimating the confounding strength $\Gamma$. Start with the pink circle at the top, answer questions in the diamond shapes and reach the leaves of the tree in blue to know which method to choose.}
    \label{fig:decision_tree}
\end{figure}

\section{Conclusion and discussion} \label{sec:conclusion}

In this comparative review, we aimed to present and evaluate the different methods to estimate or lower bound the confounding strength $\Gamma$ that appears in sensitivity models for IPW estimators (MSM and CMSM) and, to a lesser extent, in the RSM. We focused on methods that do not assign a functional form to the confounders, treatment and outcome, and compared them in an extensive simulation study in the binary treatment case. Table~\ref{tab:methods_summary} provides a synthesis of the conditions for applying each method, as well as a comparison of their strengths and weaknesses, to support informed decision-making. To complete this table, a decision tree is provided in Figure~\ref{fig:decision_tree} to help choosing a method by answering simple questions.

To our knowledge, we proposed the first algorithm and experimental results on the estimation of a lower bound on $\Gamma$ using negative control outcomes.

An examination of Table~\ref{tab:methods_summary} reveals that none of the methods can be considered entirely optimal. However, practice in epidemiology shows us that, when risk ratios are used to measure treatment effects, the E-value~\parencite{vanderweele2017sensitivity} is a widely used quantity to assess the robustness of the results to unobserved confounders. Therefore, a possible avenue for future research would be to design a counterpart to the E-value for difference in means, such as the ATE, or for expected potential outcomes. The robustness value introduced by \textcite{cinelli2020making} could open the path to such new work.

Furthermore, in Table~\ref{tab:methods_summary}, we recommend to favor certain methods when correlations with the unobserved confounders are high, but, in practice, these correlations cannot be easily estimated. Thus, developing reliable approaches for their quantification will be essential to enhance the practical utility of sensitivity analyses.

\section*{Acknowledgments}

We thank Emily Lanoy for thoughtful discussions and helpful suggestions on the framing and scope of this work. The research in this paper was supported by Sanofi and Inria (Institut National de Recherche en Informatique et en Automatique) through the CIFRE program (Convention Industrielle de Formation par la Recherche).

\section*{Conflicts of interest}

Jean-Baptiste Baitairian, Bernard Sebastien and Rana Jreich are Sanofi employees and may hold shares and/or stock options in the company. Agathe Guilloux is employed by Inria. Sandrine Katsahian is employed by Université Paris-Cité and Assistance Publique - Hôpitaux de Paris (AP-HP).

\printbibliography

\newpage

\appendix

\section{ASSUMPTIONS, PROOFS AND ADDITIONAL DETAILS}

\subsection{Assumptions} \label{app:assumptions}

We make the following assumption about the nature of the kernel introduced in Equation~\eqref{eqn:conf_estim_theta_h_t}. Common choices of $K$ include the Epanechnikov or Gaussian kernels.
\begin{assumption}[Kernel]
    $K : \mathbb{R} \to \mathbb{R}_+$ is a symmetric and integrable function, with $\int_{-1}^1 K(u) \, \mathrm{d}u = 1$.
\end{assumption}

\subsection{Proof of Proposition~\ref{prop:true_IPW_theta1}} \label{sec:proof_true_IPW_theta1}

\begin{align*}
    \theta(1) & \underset{(1)}{=} \E[\E[Y(1)|\mathbf{X}, \mathbf{U}]] \underset{(2)}{=} \E \biggl[ \frac{\E[T|\mathbf{X}, \mathbf{U}]}{\E[T|\mathbf{X}, \mathbf{U}]} \E[Y(1)|\mathbf{X},\mathbf{U}] \biggr] \\
    & \underset{(3)}{=} \E \biggl[ \frac{\E[T|\mathbf{X}, \mathbf{U}]}{e(\mathbf{X}, \mathbf{U})} \E[Y(1)|\mathbf{X},\mathbf{U}] \biggr] \underset{(4)}{=} \E \biggl[ \frac{\E[T Y(1)|\mathbf{X}, \mathbf{U}]}{e(\mathbf{X}, \mathbf{U})} \biggr] \\
    & \underset{(5)}{=} \E \biggl[ \E \biggl[ \frac{TY}{e(\mathbf{X}, \mathbf{U})} \bigg| \mathbf{X}, \mathbf{U} \biggr] \biggr] \underset{(6)}{=} \E \biggl[ \frac{TY}{e(\mathbf{X}, \mathbf{U})} \biggr]
\end{align*}
Equalities~(1) and (6) are obtained using the tower property. Equality~(2) is obtained after multiplying and dividing by the same quantity under Assumption~\ref{ass:positivity}. Equality~(3) uses the fact that $\E[T|\mathbf{X}, \mathbf{U}] = 1 \times \mathbb{P}(T=1|\mathbf{X}, \mathbf{U}) + 0 \times \mathbb{P}(T=0|\mathbf{X}, \mathbf{U}) = e(\mathbf{X}, \mathbf{U})$. Equality~(4) uses ($\mathbf{X}, \mathbf{U}$)-ignorability (Assumption~\ref{ass:XU-ignorability}). Finally, Equality~(5) uses SUTVA: $TY = T(TY(1) + (1-T)Y(0)) = T^2 Y(1) + T(1-T)Y(0) = TY(1)$.

\begin{table*}[h]
    \centering
    \caption{Summary of the references associated with each method for estimating the confounding strength $\Gamma$.}
    \label{tab:methods_references}
    \begin{tabular}{|c|l|}
        \hline
        \textbf{Method} & \makecell[c]{\textbf{References}} \\
        \hline
        \hline
        \textbf{\makecell{Caveats,\\ no estimation}} & \makecell[l]{\textcite{scharfstein1999adjusting}, \textcite{robins2002covariance},\\ \textcite{kallus2018confounding}} \\
        \hline
        \textbf{\makecell{Critical value,\\ range of $\Gamma$,\\ domain knowledge}} & \makecell[l]{\textbf{Range of $\Gamma$}\\ \textcite{tan2006distributional}, \textcite{kallus2018confounding},\\ \textcite{zhao2019sensitivity},\\ \textcite{jesson2021quantifying}, \textcite{dorn2022sharp},\\ \textcite{yadlowsky2022bounds},\\ \textcite{jesson2022scalable}, \textcite{dorn2024doubly}\\ \textbf{Critical value}\\ \textcite{tan2006distributional}, \textcite{vanderweele2017sensitivity},\\ \textcite{jesson2021quantifying},\\ \textcite{yadlowsky2022bounds}, \textcite{dorn2022sharp},\\ \textcite{dorn2024doubly},\\ \textcite{frauen2024sharp},\\ \textcite{de2024hidden}, \textcite{baitairian2025sharp}\\ \textbf{Domain knowledge}\\ \textcite{robins2002covariance}, \textcite{oprescu2023b},\\ \textcite{frauen2024sharp}} \\
        \hline
        \textbf{\makecell{Informal\\ benchmarking}} & \makecell[l]{\textbf{Leave-one-out}\\ \textcite{imbens2003sensitivity}, \textcite{kallus2018confounding}, \textcite{kallus2019interval},\\ \textcite{cinelli2020making}, \textcite{kallus2021minimax}, \textcite{dorn2022sharp},\\ \textcite{lu2023flexible}, \textcite{oprescu2023b}, \textcite{dorn2024doubly},\\ \textcite{mcclean2024calibrated},\\ \textcite{frauen2024sharp},\\ \textcite{baitairian2025sharp}, \textcite{zhang2025enhanced}\\ \textbf{Leave-multiple-out}\\ \textcite{bonvini2022sensitivity}, \textcite{mcclean2024calibrated}} \\
        \hline
        \textbf{\makecell{Lower bound\\ via RCT}} & \textcite{yadlowsky2022bounds}, \textcite{de2024hidden} \\
        \hline
        \textbf{\makecell{Lower bound\\ via NCO}} & \textcite{lipsitch2010negative}, \textcite{kallus2018confounding} \\
        \hline
        \textbf{\makecell{Sensitivity\\ contour plot}} & \textcite{hsu2013calibrating}, \textcite{zhang2020calibrated}, \textcite{cinelli2020making} \\
        \hline
    \end{tabular}
\end{table*}

\subsection{Other sensitivity models} \label{app:other_sensitivity_models}

\subsubsection{Sensitivity models for binary treatments}

Rosenbaum's Sensitivity Model~\parencite{rosenbaum2002observational} assumes that units who appear similar in terms of observed confounders $\mathbf{X}$ but are different in terms of unobserved confounders $\mathbf{U}$ may differ in their odds of being treated by at most a factor of $\Gamma \geq 1$, the confounding strength.
\begin{definition}[RSM, formulation with $\mathbf{U}$] \label{def:RSM_U}
    Let $\Gamma \geq 1$ be a fixed sensitivity parameter. Under positivity (Assumption~\ref{ass:positivity}), the RSM is defined as the set $\mathrm{RSM}_\mathbf{U}(\Gamma)$ of propensity scores such that
    \begin{align*}
        \mathrm{RSM}_\mathbf{U}(\Gamma) \coloneq \{& e(\cdot, \cdot); \; \forall (\mathbf{x}, \mathbf{u}_1, \mathbf{u}_2) \in \mathcal{X} \times \mathcal{U} \times \mathcal{U}, \\
        & \Gamma^{-1} \leq \mathrm{OR}(e(\mathbf{x}, \mathbf{u}_1), e(\mathbf{x}, \mathbf{u}_2)) \leq \Gamma \}.
    \end{align*}
\end{definition}
An equivalent formulation with potential outcomes can also be defined after introducing the notation $e_a(\cdot, \cdot) \coloneq \mathbb{P}(T=1 | \mathbf{X}=\cdot, Y(a)=\cdot)$, where $a \in \{0,1\}$.
\begin{definition}[RSM, formulation with $Y(0)$ or $Y(1)$] \label{def:RSM_Y}
    Let $\Gamma \geq 1$ be a fixed sensitivity parameter. Under positivity (Assumption~\ref{ass:positivity}), the RSM is defined as the set $\mathrm{RSM}_{Y(a)}(\Gamma)$ of propensity scores such that
    \begin{align*}
        \mathrm{RSM}_{Y(a)}(\Gamma) \coloneq \{& e_a(\cdot, \cdot); \; \forall (\mathbf{x}, y_1, y_2) \in \mathcal{X} \times \mathcal{Y} \times \mathcal{Y}, \\
        & \Gamma^{-1} \leq \mathrm{OR}(e_a(\mathbf{x}, y_1), e_a(\mathbf{x}, y_2)) \leq \Gamma \},
    \end{align*}
    where $a \in \{0,1\}$.
\end{definition}
\textcite{yadlowsky2022bounds} formally prove that the formulations with $\mathbf{U}$ (Definition~\ref{def:RSM_U}) or one of the two potential outcomes (Definition~\ref{def:RSM_Y}) are equivalent (see their Lemma~2.1).

In Section~\ref{sec:sensitivity_models}, we presented a formulation of the MSM with the unobserved confounders $\mathbf{U}$. Again, a formulation where $\mathbf{U}$ is replaced by one of the two potential outcomes can be defined and is equivalent to the one in Definition~\ref{def:MSM_U}. See Lemma~1 from \textcite{tan2024model} for a formal proof of this equivalence.
\begin{definition}[MSM, formulation with $Y(0)$ or $Y(1)$] \label{def:MSM_Y}
    Let $\Gamma \geq 1$ be a fixed sensitivity parameter. Under positivity (Assumption~\ref{ass:positivity}), the MSM is defined as the set $\mathrm{MSM}_{Y(a)}(\Gamma)$ of propensity scores such that
    \begin{align*}
        \mathrm{MSM}_{Y(a)}(\Gamma) \coloneq \{& e_a(\cdot, \cdot); \; \forall (\mathbf{x}, y) \in \mathcal{X} \times \mathcal{Y}, \\
        & \Gamma^{-1} \leq \mathrm{OR}(e_a(\mathbf{x}, y), e(\mathbf{x})) \leq \Gamma \},
    \end{align*}
    where $a \in \{0,1\}$.
\end{definition}
In their Proposition~3, \textcite{zhao2019sensitivity} show that, for $\Gamma \geq 1$, the RSM from Definition~\ref{def:RSM_Y} and the MSM from Definition~\ref{def:MSM_Y} are linked by the following inclusion: $\mathrm{MSM}_{Y(a)}(\sqrt{\Gamma}) \subseteq \mathrm{RSM}_{Y(a)}(\Gamma)$, for $a \in \{0,1\}$. The inverse inclusion is only true if the propensity scores are \textit{compatible}, in a certain way defined in their article. See also Appendix~I of \textcite{tan2024model} for additional elements on the link between the RSM and the MSM.

\subsubsection{Sensitivity models for continuous treatments}

The original Continuous Marginal Sensitivity Model from \textcite{jesson2022scalable} is given hereafter. As put forward in Proposition~2.3 by \textcite{baitairian2025sharp}, it is equivalent to the formulation in terms of unobserved confounders $\mathbf{U}$ from Definition~\ref{def:CMSM_U}. We assume that, for all $(\mathbf{x}, y) \in \mathcal{X} \times \mathcal{Y}$, the conditional density $t \mapsto f(T=t | \mathbf{X}=\mathbf{x},Y(t)=y)$ is absolutely continuous with respect to $t \mapsto f(T=t | \mathbf{X}=\mathbf{x})$.
\begin{definition}[CMSM, formulation with $Y(t)$] \label{def:CMSM_Y}
    Let $\Gamma \geq 1$ be a fixed sensitivity parameter. The CMSM is defined as the set $\mathrm{CMSM}_{Y(t)}(\Gamma)$ of generalized propensity scores such that
    \begin{align*}
        & \mathrm{CMSM}_{Y(t)}(\Gamma) \coloneq \Biggl\{ f(T=t | \mathbf{X}=\cdot, Y(t)=\cdot); \\
        & \forall (\mathbf{x}, y) \in \mathcal{X} \times \mathcal{Y}, \, \Gamma^{-1} \leq \frac{f(T=t | \mathbf{X}=\mathbf{x}, Y(t)=y)}{f(T=t | \mathbf{X}=\mathbf{x})} \leq \Gamma \Biggr\},
    \end{align*}
    where $t \in \mathcal{T}$.
\end{definition}

\subsection{Summary of the references for each method}

Table~\ref{tab:methods_references} provides a summary of the references associated with each method for estimating or lower bounding the confounding strength $\Gamma$.

\section{EXPERIMENTS}

The experiments were conducted on Amazon EC2 c6i.4xlarge instances. The code was developed under the R Statistical Software version 4.3.2~\parencite{rstatisticalsoftware}. See Section~\ref{app:libraries_licenses} for an exhaustive list of the libraries that were used and their corresponding licenses.

\subsection{Additional implementation details} \label{app:add_impl_details}

\subsubsection{Sensitivity analysis via Quantile Balancing (QB)}

When computing confidence intervals via the percentile bootstrap method, we use parallel computing on 16 CPUs to speed up computations.

\subsubsection{Informal benchmarking}

When the propensity scores are estimated with random forests, we fine-tune four hyperparameters from the \texttt{probability\_forest} function (\texttt{grf} package) via 5-fold cross-validation using a log-loss: \texttt{num.trees} (the number of trees grown in the forest), \texttt{min.node.size} (a target for the minimum number of observations in each tree leaf), \texttt{sample.fraction} (fraction of the data used to build each tree), and \texttt{mtry} (number of variables tried for each split). We try all combinations involving the parameters summarized in Table~\ref{tab:finetuning_inf_bench}.

\begin{table}[h]
    \centering
    \begin{tabular}{|c|c|}
        \hline
        \textbf{Parameter} & \textbf{Values} \\
        \hline
        \texttt{num.trees} & (30, 40, 50) \\
        \texttt{min.node.size} & (5, 10, 20) \\
        \texttt{sample.fraction} & (0.1, 0.2, 0.5) \\
        \texttt{mtry} & (2, 4, default) \\
        \hline
    \end{tabular}
    \caption{Fine-tuned hyperparameters with tested values, where ``default" is the default value of the \texttt{mtry} parameter in \texttt{probability\_forest}.}
    \label{tab:finetuning_inf_bench}
\end{table}

\subsubsection{Lower bound via RCTs}

When computing the empirical standard deviations $\hat{\sigma}$, $\hat{\sigma}_-$ and $\hat{\sigma}_+$ on 100 bootstrap samples, we use parallel computing on 16 CPUs to speed up computations.

\subsection{Details about the simulated dataset and additional results} \label{app:details_sim_data_and_results}

\subsubsection{Complete simulation setup} \label{app:complete_simulation_setup}

To compare the three methods, we extended the simulation setup from \textcite{de2024hidden} by increasing the dimensionality of $\mathbf{X}$ and $\mathbf{U}$. In our experiments, unless stated otherwise, we considered $p_\mathbf{X} = 5$ observed confounders and $p_\mathbf{U} = 2$ unobserved confounders. We used multidimensional observed confounders to be able to apply informal benchmarking and we chose
\begin{equation*}
    \mathbf{X}_\mathrm{rct} \sim \mathcal{U}(-0.9, 0.9)^{p_\mathbf{X}} \quad \text{and} \quad \mathbf{X}_\mathrm{obs} \sim \mathcal{U}(-1, 1)^{p_\mathbf{X}},
\end{equation*}
where $\mathcal{U}(a, b)^p$ is the uniform distribution on $(a, b)$ of dimension $p$, to satisfy support inclusion (Assumption~\ref{ass:support_inclusion}). For $j$ in $[\![1, p_\mathbf{U}]\!]$, conditionally on $\mathbf{X}=\mathbf{x}$, we designed each unobserved confounder $U_j$ as
\begin{equation*}
    U_j|\mathbf{X}=\mathbf{x} \sim \lambda G + (1 - \lambda) \beta_j^T \mathbf{x} = \mathcal{N} \bigl( (1-\lambda) \beta_j^T \mathbf{x}, \, \lambda^2 \bigr),
\end{equation*}
with $G \sim \mathcal{N}(0, 1)$, $0 < \lambda < 1$, and $\beta_j = (\beta_{j,1}, \dots, \beta_{j,p_\mathbf{X}}) \in \mathbb{R}^{p_\mathbf{X}}_+$. For simplicity, we considered that $\mathbf{U} = \sum_{j=1}^{p_\mathbf{U}} U_j / p_\mathbf{U}$.

For the observational data, we chose the distribution of $T$ conditionally on $\mathbf{X}$ and $\mathbf{U}$ to be a Bernoulli which satisfied the MSM with a true sensitivity parameter $\Gamma^\star$. In the observational data, we designed the true propensity score $e^\mathrm{obs}(\mathbf{X}, \mathbf{U}) = \mathbb{P}^\mathrm{obs}(T=1|\mathbf{X}, \mathbf{U})$ such that it marginalizes to the nominal propensity score $e^\mathrm{obs}(\mathbf{X}) = \mathbb{P}^\mathrm{obs}(T=1|\mathbf{X}) = \mathrm{logistic}(\delta^T \mathbf{X} + 0.5)$, with $\delta \in \mathbb{R}^{p_\mathbf{X}}$. To do so, we chose a true propensity score of the form
\begin{equation} \label{eqn:adversarial_prop_score}
    e^\mathrm{obs}(\mathbf{X}, \mathbf{U}) = l(\mathbf{X}) \cdot \mathds{1}(\mathbf{U} > t(\mathbf{X})) + u(\mathbf{X}) \cdot \mathds{1}(\mathbf{U} \leq t(\mathbf{X})),
\end{equation}
where
\begin{align*}
    & l(\mathbf{X}) = \frac{e^\mathrm{obs}(\mathbf{X})}{e^\mathrm{obs}(\mathbf{X}) + (1-e^\mathrm{obs}(\mathbf{X})) \Gamma^\star} \quad \text{and} \\
    & u(\mathbf{X}) = \frac{e^\mathrm{obs}(\mathbf{X})}{e^\mathrm{obs}(\mathbf{X}) + (1-e^\mathrm{obs}(\mathbf{X})) /\Gamma^\star}.
\end{align*}
We want to find the threshold $t(\mathbf{X})$ such that $\E[e^\mathrm{obs}(\mathbf{X}, \mathbf{U}) | \mathbf{X}] = e^\mathrm{obs}(\mathbf{X})$. Therefore, by applying the expectancy conditionally on $\mathbf{X}$ on Equation~\eqref{eqn:adversarial_prop_score} and using the previous constraint, we get
\begin{align*}
    \mathbb{P}(\mathbf{U} \leq t(\mathbf{X}) | \mathbf{X}) = \frac{e^\mathrm{obs}(\mathbf{X}) - l(\mathbf{X})}{u(\mathbf{X}) - l(\mathbf{X})}.
\end{align*}
As we assumed that $\mathbf{U} = \sum_{j=1}^{p_\mathbf{U}} U_j / p_\mathbf{U}$, the previous equation becomes
\begin{align*}
    \mathbb{P} \Bigl( \sum_j U_j / p_\mathbf{U} \leq t(\mathbf{X}) \Big| \mathbf{X} \Bigr) = \frac{e^\mathrm{obs}(\mathbf{X}) - l(\mathbf{X})}{u(\mathbf{X}) - l(\mathbf{X})}.
\end{align*}
As the distribution of $U_j|\mathbf{X}=\mathbf{x}$ is Gaussian, if we assume that the $U_j$ are mutually independent conditionally on $\mathbf{X}=\mathbf{x}$, the same applies to $\sum_j U_j / p_\mathbf{U}$ conditionally on $\mathbf{X}=\mathbf{x}$, so
\begin{align*}
    \sum_j U_j / p_\mathbf{U} \, | \, \mathbf{X}=\mathbf{x} \sim \mathcal{N} \Bigl( \sum_j (1-\lambda) \beta_j^T \mathbf{x} / p_\mathbf{U}, \, \lambda^2 / p_\mathbf{U} \Bigr).
\end{align*}
Therefore, the threshold $t(\mathbf{X})$ is the quantile of a normal distribution, which can be found by using the \texttt{qnorm} function:
\begin{align*}
    t(\mathbf{X}) = \texttt{qnorm} \biggl(& \frac{e^\mathrm{obs}(\mathbf{X}) - l(\mathbf{X})}{u(\mathbf{X}) - l(\mathbf{X})}, \\
    & \texttt{mean} = \sum_j (1-\lambda) \beta_j^T \mathbf{x} / p_\mathbf{U}, \, \texttt{sd} = \lambda / \sqrt{p_\mathbf{U}} \biggr).
\end{align*}
In the RCT data, we simply defined the true propensity score as $e^\mathrm{rct}(\mathbf{X}, \mathbf{U}) = \mathbb{P}^\mathrm{rct}(T=1|\mathbf{X}, \mathbf{U}) = 1/2$ to satisfy internal validity and positivity (Assumption~\ref{ass:internal_validity}).

Denoting $p_W$ the number of negative control outcomes, the outcome $Y$ and negative control outcome $W_k$, for $k \in [\![1, p_W]\!]$, were defined via the following linear models
\begin{align*}
    Y(T) & = (2T-1) \theta_1^T \mathbf{X} + \theta_2^T \mathbf{U} + (2T-1) \times \mathrm{ATE}^\star/2 + \varepsilon_Y, \\
    W_k & = \theta_{3, k}^T \mathbf{X} + \theta_{4, k}^T \mathbf{U} + \varepsilon_W,
\end{align*}
where $\mathrm{ATE}^\star$ is the true ATE, $(\theta_1, \theta_2, \theta_{3,k}, \theta_{4,k}) \in \mathbb{R}^{p_\mathbf{X}} \times \mathbb{R}^{p_\mathbf{U}} \times \mathbb{R}^{p_\mathbf{X}} \times \mathbb{R}^{p_\mathbf{U}}$, $\varepsilon_Y \sim \mathcal{N}(0, \sigma_Y^2)$ and $\varepsilon_W \sim \mathcal{N}(0, \sigma_W^2)$. Finally, to generate data that correspond to the three cases described in Figure~\ref{fig:negative_control_outcome} (ideal and approximately $\mathbf{U}$-comparable negative control outcomes), we removed one unobserved confounder---without loss of generality, the first one---from the generation process of $Y$ in case 1 (Figure~\ref{fig:negative_control_outcome}b), and one unobserved confounder from the generation process of $W$ in case 2 (Figure~\ref{fig:negative_control_outcome}c).

\begin{remark} \label{rem:corr_XU_and_T}
    The design of $e^\mathrm{obs}(\mathbf{X}, \mathbf{U})$ and $U_j|\mathbf{X}=\mathbf{x}$ imposes that, when the correlation between $\mathbf{X}$ and $\mathbf{U}$ varies through $\lambda$, $\mathbf{U}$ and $t(\mathbf{X})$ vary with the same factor $\lambda$ and thus, if $\mathbf{U} > t(\mathbf{X})$ for a certain value of $\lambda = \lambda_1$, it keeps the same order when $\lambda = \lambda_2 \neq \lambda_1$. Therefore, varying the correlation between $\mathbf{X}$ and $\mathbf{U}$ has no impact on $T$, even if this seems unnatural.
\end{remark}

In our implementation, for reproducibility purpose, we set the random seed to 1 and, unless stated otherwise, we used the parameters summarized in Table~\ref{tab:simu_param_values}.

\begin{table*}[h]
    \centering
    \caption{Parameter values used in the simulation setup to generate Figures~\ref{fig:sa_1} to \ref{fig:sa_14}. $\mathcal{U}(a, b)^p$ means that the vector of size $p$ was generated randomly according to a uniform distribution on $(a, b)$.}
    \label{tab:simu_param_values}
    \begin{adjustbox}{max width=\textwidth}
    \begin{tabular}{|l|lllllll|}
        \hline
        \textbf{Param.} & \textbf{Exp. 1} & \textbf{Exp. 2} & \textbf{Exp. 3} & \textbf{Exp. 4} & \textbf{Exp. 5} & \textbf{Exp. 6} & \textbf{Exp. 7} \\
        \hline
        $p_\mathbf{X}$ & 5 & 5 & 5 & 5 & 5 & 5 & 5 \\
        $p_\mathbf{U}$ & 2 & 2 & 2 & 2 & 2 & 2 & 2 \\
        $p_W$ & 2 & 2 & 2 & 2 & 2 & 2 & 2 \\
        $n_\mathrm{rct}$ & 2000 & 2000 & 2000 & 2000 & 2000 & 2000 & 2000 \\
        $n_\mathrm{obs}$ & 2000 & 2000 & 2000 & 2000 & 2000 & 2000 & 2000 \\
        $\mathrm{ATE}^\star$ & 0.25 & 0.25 & 0.25 & 0.25 & 0.25 & 0.25 & 0.25 \\
        $\Gamma^\star$ & 8 & 8 & 8 & 8 & 8 & 8 & 5 \\
        $\lambda$ & 1 & 1 & 1 & 0.85 & 0.85 & 0.2 & 1 \\
        $\beta_j$ & $\mathcal{U}(1.5, 2)^{p_\mathbf{X}}$ & $\mathcal{U}(1.5, 2)^{p_\mathbf{X}}$ & $\mathcal{U}(1.5, 2)^{p_\mathbf{X}}$ & $\mathcal{U}(1.5, 2)^{p_\mathbf{X}}$ & $\mathcal{U}(1.5, 2)^{p_\mathbf{X}}$ & $\mathcal{U}(1.5, 2)^{p_\mathbf{X}}$ & $\mathcal{U}(1.5, 2)^{p_\mathbf{X}}$ \\
        $\theta_1$ & $\mathcal{U}(0.1, 1)^{p_\mathbf{X}}$ & $\mathcal{U}(0.1, 1)^{p_\mathbf{X}}$ & $\mathcal{U}(0.1, 1)^{p_\mathbf{X}}$ & $\mathcal{U}(0.1, 1)^{p_\mathbf{X}}$ & $\mathcal{U}(0.1, 1)^{p_\mathbf{X}}$ & $\mathcal{U}(0.1, 1)^{p_\mathbf{X}}$ & $\mathcal{U}(0.1, 1)^{p_\mathbf{X}}$ \\
        $\theta_2$ & $\mathcal{U}(0.1, 0.2)^{p_\mathbf{U}}$ & $\mathcal{U}(0.01, 0.02)^{p_\mathbf{U}}$ & $\mathcal{U}(0.5, 1)^{p_\mathbf{U}}$ & $\mathcal{U}(0.01, 0.02)^{p_\mathbf{U}}$ & $\mathcal{U}(0.7, 1)^{p_\mathbf{U}}$ & $\mathcal{U}(0.7, 1)^{p_\mathbf{U}}$ & $\mathcal{U}(0.1, 0.2)^{p_\mathbf{U}}$ \\
        $\theta_{3, k}$ & $\mathcal{U}(0.1, 1)^{p_\mathbf{X}}$ & $\mathcal{U}(0.1, 1)^{p_\mathbf{X}}$ & $\mathcal{U}(0.1, 1)^{p_\mathbf{X}}$ & $\mathcal{U}(0.1, 1)^{p_\mathbf{X}}$ & $\mathcal{U}(0.1, 1)^{p_\mathbf{X}}$ & $\mathcal{U}(0.1, 1)^{p_\mathbf{X}}$ & $\mathcal{U}(0.1, 1)^{p_\mathbf{X}}$ \\
        $\theta_{4, k}$ & $\mathcal{U}(0.05, 0.1)^{p_\mathbf{U}}$ & $\mathcal{U}(0.01, 0.02)^{p_\mathbf{U}}$ & $\mathcal{U}(0.5, 1)^{p_\mathbf{U}}$ & $\mathcal{U}(0.01, 0.02)^{p_\mathbf{U}}$ & $\mathcal{U}(0.7, 1)^{p_\mathbf{U}}$ & $\mathcal{U}(0.7, 1)^{p_\mathbf{U}}$ & $\mathcal{U}(0.05, 0.1)^{p_\mathbf{U}}$ \\
        $\delta$ & $\mathcal{U}(0.1, 0.5)^{p_\mathbf{X}}$ & $\mathcal{U}(0.1, 0.5)^{p_\mathbf{X}}$ & $\mathcal{U}(0.1, 0.5)^{p_\mathbf{X}}$ & $\mathcal{U}(0.1, 0.5)^{p_\mathbf{X}}$ & $\mathcal{U}(0.1, 0.5)^{p_\mathbf{X}}$ & $\mathcal{U}(0.1, 0.5)^{p_\mathbf{X}}$ & $\mathcal{U}(0.1, 0.5)^{p_\mathbf{X}}$ \\
        $\sigma_Y$ & 0.1 & 0.1 & 0.1 & 0.1 & 0.1 & 0.1 & 0.1 \\
        $\sigma_W$ & 0.1 & 0.1 & 0.1 & 0.1 & 0.1 & 0.1 & 0.1 \\
        \hline
        \hline
        \textbf{Param.} & \textbf{Exp. 8} & \textbf{Exp. 9} & \textbf{Exp. 10} & \textbf{Exp. 11} & \textbf{Exp. 12} & \textbf{Exp. 13} & \textbf{Exp. 14} \\
        \hline
        $p_\mathbf{X}$ & 5 & 5 & 10 & 5 & 5 & 5 & 5 \\
        $p_\mathbf{U}$ & 2 & 2 & 2 & 3 & 3 & 3 & 2 \\
        $p_W$ & 2 & 2 & 2 & 2 & 2 & 2 & 2 \\
        $n_\mathrm{rct}$ & 2000 & 500 & 2000 & 2000 & 2000 & 200 & 2000 \\
        $n_\mathrm{obs}$ & 2000 & 500 & 2000 & 2000 & 2000 & 2000 & 2000 \\
        $\mathrm{ATE}^\star$ & 0.25 & 0.25 & 0.25 & 0.25 & 0.25 & 0.25 & 0.25 \\
        $\Gamma^\star$ & 20 & 8 & 8 & 8 & 8 & 8 & 8 \\
        $\lambda$ & 1 & 1 & 1 & 1 & 1 & 0.85 & 1 \\
        $\beta_j$ & $\mathcal{U}(1.5, 2)^{p_\mathbf{X}}$ & $\mathcal{U}(1.5, 2)^{p_\mathbf{X}}$ & $\mathcal{U}(1.5, 2)^{p_\mathbf{X}}$ & $\mathcal{U}(1.5, 2)^{p_\mathbf{X}}$ & $\mathcal{U}(1.5, 2)^{p_\mathbf{X}}$ & $\mathcal{U}(1.5, 2)^{p_\mathbf{X}}$ & $\mathcal{U}(1.5, 2)^{p_\mathbf{X}}$ \\
        $\theta_1$ & $\mathcal{U}(0.1, 1)^{p_\mathbf{X}}$ & $\mathcal{U}(0.1, 1)^{p_\mathbf{X}}$ & $\mathcal{U}(0.1, 1)^{p_\mathbf{X}}$ & $\mathcal{U}(0.1, 1)^{p_\mathbf{X}}$ & $\mathcal{U}(0.1, 1)^{p_\mathbf{X}}$ & $\mathcal{U}(0.1, 1)^{p_\mathbf{X}}$ & $\mathcal{U}(0.1, 1)^{p_\mathbf{X}}$ \\
        $\theta_2$ & $\mathcal{U}(0.1, 0.2)^{p_\mathbf{U}}$ & $\mathcal{U}(0.1, 0.2)^{p_\mathbf{U}}$ & $\mathcal{U}(0.1, 0.2)^{p_\mathbf{U}}$ & $\mathcal{U}(0.1, 0.2)^{p_\mathbf{U}}$ & $\mathcal{U}(0.1, 0.2)^{p_\mathbf{U}}$ & $\mathcal{U}(0.1, 0.2)^{p_\mathbf{U}}$ & $\mathcal{U}(0.1, 0.2)^{p_\mathbf{U}}$ \\
        $\theta_{3, k}$ & $\mathcal{U}(0.1, 1)^{p_\mathbf{X}}$ & $\mathcal{U}(0.1, 1)^{p_\mathbf{X}}$ & $\mathcal{U}(0.1, 1)^{p_\mathbf{X}}$ & $\mathcal{U}(0.1, 1)^{p_\mathbf{X}}$ & $\mathcal{U}(0.1, 1)^{p_\mathbf{X}}$ & $\mathcal{U}(0.1, 1)^{p_\mathbf{X}}$ & $\mathcal{U}(0.1, 1)^{p_\mathbf{X}}$ \\
        $\theta_{4, k}$ & $\mathcal{U}(0.05, 0.1)^{p_\mathbf{U}}$ & $\mathcal{U}(0.05, 0.1)^{p_\mathbf{U}}$ & $\mathcal{U}(0.05, 0.1)^{p_\mathbf{U}}$ & $\mathcal{U}(0.05, 0.1)^{p_\mathbf{U}}$ & $\mathcal{U}(0.05, 0.1)^{p_\mathbf{U}}$ & $\mathcal{U}(0.05, 0.1)^{p_\mathbf{U}}$ & $\mathcal{U}(0.05, 0.1)^{p_\mathbf{U}}$ \\
        $\delta$ & $\mathcal{U}(0.1, 0.5)^{p_\mathbf{X}}$ & $\mathcal{U}(0.1, 0.5)^{p_\mathbf{X}}$ & $\mathcal{U}(0.1, 0.5)^{p_\mathbf{X}}$ & $\mathcal{U}(0.1, 0.5)^{p_\mathbf{X}}$ & $\mathcal{U}(0.1, 0.5)^{p_\mathbf{X}}$ & $\mathcal{U}(0.1, 0.5)^{p_\mathbf{X}}$ & $\mathcal{U}(0.1, 0.5)^{p_\mathbf{X}}$ \\
        $\sigma_Y$ & 0.1 & 0.1 & 0.1 & 0.1 & 0.1 & 0.1 & 0.1 \\
        $\sigma_W$ & 0.1 & 0.1 & 0.1 & 0.1 & 0.1 & 0.1 & 0.1 \\
        \hline
    \end{tabular}
    \end{adjustbox}
\end{table*}

\subsubsection{Additional results} \label{app:add_results}

\begin{figure}
    \centering
    \includegraphics[width=\linewidth]{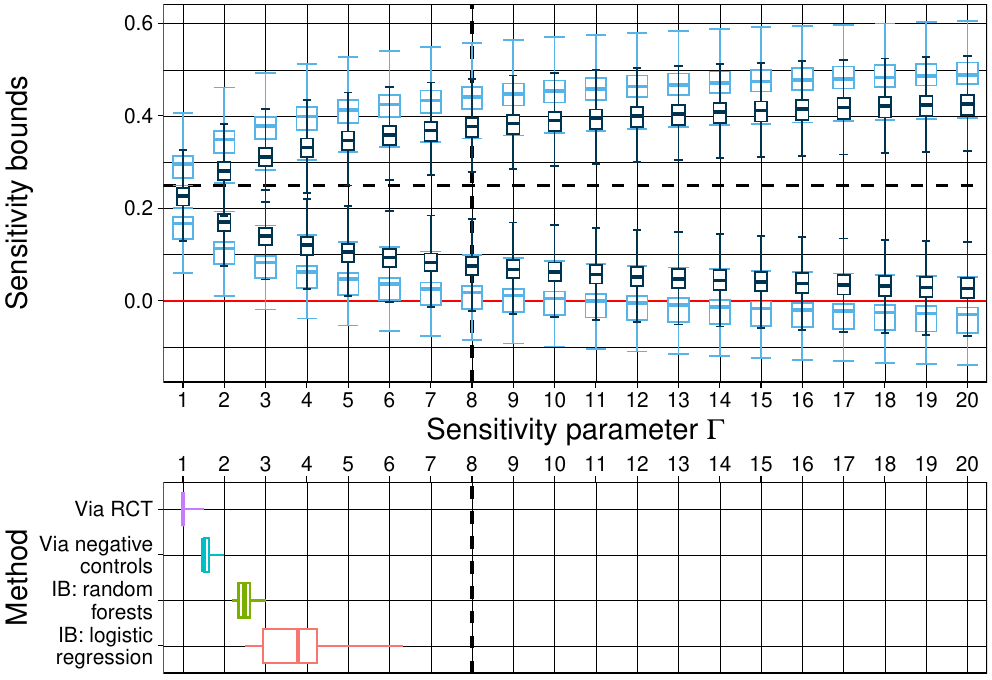}
    \caption{Sensitivity analysis n°2. IB: Informal Benchmarking.}
    \label{fig:sa_2}
\end{figure}

\begin{figure}
    \centering
    \includegraphics[width=\linewidth]{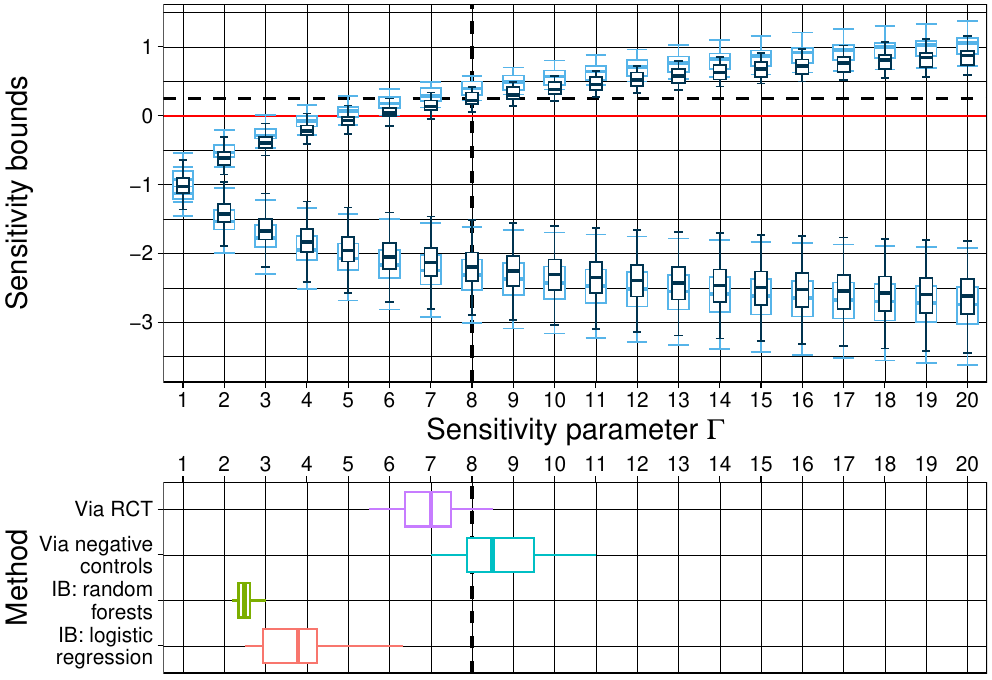}
    \caption{Sensitivity analysis n°3. IB: Informal Benchmarking.}
    \label{fig:sa_3}
\end{figure}

\begin{figure}
    \centering
    \includegraphics[width=\linewidth]{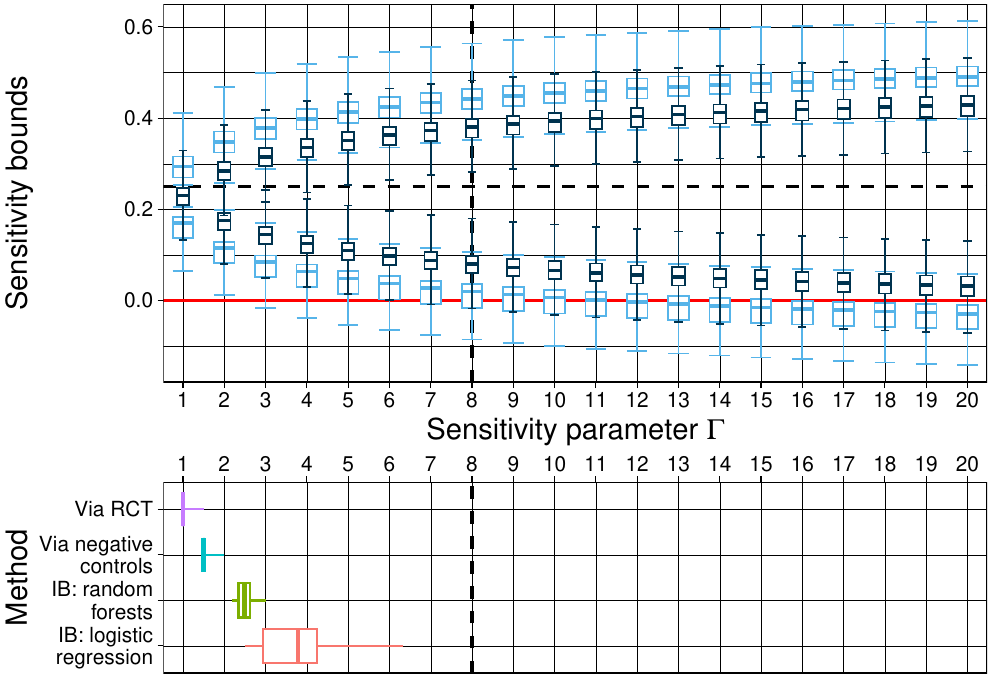}
    \caption{Sensitivity analysis n°4. IB: Informal Benchmarking.}
    \label{fig:sa_4}
\end{figure}

\begin{figure}
    \centering
    \includegraphics[width=\linewidth]{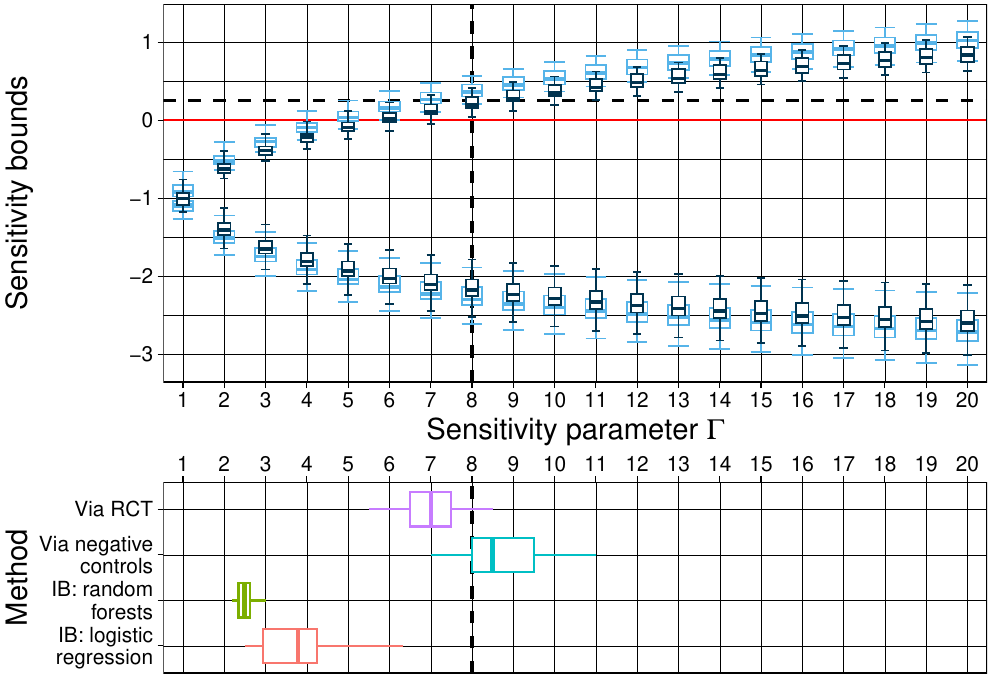}
    \caption{Sensitivity analysis n°5. IB: Informal Benchmarking.}
    \label{fig:sa_5}
\end{figure}

\begin{figure}
    \centering
    \includegraphics[width=\linewidth]{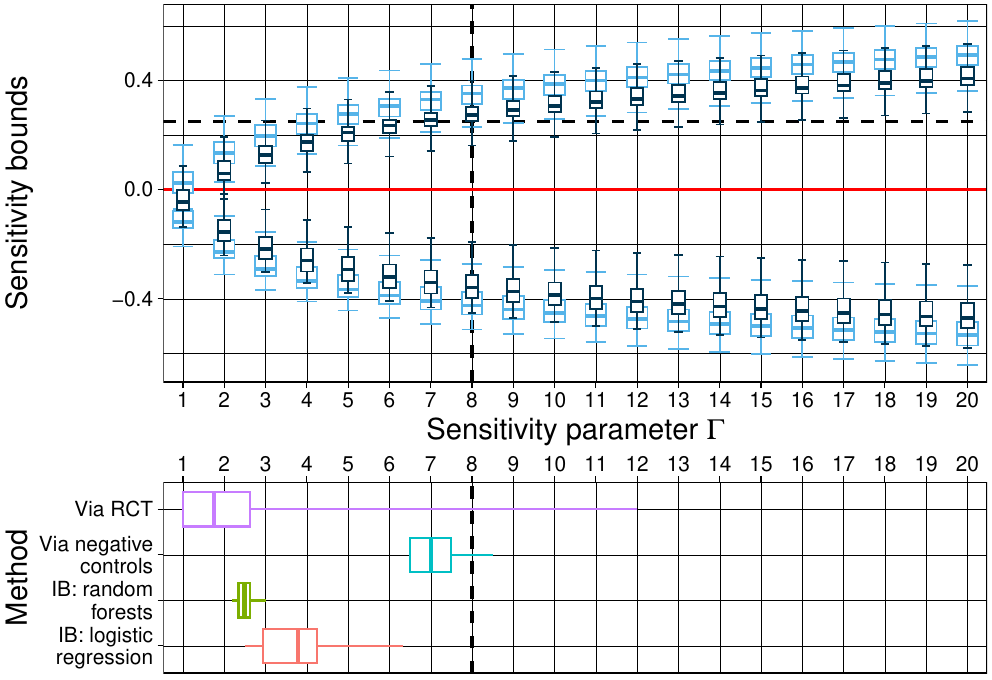}
    \caption{Sensitivity analysis n°6. IB: Informal Benchmarking.}
    \label{fig:sa_6}
\end{figure}

\begin{figure}
    \centering
    \includegraphics[width=\linewidth]{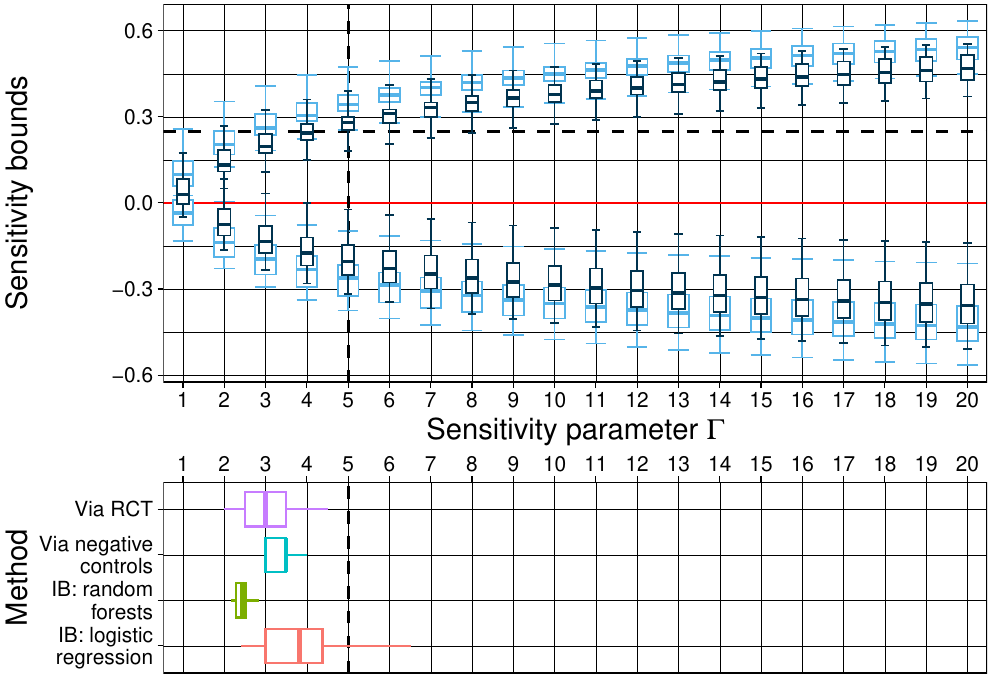}
    \caption{Sensitivity analysis n°7. IB: Informal Benchmarking.}
    \label{fig:sa_7}
\end{figure}

\begin{figure}
    \centering
    \includegraphics[width=\linewidth]{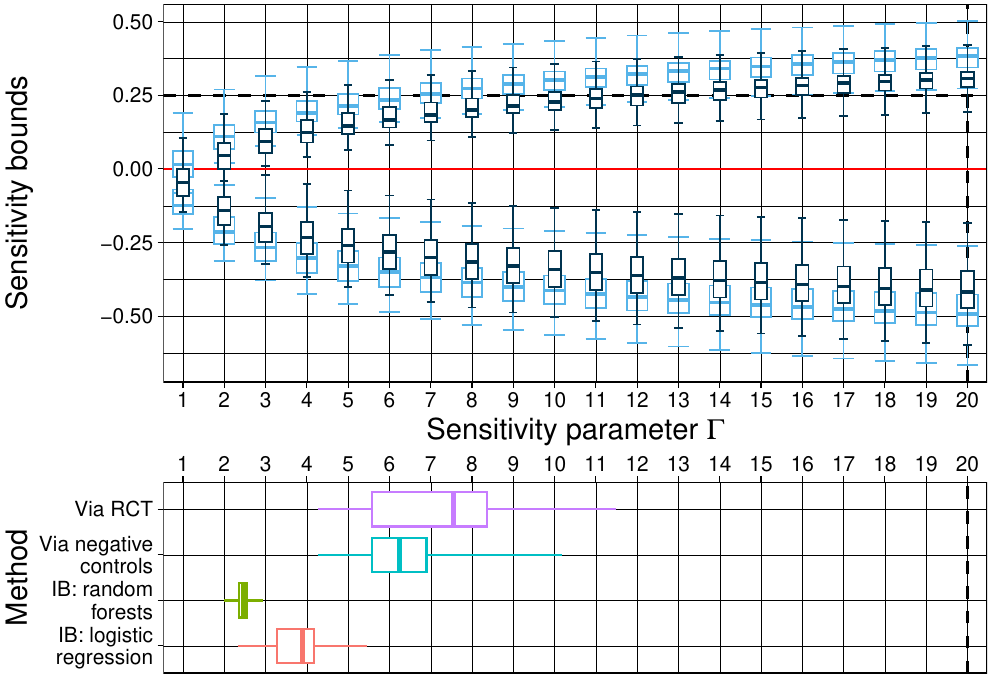}
    \caption{Sensitivity analysis n°8. IB: Informal Benchmarking.}
    \label{fig:sa_8}
\end{figure}

\begin{figure}
    \centering
    \includegraphics[width=\linewidth]{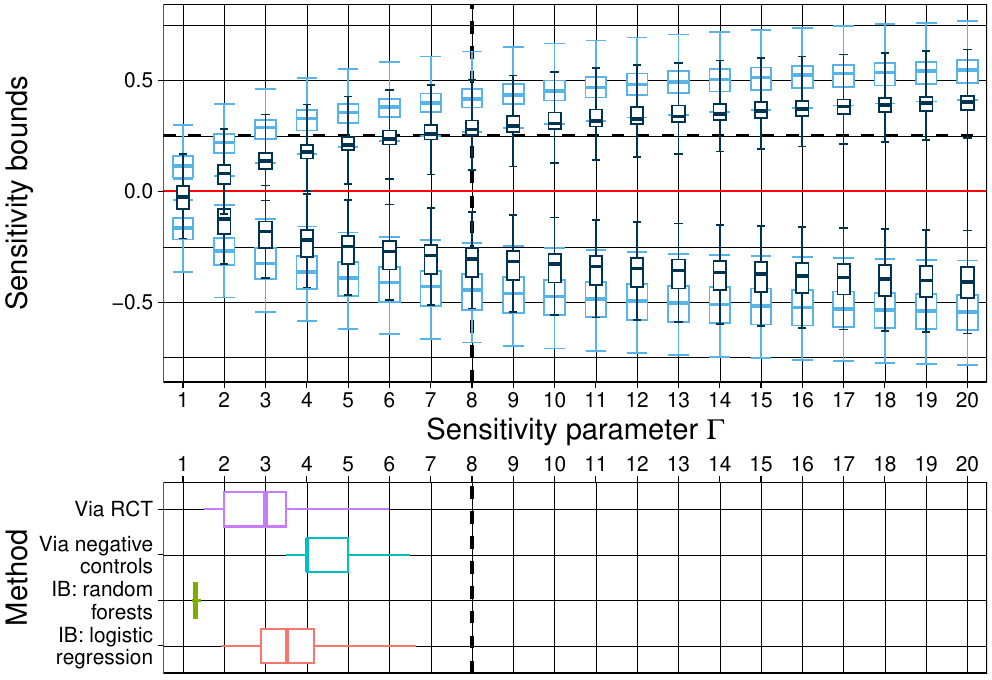}
    \caption{Sensitivity analysis n°9. IB: Informal Benchmarking.}
    \label{fig:sa_9}
\end{figure}

\begin{figure}
    \centering
    \includegraphics[width=\linewidth]{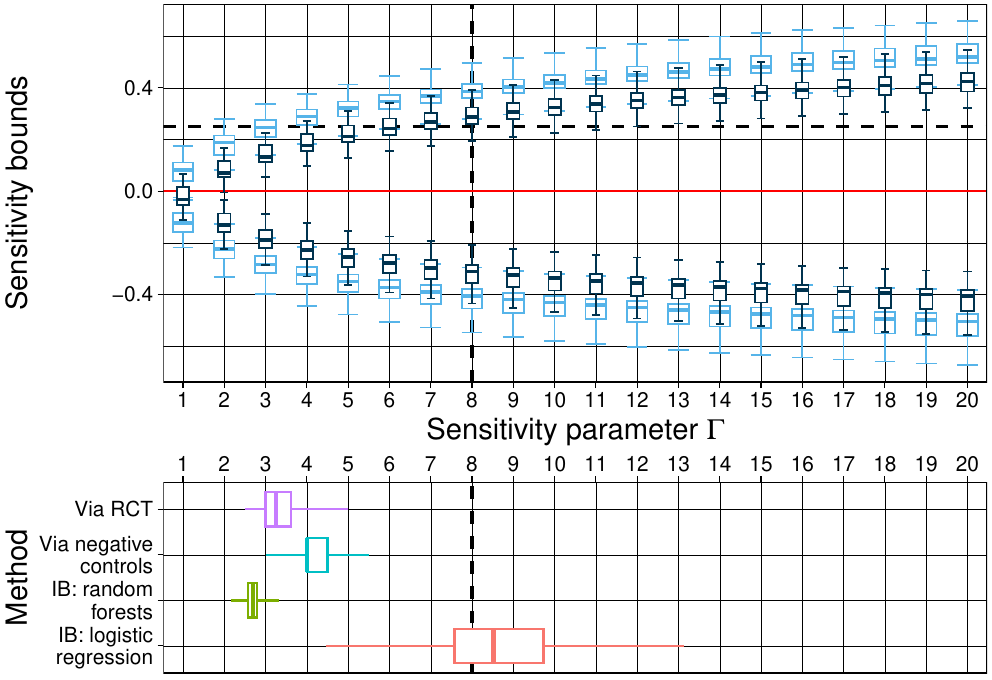}
    \caption{Sensitivity analysis n°10. IB: Informal Benchmarking.}
    \label{fig:sa_10}
\end{figure}

\begin{figure}
    \centering
    \includegraphics[width=\linewidth]{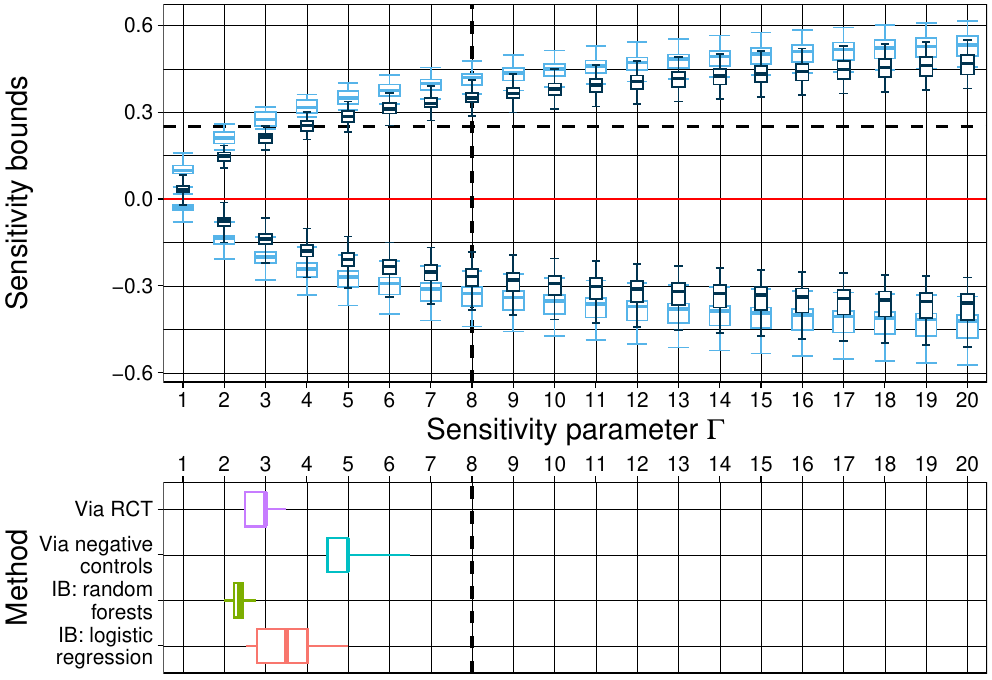}
    \caption{Sensitivity analysis n°11. IB: Informal Benchmarking.}
    \label{fig:sa_11}
\end{figure}

\begin{figure}
    \centering
    \includegraphics[width=\linewidth]{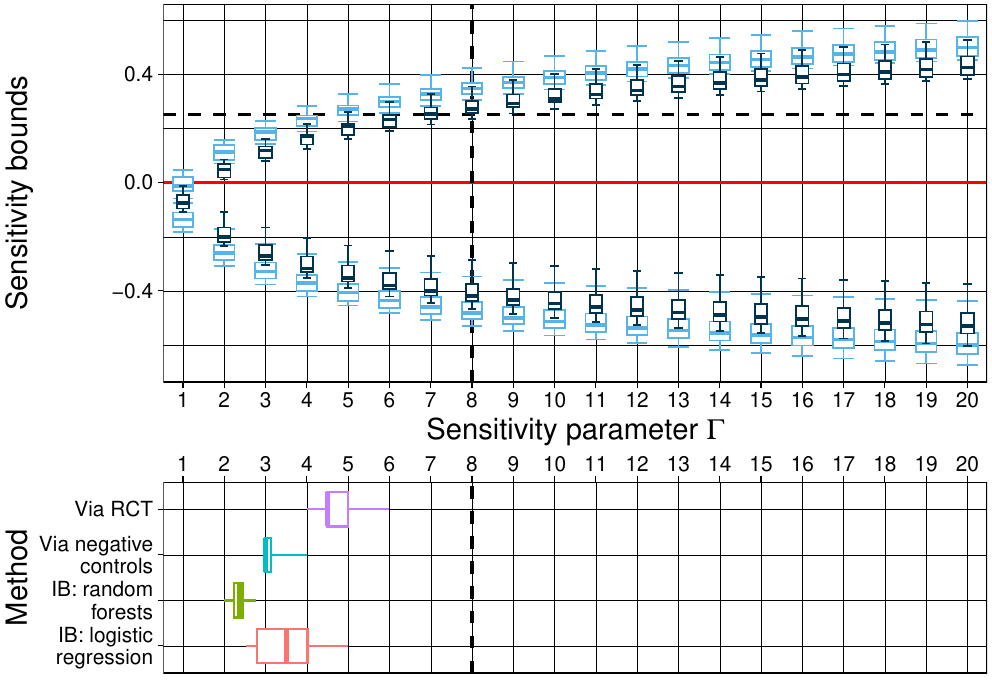}
    \caption{Sensitivity analysis n°12. IB: Informal Benchmarking.}
    \label{fig:sa_12}
\end{figure}

\begin{figure}
    \centering
    \includegraphics[width=\linewidth]{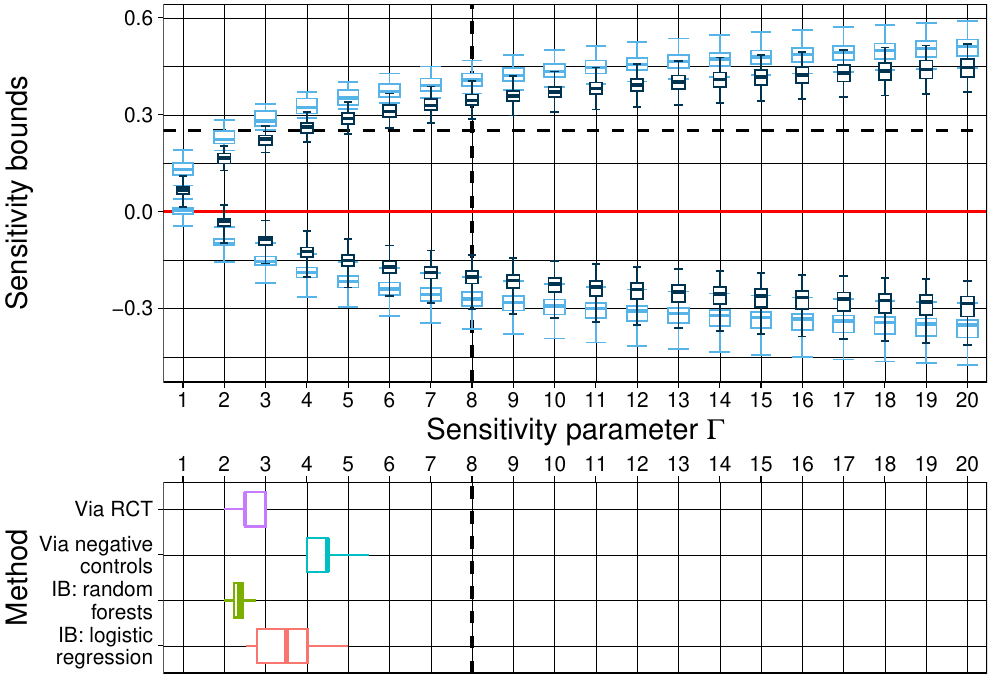}
    \caption{Sensitivity analysis n°13. IB: Informal Benchmarking.}
    \label{fig:sa_13}
\end{figure}

\begin{figure}
    \centering
    \includegraphics[width=\linewidth]{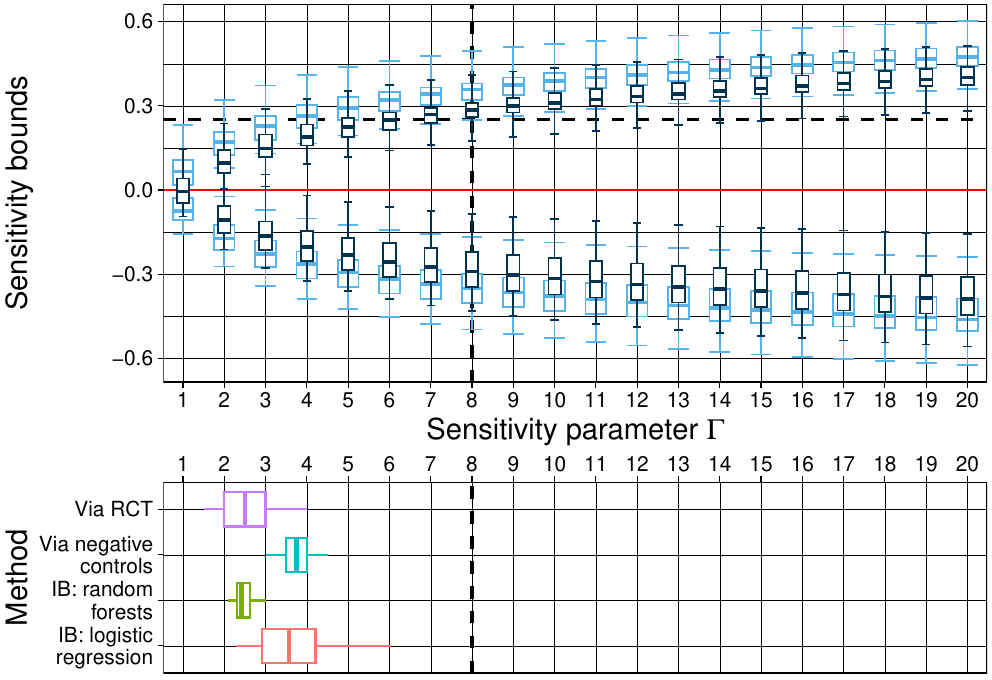}
    \caption{Sensitivity analysis n°14. IB: Informal Benchmarking.}
    \label{fig:sa_14}
\end{figure}

In this section, we represent graphically the results given in Table~\ref{tab:estimated_gamma}. In Figures~\ref{fig:sa_2} to \ref{fig:sa_14}, the sensitivity bounds for the ATE as a function of the sensitivity parameter $\Gamma$ are represented above. As 20 Monte-Carlo samples were used, we represent the bounds with boxplots. Boxplots in light blue correspond to the CIs, whereas boxplots in dark blue correspond to the PEIs. The red horizontal line indicates the null effect. The dotted horizontal line is the true ATE, $\mathrm{ATE}^\star = 0.25$, and the dotted vertical line is the true sensitivity parameter $\Gamma^\star = 8$ to estimate. Below, for each method, we used boxplots to show the estimated sensitivity parameters on the same 20 Monte-Carlo samples. In this representation, the boxplots include the outliers, i.e.\ they go from the minimum to the maximum estimated value.

Table~\ref{tab:robustness_complete} completes Table~\ref{tab:robustness} by providing percentages of inclusion of the null effect and of the true ATE in the CIs for all experiments described in Table~\ref{tab:simul_cases}.

\begin{table*}[ht]
    \centering
    \caption{Percentage of inclusion of the null effect and of the true ATE, $\mathrm{ATE}^\star$, in the CIs after performing sensitivity analyses via QB when $\Gamma = 1$, $\hat{\Gamma}$ (median), and $\Gamma^\star$ on 20 Monte-Carlo samples for each experiment and method in Table~\ref{tab:simul_cases}.}
    \label{tab:robustness_complete}
    \begin{tabular}{|c|cccccc|cccccc|}
        \hline
        \textbf{Exp.} & \multicolumn{6}{|c|}{\textbf{Null effect}} & \multicolumn{6}{|c|}{$\mathrm{\mathbf{ATE}}^\star$} \\
        \textbf{\#} & $\Gamma = 1$ & $\hat{\Gamma}_\mathrm{IB,\,log}$ & $\hat{\Gamma}_\mathrm{IB,\,rf}$ & $\hat{\Gamma}_\mathrm{LB}^\mathrm{RCT}$ & $\hat{\Gamma}_\mathrm{LB}^\mathrm{NC}$ & $\Gamma = \Gamma^\star$ & $\Gamma = 1$ & $\hat{\Gamma}_\mathrm{IB,\,log}$ & $\hat{\Gamma}_\mathrm{IB,\,rf}$ & $\hat{\Gamma}_\mathrm{LB}^\mathrm{RCT}$ & $\hat{\Gamma}_\mathrm{LB}^\mathrm{NC}$ & $\Gamma = \Gamma^\star$ \\ 
        \hline
        \hline
        1 & 85 & 100 & 100 & 100 & 100 & 100 & 0 & 55 & 15 & 55 & 55 & 95 \\
        2 & 0 & 10 & 5 & 0 & 0 & 40 & 90 & 100 & 100 & 90 & 95 & 100 \\
        3 & 0 & 10 & 0 & 100 & 100 & 100 & 0 & 0 & 0 & 60 & 95 & 90 \\
        \hline
        4 & 0 & 10 & 5 & 0 & 0 & 35 & 80 & 100 & 100 & 90 & 95 & 100 \\
        5 & 0 & 10 & 0 & 100 & 100 & 100 & 0 & 0 & 0 & 55 & 95 & 90 \\
        6 & 55 & 100 & 100 & 100 & 100 & 100 & 0 & 30 & 5 & 5 & 95 & 95 \\
        \hline
        7 & 65 & 100 & 100 & 100 & 100 & 100 & 5 & 90 & 45 & 60 & 80 & 100 \\
        8 & 55 & 100 & 100 & 100 & 100 & 100 & 0 & 10 & 5 & 60 & 50 & 100 \\
        9 & 85 & 100 & 95 & 100 & 100 & 100 & 10 & 70 & 15 & 60 & 90 & 100 \\
        10 & 95 & 100 & 100 & 100 & 100 & 100 & 0 & 100 & 25 & 55 & 75 & 100 \\
        \hline
        11 & 90 & 100 & 100 & 100 & 100 & 100 & 0 & 100 & 40 & 85 & 100 & 100 \\
        12 & 40 & 100 & 100 & 100 & 100 & 100 & 0 & 5 & 0 & 55 & 0 & 100 \\
        13 & 45 & 100 & 100 & 100 & 100 & 100 & 0 & 100 & 45 & 55 & 100 & 100 \\
        \hline
        14 & 85 & 100 & 100 & 100 & 100 & 100 & 0 & 45 & 15 & 20 & 55 & 100 \\
        \hline
    \end{tabular}
\end{table*}

\subsection{Libraries and Licenses} \label{app:libraries_licenses}

Libraries from Table~\ref{tab:libraries_licenses} are provided for reproducibility purpose. Libraries from Table~\ref{tab:libraries_licenses_implementation} were used in the proposed \texttt{R} implementation.

\begin{table*}[h]
    \centering
    \caption{Libraries and corresponding licenses for reproducibility of the experimental results.}
    \label{tab:libraries_licenses}
    \begin{tabular}{|c|c|c|c|c|}
        \hline
        \textbf{Library} & \textbf{Authors} & \textbf{Version} & \textbf{License} & \textbf{Purpose} \\
        \hline
        \hline
        \texttt{quantreg} & \textcite{quantreg} & 5.97 & GPL-2 / GPL-3 & Quantile regression \\
        \texttt{grf} & \textcite{grf} & 2.3.2 & GPL-3 & Random forests \\
        \hline
        \texttt{doSNOW} & \textcite{doSNOW} & 1.0.20 & GPL-2 & Parallel computing \\
        \texttt{foreach} & \textcite{foreach} & 1.5.2 & Apache License 2.0 & Parallel computing \\
        \texttt{ggplot2} & \textcite{ggplot2} & 3.4.4 & MIT & Plots \\
        \texttt{latex2exp} & \textcite{latex2exp} & 0.9.6 & MIT & Plots \\
        \texttt{cowplot} & \textcite{cowplot} & 1.1.2 & GPL-2 & Plots \\
        \hline
    \end{tabular}
\end{table*}

\begin{table*}[h]
    \centering
    \caption{Libraries and corresponding licenses used in the provided implementation.}
    \label{tab:libraries_licenses_implementation}
    \begin{tabular}{|c|c|c|c|c|}
        \hline
        \textbf{Library} & \textbf{Authors} & \textbf{Version} & \textbf{License} & \textbf{Purpose} \\
        \hline
        \hline
        \texttt{foreach} & \textcite{foreach} & 1.5.2 & Apache License 2.0 & Parallel computing \\
        \texttt{ggplot2} & \textcite{ggplot2} & 3.4.4 & MIT & Plots \\
        \texttt{latex2exp} & \textcite{latex2exp} & 0.9.6 & MIT & Plots \\
        \hline
    \end{tabular}
\end{table*}

\end{document}